%% file: colm2026_conference.tex
\documentclass{article} 
\usepackage[final]{colm2026_conference}

\usepackage{microtype}
\usepackage{hyperref}
\usepackage{url}
\usepackage{booktabs}
\usepackage{graphicx}
\usepackage[table]{xcolor}
\usepackage{makecell}
\usepackage{subcaption}
\usepackage{nicematrix}
\usepackage[most]{tcolorbox}
\usepackage{fvextra}

\usepackage{lineno}

\definecolor{darkblue}{rgb}{0, 0, 0.5}
\hypersetup{colorlinks=true, citecolor=darkblue, linkcolor=darkblue, urlcolor=darkblue}

\newtcolorbox[list inside=prompt,auto counter,number within=section]{prompt}[1][]{
    colbacktitle=black!60,
    fonttitle=\small,
    coltitle=white,
    fontupper=\footnotesize,
    boxsep=3pt,
    left=0pt,
    right=0pt,
    top=0pt,
    bottom=0pt,
    boxrule=1pt,
    #1,
    breakable,              
}

\newcommand{\camready}[1]{{#1}}

\title{Large Language Models Align with the Human Brain during Creative Thinking}

\newcommand{\authsep}{\hspace{0.8em}}

\author{Mete Ismayilzada\textsuperscript{1,2} 
\authsep Simone A. Luchini\textsuperscript{4,5}
\authsep Abdulkadir Gokce\textsuperscript{1} 
\authsep Badr AlKhamissi\textsuperscript{1} \\
\textbf{Antoine Bosselut}\textsuperscript{1} 
\authsep \textbf{Antonio Laverghetta Jr.}\textsuperscript{3} 
\authsep \textbf{Lonneke van der Plas}\textsuperscript{\textdagger2} \\
\textbf{Roger E. Beaty}\textsuperscript{\textdagger5} \\
\textsuperscript{1}EPFL, 
\textsuperscript{2}Università della Svizzera italiana (USI), 
\textsuperscript{3}Wesleyan University,\\
\textsuperscript{4}Paris Brain Institute (ICM),
\textsuperscript{5}Pennsylvania State University
}

\begingroup
\footnotetext{Equal supervision}
\endgroup

\begin{document}

\ifcolmsubmission
\linenumbers
\fi

\maketitle

\begin{abstract}
\input{00_abstract}
\end{abstract}

\section{Introduction}
\input{01_introduction}
\section{Related Work}
\input{02_related_work}

\section{Methodology}
\input{03_methodology}

\section{Experimental Setup}
\input{04_experiments}

\section{Results}
\input{05_results}

\section{Discussion}
\input{06_discussion}
\section*{Conclusion}
\input{07_conclusion}



\section*{Acknowledgements}
\input{10_acknowledgements}

\bibliography{colm2026_conference}
\bibliographystyle{colm2026_conference}

\newpage
\appendix

\input{11_appendix}

\end{document}

%% file: 00_abstract.tex
Creative thinking is a fundamental aspect of human cognition, and divergent thinking—the capacity to generate novel and varied ideas—is widely regarded as its core generative engine. Large language models (LLMs) have recently demonstrated impressive performance on divergent thinking tests and prior work has shown that models with higher task performance tend to be more aligned to human brain activity. However, existing brain-LLM alignment studies have focused on passive, non-creative tasks. Here, we explore brain alignment during creative thinking using fMRI data from 170 participants performing the Alternate Uses Task (AUT). We extract representations from LLMs varying in size (270M–72B) and measure alignment to brain responses via Representational Similarity Analysis (RSA), targeting the creativity-related default mode and frontoparietal networks. We find that brain-LLM alignment is positively associated with model size (default mode network only) and with idea originality (both networks), with these relationships clearest early in the creative process. We further find that post-training objectives are associated with functionally selective differences in alignment: a creativity-optimized \textsc{Llama-3.1-8B-Instruct} retains alignment with high-creativity neural responses while lacking the positive low-creativity alignment present in other variants, but a reasoning-trained variant shows the opposite asymmetry, with negative alignment to high-creativity responses. Together, these results suggest that brain alignment offers an informative lens on how post-training relates to the neural geometry of human creative thought.

%% file: 01_introduction.tex
Creative thinking stands as one of the most distinctive and complex capacities of the human mind, underpinning scientific discovery, artistic expression, and everyday problem-solving \citep{torrance1974torrance, guilford1967nature, duncker1948problem}. Within the broader landscape of human cognition, creativity is commonly decomposed into two complementary modes: \textit{convergent thinking}, which involves narrowing down multiple possibilities to arrive at a single, well-defined solution, and \textit{divergent thinking}, which involves generating a wide range of novel ideas or associations from a single starting point \citep{guilford1967nature, mednick1962associative}. While both modes contribute to creative cognition, divergent thinking is widely considered the more foundational driver of creativity, as it captures the generative, open-ended exploration that gives rise to original thought \citep{runco2012divergent}. Its study has therefore attracted greater attention in both cognitive psychology and neuroscience, with researchers seeking to understand not only what makes humans capable of such flexible ideation, but also what neural substrates and cognitive mechanisms support it \citep{beaty2018robust, abraham2013promises, beaty2016creative}.

\begin{figure}[t]
\includegraphics[width=\linewidth,trim={1cm 1cm 0cm 1cm}]{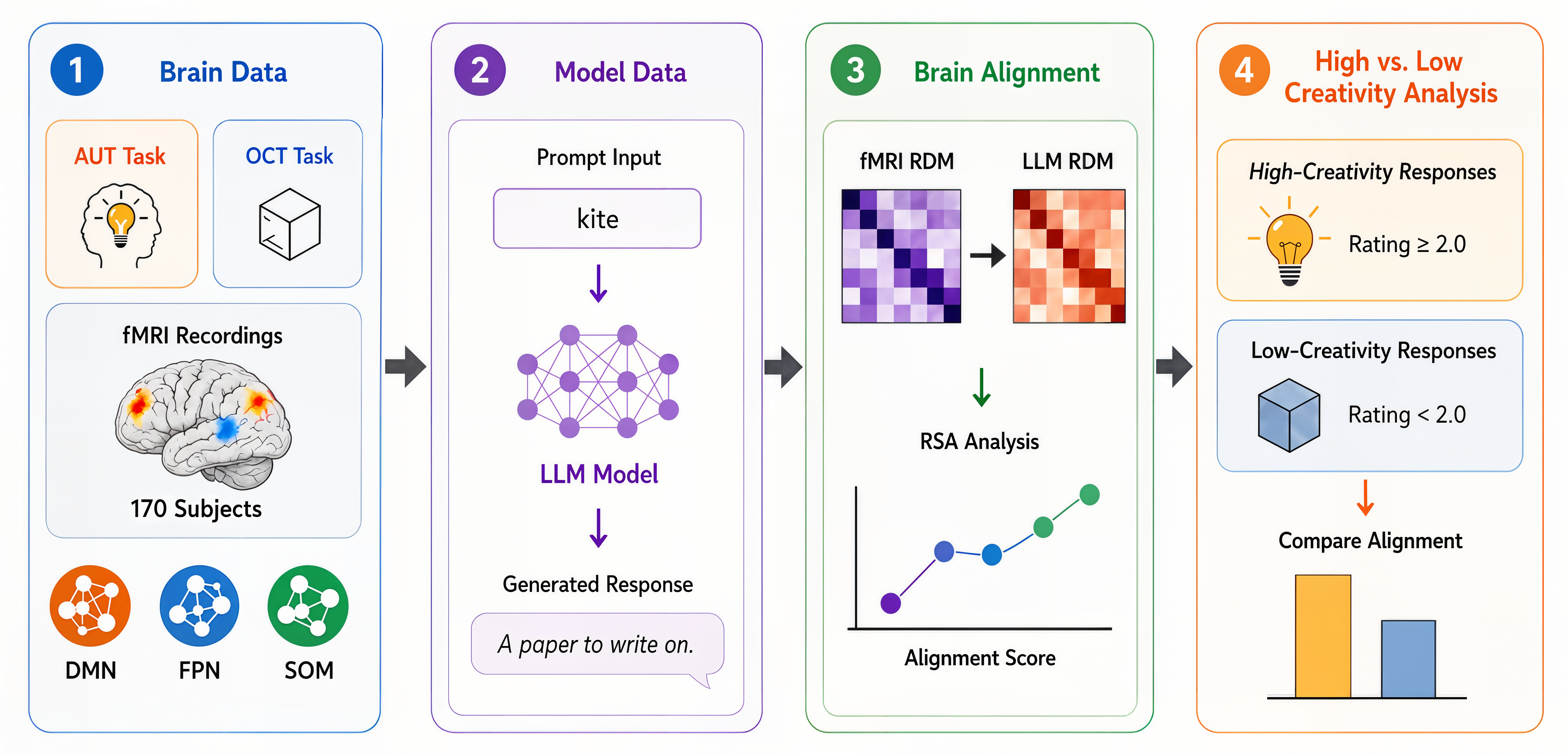}
\centering
\caption{\textbf{Our high-level brain alignment methodology.} Participants perform a creative task (AUT) and a matched non-creative control (OCT) during fMRI, while LLMs receive the same instructions and stimuli. We extract layer-wise activations at the prompt stage and after response generation, and compute alignment as the rank correlation between per-subject fMRI and LLM representational dissimilarity matrices (RDMs). Alignment is evaluated in the default mode (DMN) and frontoparietal (FPN) networks, plus the somatomotor network (SOM) as a control, and then compared between high- and low-creativity human responses.}
\label{fig:method}
\end{figure}

In recent years, large language models (LLMs) have rapidly expanded their reach across a wide range of cognitive tasks, with divergent thinking among the abilities they now exhibit \citep{ismayilzada2024creativity, zhao2023survey}. Models have shown impressive results on established creativity benchmarks such as the Alternative Uses Test (AUT), which measures the ability to generate diverse and unconventional uses for everyday objects, in some cases reaching or exceeding average human performance. \citep{goes2023pushing, stevenson2022putting, bellemare2026divergent}. Alongside these behavioral results, a growing body of work has established that LLMs with higher task performance tend to exhibit greater alignment with the human language system \citep{schrimpf2021neural, goldstein2022shared, caucheteux2022brains}. However, prior work has predominantly focused on passive language processing tasks, in which participants read or listen to naturalistic text \citep{aw2023instruction, alkhamissi2025language, lebel2023natural} or, recently, simple word association \citep{kwon2024assessing} and abstract reasoning tasks \citep{pinier2025large}. Divergent thinking, however, is a higher-order cognitive function often employing a broad and distributed network of brain regions well beyond the classical language system \citep{liu2024neural}. Whether more creative LLMs are correspondingly more aligned to the neural representations underlying human creative cognition, therefore, remains an open and compelling question, and one that existing brain-LLM alignment frameworks have yet to address.

In this work, we take a first step toward addressing this question by systematically investigating brain-LLM alignment in the context of creative thinking using the AUT task. We leverage fMRI data collected from \citet{beaty2018robust}, in which participants performed both a creative task, namely the AUT, and a matched non-creative control task, namely the Object Characteristics Task (OCT), which involves generating typical characteristics of everyday objects \citep{beaty2018robust}. Similarly, we extract model representations to the same stimuli in these tasks from a range of LLMs and measure their alignment to the corresponding brain responses using Representational Similarity Analysis (RSA) \citep{kriegeskorte2008representational}. On the neural side, we examine alignment across several canonical regions of interest involved in creative thinking, namely, the Default Mode Network (DMN) and the Frontoparietal Network (FPN) \citep{luchini2025role, liu2024neural}.
\camready{
Our results reveal that brain alignment during creative thinking is stage-dependent: alignment is positively associated with model size and creative task performance when measured at the prompt stage, but this relationship weakens once models generate responses. We further find that post-training objectives are associated with functionally selective differences in alignment, with creativity-optimized and reasoning-chain training exhibiting opposite asymmetries relative to the neural geometry of human creative thought. To the best of our knowledge, this is the first study to directly examine brain-LLM alignment in the context of an active creative thinking task, and we hope our findings encourage deeper investigation into the neural plausibility of creative cognition in language models. 

}

%% file: 02_related_work.tex
\paragraph{Brain-LLM Alignment} 
A growing body of work has established that the internal representations of large language models (LLMs) exhibit meaningful similarity to neural activity recorded during vision \citep{yamins2014performance, khaligh2014deep, cichy2016comparison, gokce2024scaling}, auditory \citep{koumura2023human, kell2018task, tuckute2023many}, and human language processing \citep{schrimpf2021neural, goldstein2022shared, caucheteux2022brains, tuckute2024driving, binz2025foundation} \camready{as well as during comprehension of computer programs \citep{srikant2022convergent}}. These studies have demonstrated that LLM representations can linearly predict brain responses to the same linguistic stimuli across a variety of neuroimaging paradigms, including reading naturalistic stories and sentences. More recent work has extended these findings by examining how specific model properties relate to brain alignment, finding strong correlations with model size and world knowledge, and demonstrating that instruction-tuning further improves alignment beyond what is achieved by pretraining alone \citep{aw2023instruction}. Parallel efforts have examined how brain alignment evolves over the course of LLM training
\citep{alkhamissi2025language}. However, it is notable that prior alignment work has been conducted largely using passive language paradigms in which participants read or listen to naturalistic text. Only recently, some works have examined the brain alignment in active cognitive reasoning tasks such as novel word association generation \citep{kwon2024assessing} and abstract reasoning \citep{pinier2025large}. Closest to our work is \citet{kwon2024assessing}, which has evaluated alignment across two task conditions: A creative condition where participants had to generate original associates to a target word, and a non-creative condition where participants would generate appropriate associations. Findings have revealed that LLM activations align with human brain activity during creative word associations, but not during appropriate word associations, indicating that complex language production is crucial to alignment. However, to the best of our knowledge, no prior work has examined brain alignment during active creative thinking. We address this gap by studying brain–LLM alignment in an active creative task (AUT task), extending emerging efforts that probe alignment in settings requiring cognitive flexibility.

\paragraph{LLMs and Creative Thinking}
Beyond passive language comprehension, LLMs have recently been shown to exhibit impressive performance on tasks requiring creative and divergent thinking \citep{ismayilzada2024creativity}. Studies evaluating models on creativity tasks such as the AUT have found that frontier models can generate responses that are rated as highly original, sometimes matching or exceeding average human performance \citep{goes2023pushing, stevenson2022putting, bellemare2026divergent}. Prior work has proposed various techniques to improve creative thinking and problem-solving in LLMs, such as creative prompting \citep{nguyen2025divergent, tian2024macgyver, lu2025benchmarking} and fine-tuning/preference optimization strategies \citep{puerto2025fine, ismayilzada-etal-2025-creative}. Overall, progress in this field has mainly focused on evaluating and improving divergent creative thinking in LLMs; however, the question of whether high performance on creative thinking tasks is accompanied by representational similarity to the neural systems that support human creativity remains unexplored. We aim to investigate this question of brain-LLM alignment in the present work.

%% file: 03_methodology.tex
\subsection{Brain Data}
\label{sec:brain-data}
We use the fMRI recordings of human subjects while they solve AUT and OCT tasks from \citet{beaty2018robust}. This brain dataset contains fMRI recordings from $170$ healthy subjects. Subjects were given the following instructions, respectively, for each task, and then were shown a stimulus one at a time. 

\begin{prompt}[title={AUT Instructions}, label=prompt:aut_instructions]
\begin{Verbatim}[breaklines, breaksymbol={}]
Think of creative uses for objects. Report the most original idea. Keep thinking of ideas during the thinking period. Quality is more important than quantity.
\end{Verbatim}
\end{prompt}

\begin{prompt}[title={OCT Instructions}, label=prompt:oct_instructions]
\begin{Verbatim}[breaklines, breaksymbol={}]
Think of the physical properties of objects. Report the most prominent characteristic. Keep thinking of ideas during the thinking period.
\end{Verbatim}
\end{prompt}

After a brief period of thinking, their responses, along with fMRI recordings, were collected for each stimulus. In total, subjects solved both tasks for $46$ stimuli, where each subject was exposed to on average $23$ stimuli for each task. For example, given the stimulus \textit{``kite''}, one subject responded with \textit{``a paper to write on''} and \textit{``rectangular''} respectively for AUT and OCT tasks. 

We first perform data cleaning based on the subject responses and, as a result, remove data for $8$ participants: those who failed to follow the task instructions, those who failed to provide a response, and those whose preprocessed brain data had a different voxel dimensionality from the remaining subjects, which prevented their inclusion in the group-level RDM analyses. Next, we apply standard fMRI data preprocessing (e.g., removing confounders, detrending, standardizing, and signal filtering) using \texttt{nilearn}\footnote{https://nilearn.github.io} library, which results in ``clean'' (but raw) fMRI brain activations \citep{matthews2004functional}. Rather than using these raw activations, we follow standard practice and estimate single-trial beta maps using a General Linear Model (GLM) \citep{draper1998applied}, and conduct all subsequent brain-LLM alignment analyses on these estimated neural response patterns. We extract brain data from across three functional brain networks using the \citet{yeo2011organization} parcellation: the Default Mode Network (DMN), the Frontoparietal Network (FPN), and the Somatomotor Network (SOM). The DMN and FPN were selected given their well-established roles in creative cognition \citep{beaty2016creative, luchini2025role, liu2024neural}, while the SOM was selected given its limited involvement in higher-order cognitive processes, making it a suitable control network.

\subsection{Model Data}
\label{sec:model-data}
We provide the same instructions\footnote{We only replace the word ``objects'' with the stimulus name since we evaluate language-only models.} and the stimuli to an LLM and extract its intermediate layer activations as the model representations of the stimuli. 
\camready{
Matching the model's prompt to the instruction given to human participants is important in brain-LLM alignment studies, since alignment is computed against the neural responses elicited by that instruction; we therefore hold prompt content fixed across conditions in our primary analyses. Task-dependency is instead tested through the OCT control, which matches the AUT in stimulus, structure, and language, differing only in whether the goal is creative generation or object-property reporting \citep{beaty2018robust}. We note that the OCT is a more stringent control than a non-language placeholder prompt would be, as it shares substantially more surface structure with the AUT ($42\%$ token overlap).
}
Prior brain-LLM alignment studies typically extract model representations for the input prompt alone, which is sufficient for passive language tasks where no generation occurs. However, this is not sufficient for our task, which involves active generation, so we extract representations at both the prompt stage and after the model has generated a response, allowing us to capture the full generative process that more closely mirrors the demands of the AUT task. 
\camready{We note that this setup is similar to \citet{kwon2024assessing} and \citet{oota2025correlating}, except that we examine alignment at both the prompt and the generation stage, while they consider the generation stage only.}

\begin{figure}[t]
\includegraphics[width=\linewidth]{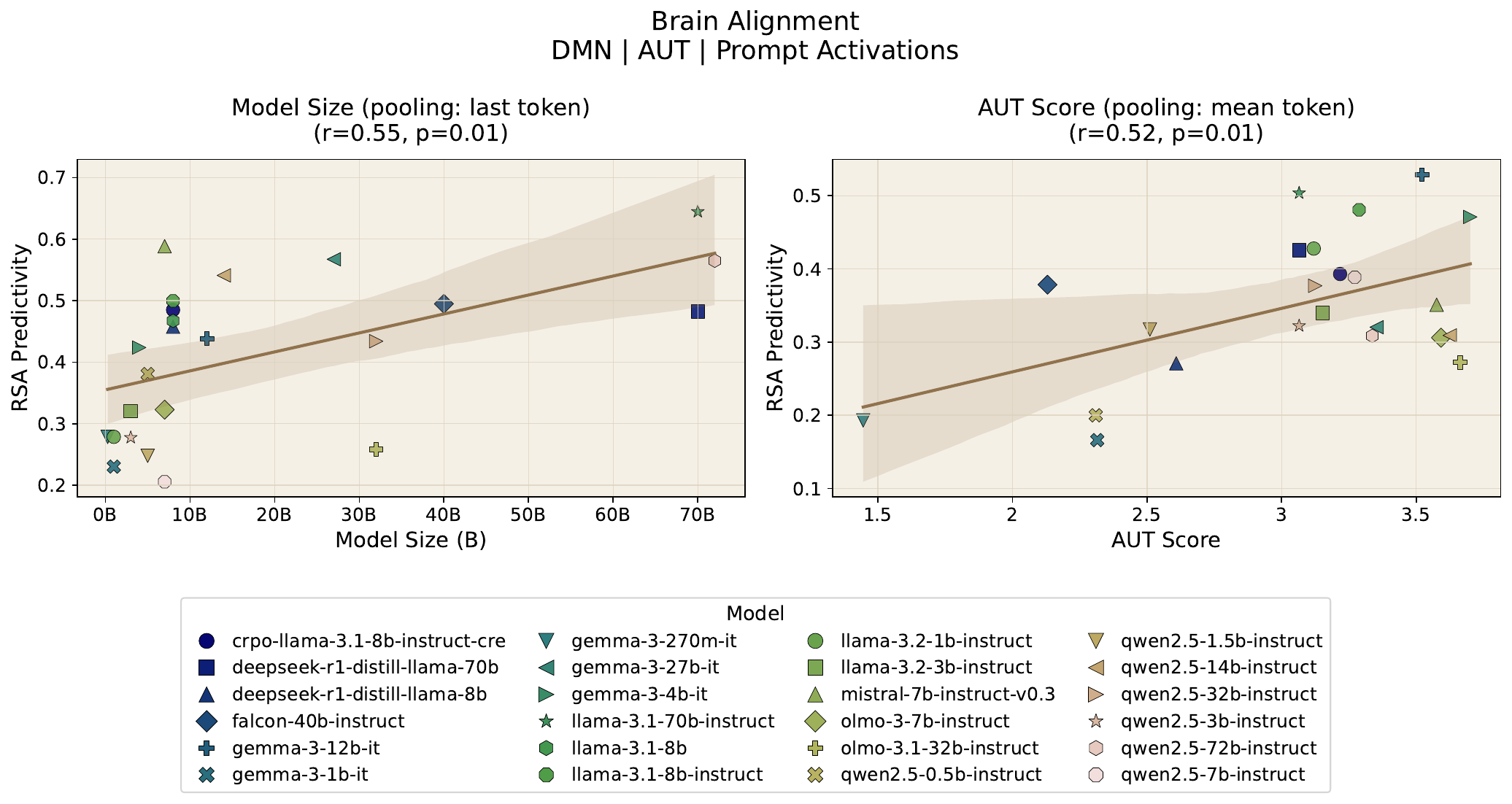}
\centering
\caption{\textbf{Default Mode Network (DMN) AUT brain alignment results by model size and task performance using prompt-only model activations.} $r$ and $p$ correspond to the Pearson correlation coefficient and p-value.}
\label{fig:alignment-aut-yeo-dmn-prompt}
\end{figure}

\subsection{Brain Alignment}
\label{sec:brain-alignment}
Brain alignment refers to the degree of similarity between the internal representations of an LLM and human neural activity \citep{schrimpf2018brain}. We evaluate alignment using Representational Similarity Analysis (RSA) \citep{kriegeskorte2008representational} \camready{following the pipeline conventions of \citet{nili2014toolbox}}. We note that while some past works have used linear predictivity (i.e., fitting a regression model to predict voxel-wise fMRI responses) \citep{schrimpf2018brain,schrimpf2021neural}, this metric typically requires a large sample of stimuli responses to be effective. While our data is large in terms of the number of subjects, the number of stimulus samples per subject is insufficient to train a meaningful regression model. Therefore, we opt not to use it in our alignment computation \camready{(full details in Appendix \ref{sec:rsa-predictivity}).}

We compute alignment at the level of individual subjects and report the median across subjects. To account for the inherent noise in fMRI measurements, we estimate a noise ceiling per subject, which represents the maximum alignment score achievable given the reliability of the neural data. Raw alignment scores are then divided by their corresponding noise ceilings, yielding noise-ceiling-normalized scores that allow for more meaningful comparisons across brain regions and models. We evaluate alignment at every layer of each LLM and take the maximum value across layers as the model's overall brain alignment score, following \citet{schrimpf2018brain}. This layer-selection procedure is motivated by the observation that different layers of a network capture different levels of representational abstraction and allow each model to be assessed at its most brain-aligned processing stage \citep{elhage2021mathematical, tenney2019bert, jawahar-etal-2019-bert}.

\subsection{Post-training Objectives and Creativity}
\label{sec:high-low}
Prior work has shown that instruction tuning improves LLM alignment with the human language system \citep{aw2023instruction}, raising the question of whether different post-training approaches, such as those targeting creativity or reasoning, can selectively modulate alignment with neural responses associated with creative ideation. To explore this, we partition the human brain data into high- and low-creativity populations based on the mean human creativity rating assigned to each response. For each AUT stimulus, participant responses were rated on their creativity by four independent human raters ($ICC =0.75$, good inter-rater agreement), and we followed a standard practice to aggregate these scores by averaging to yield a single creativity score per response \citep{organisciak2023beyond, amabile1982social}. 
\camready{Since responses are rated on a scale of 1 to 5 and the mean rating distribution is right-skewed (Appendix Figure \ref{fig:dist-mean-rating}), the choice of cutoff separating low- from high-creativity responses is not uniquely determined. We therefore report results across three cutoffs that roughly balance split sizes: $1.75$, $2.0$, and $2.25$, where low- and high-creativity responses are those with $rating < cutoff$ and $rating \ge cutoff$ respectively. We use $2.0$ in the main text and report the remaining two as robustness checks (Appendix \ref{sec:detailed-high-low}).}

\begin{figure}[t]
\includegraphics[width=\linewidth]{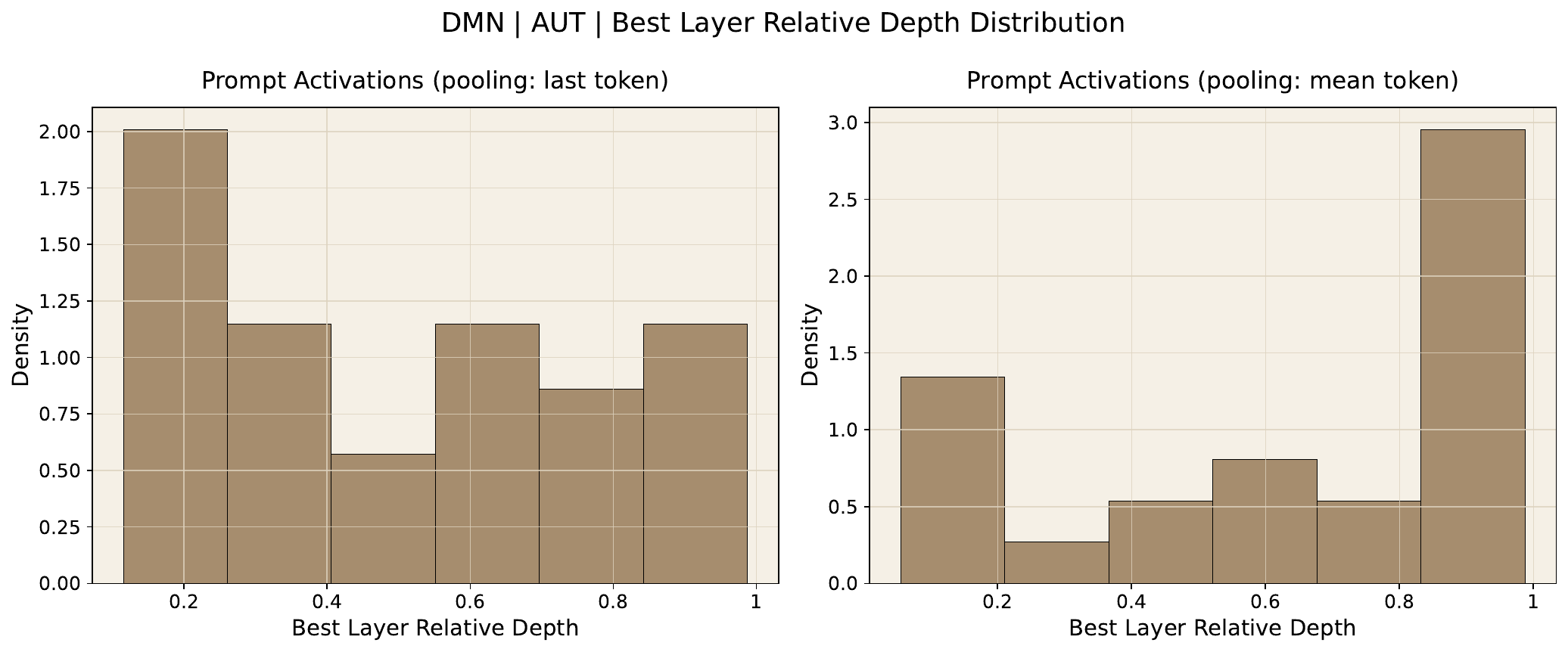}
\centering
\caption{The distribution of best model layer (measured by relative depth) for DMN AUT brain alignment on prompt-only model activations. }
\label{fig:alignment-best-layer-dist-aut-yeo-dmn}
\end{figure}


%% file: 04_experiments.tex
\subsection{Models}
\label{sec:models}
We consider several open-source instruction following models varying in their size: \textsc{Gemma-3-\{270M, 1B, 4B, 12B, 27B\}-it} \citep{kamath2025gemma}, \textsc{Llama-3.1-\{8B, 70B\}-Instruct}, \textsc{Llama-3.2-\{1B, 3B\}-Instruct} \citep{grattafiori2024llama}, \textsc{Olmo-3-7B-Instruct}, \textsc{Olmo-3.1-32B-Instruct} \citep{olmo2025olmo3}, \textsc{Falcon-40B-Instruct} \citep{almazrouei2023falcon}, \textsc{DeepSeek-R1-Distill-Llama-70B} \citep{deepseekai2025deepseekr1incentivizingreasoningcapability}, \textsc{Mistral-7B-Instruct-v0.3} \citep{jiang2023mistral} and \textsc{Qwen2.5-\{0.5B, 1.5B, 3B, 14B, 32B, 72B\}-Instruct} \citep{qwen2.5, qwen2}. 
Additionally, to study the alignment difference between high vs. low creativity responses and the effect of post-training techniques, we consider several variants of \textsc{Llama-3.1-8B-Instruct}: \textsc{Llama-3.1-8B} which is the base pretrained model, \textsc{CrPO-Llama-3.1-8B-Instruct-cre} that has been trained to align with human creative preferences and optimized for multiple dimensions of
creativity, namely, novelty, surprise, diversity, and quality
\citep{ismayilzada-etal-2025-creative}, and \textsc{DeepSeek-R1-Distill-Llama-8B} that has been finetuned with the reasoning outputs of \textsc{DeepSeek-R1}
\citep{deepseekai2025deepseekr1incentivizingreasoningcapability}.

\camready{
For the effects of the post-training objective analysis, to isolate the contribution of the specific creativity optimization technique from that of the training data and the fine-tuning paradigm, we additionally consider two ablations that share both the base model and the training data with \textsc{CrPO-Llama-3.1-8B-Instruct-cre}: \textsc{CrPO-SFT-Llama-3.1-8B-Instruct}, which is \textsc{Llama-3.1-8B-Instruct} supervised-finetuned on the MuCE creativity dataset, and \textsc{CrPO-DPO-Llama-3.1-8B-Instruct}, which is the same base preference-tuned on MuCE using vanilla DPO \citep{ismayilzada-etal-2025-creative}.

Finally, to test whether the observed post-training effects generalize beyond a single base model family, we replicate the post-training analysis on a set of \textsc{Qwen2.5} models: \textsc{Qwen2.5-7B} (base) and \textsc{Qwen2.5-7B-Instruct} serve as matched within-family baselines, \textsc{Qwen2.5-Math-7B} is finetuned on a mathematics corpus \citep{yang2024qwen25math}, and \textsc{DeepSeek-R1-Distill-Qwen-7B} further finetunes \textsc{Qwen2.5-Math-7B} with \textsc{DeepSeek-R1} reasoning traces \citep{deepseekai2025deepseekr1incentivizingreasoningcapability}. 
}

To ensure high creativity and diversity of outputs, we sample four generations for each prompt with \textit{temperature=0.7} and \textit{top\_p=0.95}.

\subsection{Alignment Factors}
To analyze the factors driving brain–LLM alignment, we compute the Pearson correlation between alignment scores and two model attributes: parameter count and performance on the AUT task. 
For evaluating AUT performance, we employ the \textsc{Gemini-3-Flash} \citep{gemini3google} model as LLMs have been shown to accurately predict human creativity on the AUT \citep{saretzki2026investigating, organisciak2023beyond}.
We slightly adapt the original prompt by instructing the model to produce only a few words, ensuring that it outputs a single, most original idea without additional text. This adjustment makes the outputs more comparable to human responses, which are typically concise, and enables consistent scoring (Appendix Tables \ref{prompt:aut_instructions_scoring}, \ref{prompt:oct_instructions_scoring})

\subsection{Model Representations}
Since the model representations are extracted per token, past work has considered two main pooling strategies: extracting representations from the \textit{last token} or averaging
over all token representations (i.e., \textit{mean token}). We consider both approaches in our work and indicate the type of pooling used for each analysis.

\camready{
Because the AUT prompt instructs the model to \textit{think of} creative uses, our models produce reasoning-like chains in their responses rather than a bare answer. Outputs from \textsc{DeepSeek-R1-Distill} variants additionally contain explicit \texttt{<think>...</think>} reasoning traces. At the response stage, we extract activations from the full set of response tokens, so the resulting representations capture this reasoning content rather than the final answer alone. Generation is capped at 1024 tokens for all models; per-model generation length statistics are reported in Appendix Table \ref{tab:generation_lengths}.

While we hold prompt content fixed in our primary analyses (Section \ref{sec:model-data}), we additionally evaluate two baseline conditions that remove the task instruction: an \textit{empty-prompt} condition, in which the AUT prompt is replaced by an empty input, and a \textit{placeholder-prompt} condition, in which it is replaced by a sequence of \texttt{\#} characters matched to its token length.
}

%% file: 05_results.tex
\begin{figure}[t]
\includegraphics[width=\linewidth]{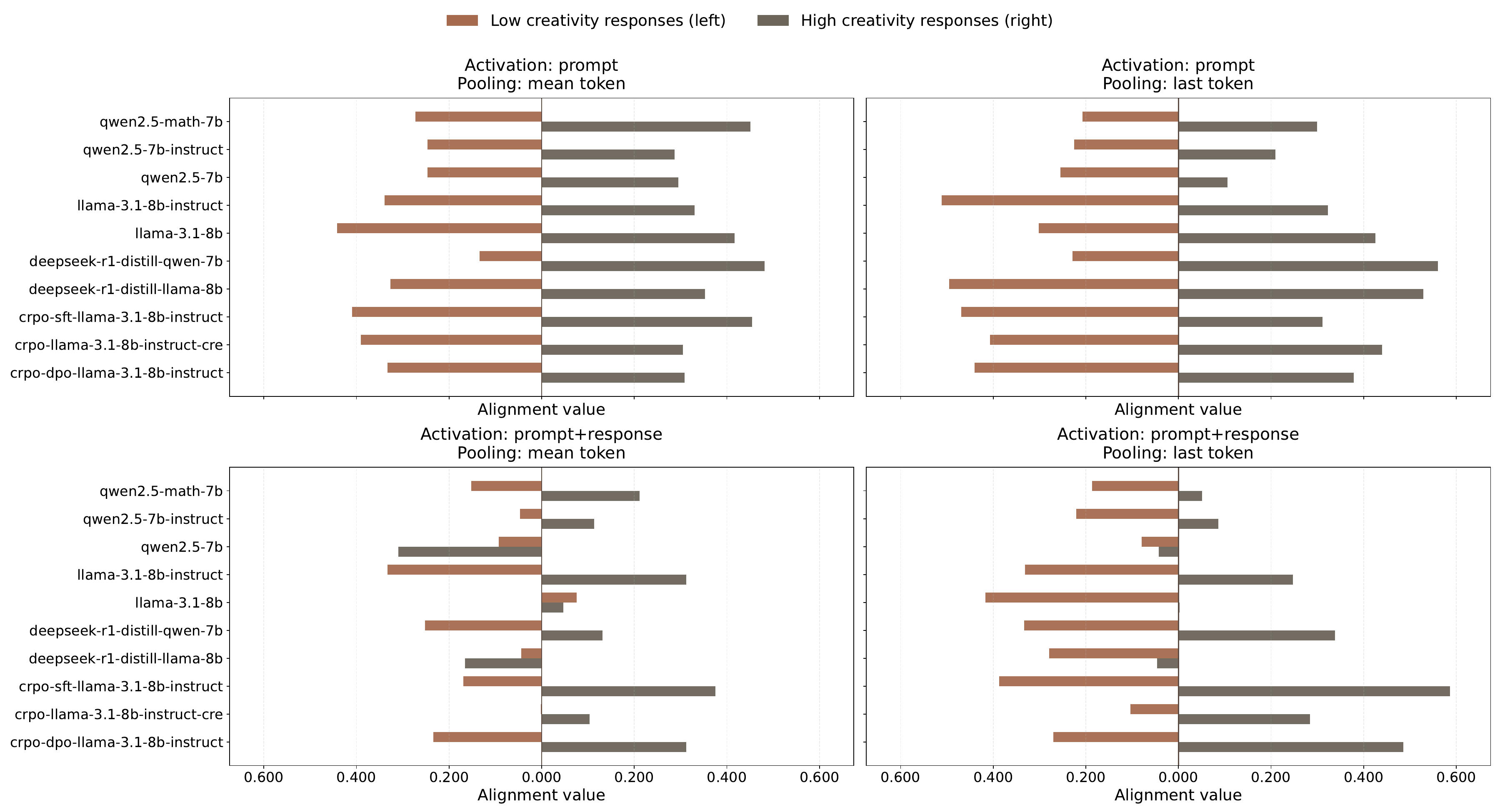}
\centering
\caption{\textbf{Default Mode Network (DMN) AUT brain alignment results for high and low creativity response populations (rating cutoff = 2.0).} The left red and right gray bars indicate positive alignment with high and low creativity responses, respectively. Flipped bars indicate negative alignment. Results for rating cutoffs $1.75$ and $2.25$ can be found in Appendix Figures \ref{fig:alignment-high-low-175} and \ref{fig:alignment-high-low-225} respectively.}
\label{fig:alignment-high-low}
\end{figure}

\paragraph{LLMs Align with the Human Brain During Creative Thinking}
Figure \ref{fig:alignment-aut-yeo-dmn-prompt} shows the brain alignment between the Default Mode Network (DMN) and model representations extracted during early cue processing — that is, after the prompt has been presented but before any response has been generated. Notably, the DMN has been directly implicated in creative cognition and is considered a core neural substrate of creative thinking \citep{luchini2025role}. We observe moderate and consistent alignment across models, and crucially, this \textbf{alignment is positively correlated with both model size} ($r=0.55, p<0.05$) and \textbf{creative task performance} as measured by the AUT score ($r=0.52, p<0.05$). This suggests that larger and more creatively capable models more closely mirror the neural representations underlying human divergent thinking during the initial stages of ideation. \camready{We note that the two relationships emerge under different pooling strategies: the size correlation is obtained with last-token pooling and the AUT-score correlation with mean-token pooling, with each attribute showing a weaker relationship under the alternative pooling (Appendix Figures \ref{fig:alignment-aut-yeo-dmn-prompt-lt}, \ref{fig:alignment-aut-yeo-dmn-prompt-mt}).} 

However, when alignment is measured using representations extracted after the model has generated its response, alignment values become more variable across models, and the correlations with both model size and task performance weaken (Appendix Figures \ref{fig:alignment-aut-yeo-dmn-resp-lt}, \ref{fig:alignment-aut-yeo-dmn-resp-mt}). This decoupling may reflect the fact that models tend to converge on similar responses regardless of scale \citep{wenger2025we, ismayilzada2024evaluating}, or alternatively, that generated responses increasingly diverge from human responses in length, structure, and quality as model scale grows. 
\camready{
We note that this weakening is observed when pooling across model families: within the Gemma-3 family, where pretraining data and architecture are held approximately fixed, the size and AUT-score correlations remain positive at the
response stage (Appendix \ref{sec:within-family}). This suggests that the loss of correlation at the response stage may partly reflect heterogeneity across families rather than a property of the generation stage itself.
}
We also replicate some of these patterns (i.e. correlation with AUT score) for the Frontoparietal Network (FPN) (Appendix Figure \ref{fig:alignment-aut-yeo-fpn-prompt-mt}). To assess whether these effects are specific to creative cognition or reflect more general properties of model or brain representations, we repeat the alignment analyses using the non-creative OCT task and a brain network not typically associated with creativity, namely the Somatomotor network (SOM) \citep{ji2019mapping}. We find that neither model size nor task performance correlates significantly with alignment in either of these conditions across pooling strategies (Appendix Figures \ref{fig:alignment-aut-yeo-som-prompt-lt}-\ref{fig:alignment-oct-yeo-dmn-mt}), providing an important double dissociation that strengthens the interpretation that the observed alignment effects are driven specifically by the conjunction of a creativity-relevant task and a creativity-relevant brain network.
\camready{
We further verify that these effects are not attributable to differences in generation length across models. Although generated response length varies across our model set (Appendix Table \ref{tab:generation_lengths}), neither mean nor median generation length correlates significantly with DMN alignment during the AUT (Appendix Figures \ref{fig:alignment-aut-yeo-dmn-mean-genlen-lt}-\ref{fig:alignment-aut-yeo-dmn-median-genlen-mt}), indicating that the alignment patterns we report do not simply track how much text each model produces. As a complementary analysis, we also compare absolute alignment levels across these conditions (Appendix \ref{sec:condition-comparisons}), which are reliably higher in the DMN than in the SOM at the prompt stage, but comparable between the AUT and OCT conditions in most configurations.

Finally, we examine the two baseline conditions that remove the task instruction (Section \ref{sec:model-data}). Neither the \textit{empty-prompt} nor the
\textit{placeholder-prompt} condition yields significant correlations between alignment and model size or task performance, across any combination of stage and token pooling (Appendix Figures \ref{fig:alignment-aut-yeo-dmn-empty-prompt-lt}-\ref{fig:alignment-aut-yeo-dmn-plch-prompt-resp-mt}). Alignment in the creativity-relevant networks therefore depends on the model receiving the same task instructions given to human participants, rather than arising from generic input-encoding effects.
}

\camready{
\paragraph{Peak Alignment Depth Depends on Token Pooling}
Different layers of a language model capture qualitatively different aspects of linguistic and cognitive processing: earlier layers tend to encode lexical and syntactic properties, while later layers are associated with more abstract, task-specific
representations \citep{tenney2019bert, jawahar-etal-2019-bert}, and prior work has linked these layer-wise differences to brain alignment directly \citep{oota2023joint, he2026far}. We therefore asked at what depth of the network alignment with creativity-relevant brain networks peaks, computing for each model the relative depth of its best-aligned layer (best layer index divided by total number of layers).

Figure \ref{fig:alignment-best-layer-dist-aut-yeo-dmn} shows that the answer depends on how token representations are pooled. Under last-token pooling, peak-alignment depths are modal in
the earliest portion of the network (relative depth $\approx 0.15$--$0.25$) and otherwise spread broadly across the remaining depths. Under mean-token pooling, the distribution is bimodal, with a pronounced mode in the final layers (relative depth $> 0.85$) and a smaller secondary mode at the earliest depths, and comparatively little mass in between. We further find no reliable relationship between the relative depth at which alignment peaks and the magnitude of that alignment (Appendix Figures \ref{fig:alignment-best-layer-aut-yeo-dmn-lt}-\ref{fig:alignment-best-layer-aut-yeo-dmn-resp-mt}). Taken together, these results indicate that the depth at which LLM representations best match creativity-related neural geometry is not
a stable property of the model, but is contingent on the representation-extraction choice. 

Finally, we note that our layer-selection procedure takes the maximum alignment across all layers of each model \citep{schrimpf2018brain}, which yields upper bounds rather than unbiased
estimates; we quantify this optimism bias via cross-validated layer selection in Appendix
\ref{sec:layer-selection}, and find that it does not scale with layer count and that the
size correlation reported above survives its removal.
}

\camready{
\paragraph{Post-Training Objectives Are Associated with Differential Alignment to High- and Low-Creativity Neural Responses}
To examine how different post-training approaches relate to brain alignment, we partition the brain data into high and low creativity populations as described in Section \ref{sec:high-low} and measure alignment separately for each population across variants of \textsc{Llama-3.1-8B} and \textsc{Qwen2.5-7B} (Section \ref{sec:models}). Results are shown in Figure \ref{fig:alignment-high-low}, which reports the $2.0$ rating cutoff; the same patterns largely hold under the $1.75$ and $2.25$ cutoffs (Appendix Figures \ref{fig:alignment-high-low-175} and \ref{fig:alignment-high-low-225}). During early prompt processing, all model variants show broadly comparable alignment with both populations. Once models begin generating responses, however, the alignment patterns diverge across training conditions. Comparing \textsc{Llama-3.1-8B-Instruct} with its creativity-optimized counterpart \textsc{CrPO-Llama-3.1-8B-Instruct-cre}, the instruct model shows positive alignment with both populations, whereas the CrPO model shows positive alignment with the high-creativity population but no reliable positive alignment with the low-creativity one, a selective pattern consistent with its training objective. \textsc{DeepSeek-R1-Distill-Llama-8B} exhibits the opposite asymmetry to CrPO: negative alignment with high-creativity responses and no reliable negative alignment with low-creativity ones, consistent with reasoning-chain training being associated with representations further from the neural geometry of creative ideation.

Two additional comparisons help localize these effects. First, the CrPO pattern appears specific to the full optimization method rather than to its training data or fine-tuning paradigm: \textsc{CrPO-SFT-Llama-3.1-8B-Instruct} and
\textsc{CrPO-DPO-Llama-3.1-8B-Instruct}, which share both the base model and the MuCE dataset with \textsc{CrPO-Llama-3.1-8B-Instruct-cre}, both show positive alignment with the low-creativity population, and only the full CrPO variant lacks it. Second, the reasoning-training pattern partially generalizes beyond the Llama family: \textsc{DeepSeek-R1-Distill-Qwen-7B} and \textsc{Qwen2.5-Math-7B} both show
positive alignment with the low-creativity population, while the vanilla \textsc{Qwen2.5-7B} and \textsc{Qwen2.5-7B-Instruct} baselines show no consistent lean toward either population.
}

%% file: 06_discussion.tex
\camready{
In summary, our results show that 1) LLM alignment with the human brain is positively associated with model size and with performance on divergent thinking assessments, 2) the depth at which alignment peaks depends on the token-pooling strategy, and does not predict alignment magnitude, and 3) post-training objectives are associated with systematic differences in alignment, with the largest elevations observed for objectives explicitly designed to push models toward greater creativity in responses. Notably, these patterns are stage-dependent: the relationship between model properties and brain alignment is clearest during prompt processing and weakens once models begin generating responses, suggesting that the representational dynamics of language generation diverge from human creative cognition in ways that are not simply resolved by scaling. Our post-training results further indicate that the choice of training objective is associated with selective differences in how model representations relate to the neural geometry of creative thought: reasoning-chain training is associated with negative alignment to high-creativity neural responses, while creativity-optimized training shows the opposite asymmetry, retaining alignment with high-creativity responses while lacking the positive low-creativity alignment.

Our post-training results carry implications for how language models are fine-tuned in future work. Post-training strategies and datasets have thus far emphasized tasks requiring \textit{convergent thinking}, such as solving difficult math problems or coding tasks
\citep{deepseekai2025deepseekr1incentivizingreasoningcapability, lightman2023let, chen2021evaluating}, because they are easily evaluated using a human-provided ground truth or an LLM judge. While this has proven effective for developing models that can solve tasks which emphasize convergent thinking --- reasoning toward a single correct answer or solution --- our findings raise the possibility that it may simultaneously come at a cost to the ability of LLMs to come up with many and varied ideas, the hallmark of divergent thinking \citep{runco2012divergent}. Specifically, the reasoning-trained models in our set showed alignment patterns shifted toward low-creativity neural responses relative to their creativity-optimized counterparts, consistent with the possibility that reasoning-oriented post-training moves representations toward the neural geometry associated with lower creativity.

Together, these findings suggest that brain alignment may offer an informative lens for evaluating and developing language models in the context of creative cognition --- one that goes beyond behavioral benchmarks and captures something closer to the computational principles underlying human creative thought. Given the crucial role of creativity in fostering scientific, engineering, and artistic advances \citep{torrance1974torrance, guilford1967nature}, developing LLMs capable of creativity when tasked with challenging, long-horizon, or ambiguously scoped problems is an important goal. Our findings suggest that post-training exclusively with convergent-thinking tasks warrants closer scrutiny for its effects on divergent thinking.
}

%% file: 07_conclusion.tex
We presented the first investigation of brain-LLM alignment in the context of divergent creative thinking, using fMRI data from participants performing the Alternative Uses Test alongside a non-creative control task. \camready{Our results reveal that alignment with creativity-relevant brain networks is positively associated with model size and creative task performance during prompt processing, but that this relationship weakens once models generate responses, suggesting that the generative process introduces representational dynamics that diverge from human creative cognition. We further find that post-training objectives are associated with selective differences in alignment: creativity-optimized training retains alignment with high-creativity neural responses while lacking the positive low-creativity alignment present in other variants, whereas reasoning-chain training shows the opposite asymmetry, with negative alignment to high-creativity responses. Together, these findings suggest that the correspondence between LLM representations and the neural geometry of human creative thought is nuanced and sensitive to both the stage of processing examined and the nature of the training objective. We hope these results encourage further investigation into the neural plausibility of creative cognition in language models.}


%% file: 10_acknowledgements.tex
LP gratefully acknowledges the support of the Swiss National Science Foundation (grant 205121\_207437: C - LING). R.E.B. is supported by Amazon and grants from the National Science Foundation (DRL-2400782; DUE-2155070). AB also gratefully acknowledges the support of the Swiss National Science Foundation (No. 215390), the European Research Council (Starting grant no. 101222478, RESPECT-LM), the AI2050 program at Schmidt Sciences (Grant \#G-25-69783), Sony Group Corporation, and the Swiss National Supercomputing Center (CSCS) in the form of an infrastructure engineering and development project. 

%% file: 11_appendix.tex
\camready{

\input{08_limitations}

\section{RSA Predictivity}
\label{sec:rsa-predictivity}
To compute the RSA predictivity, we compare the geometry of representational dissimilarity matrices (RDMs) constructed from LLM representations and fMRI responses, respectively. LLM RDMs are constructed using Pearson correlation distance ($1-r$) between pairs of stimulus representations. On the brain side, we use all voxels within each Yeo parcel without further voxel selection; each subject's brain is resampled into the same MNI space as the Yeo parcellation, so no additional processing is required to apply the parcellation to individual brains. RDMs are constructed per subject rather than pooled across subjects, and the two RDMs are compared via Spearman rank correlation between their upper triangles. Because this comparison metric is signed, alignment scores range over $[-1, 1]$, with negative values indicating anti-alignment with the target population's representational geometry. Following \citet{nili2014toolbox}, each subject's noise ceiling is the Spearman correlation between that subject's RDM and the mean RDM computed from the remaining $N-1$ subjects.

\section{Detailed Post-training Results}
\label{sec:detailed-high-low}
Tables \ref{tab:hl_alignment_model_resp_nc0_r1.75}, \ref{tab:hl_alignment_model_resp_nc0_r2} and \ref{tab:hl_alignment_model_resp_nc0_r2.25} provide raw alignment medians with 95\% bootstrapped CIs for both token pooling modes (last-token, mean-token) and all three rating cutoff thresholds: $1.75$ ($N_{low}=1{,}594$, $N_{high}=2{,}034$), $2.0$ ($N_{low}=2{,}147$, $N_{high}=1{,}481$), and $2.25$ ($N_{low}=2{,}588$, $N_{high}=1{,}040$). The ``alignment'' column marks each cell as $+$ (CI lies largely on the positive side), $-$ (CI lies largely on the negative side), or $\pm$ otherwise; we use $\pm0.1$ as the practical boundary since alignment values within $0.1$ of zero are negligible. Prompt-stage results are omitted for brevity (all $\pm$).

\section{Robustness of the Alignment Correlations}
\label{sec:correlation-robustness}

The brain alignment correlations are computed over $n=24$ models. At this sample size, Pearson correlations can be influenced by individual observations, and a single point estimate with an uncorrected $p$-value understates this fragility. We therefore compute three additional diagnostics for every cell in our analysis grid: a leave-one-out (LOO) jackknife over models, the Spearman rank correlation alongside Pearson $r$, and a bias-corrected and accelerated (BCa) bootstrapped $95\%$ confidence
interval. Below we summarize our findings:

\paragraph{Primary correlations.}
The DMN--size correlation at the prompt stage is robust across all three diagnostics:
$r=+0.55$, BCa CI $[+0.24, +0.77]$, LOO range $[+0.45, +0.61]$, and Spearman
$\rho=+0.65$ ($p<0.001$). The rank structure is therefore at least as strong as the
linear fit, and no single model carries the effect: removing the most influential
observation (\textsc{Llama-3.1-70B-Instruct}) reduces $r$ only to $+0.45$. The AUT-score
correlations are positive but less well estimated. In the DMN, $r=+0.52$ with BCa CI
$[+0.07, +0.76]$ and LOO range $[+0.40, +0.58]$; in the FPN, $r=+0.46$ with a BCa CI that
straddles zero, $[-0.06, +0.77]$, and LOO range $[+0.29, +0.59]$. In both cases Spearman
$\rho$ is substantially lower than Pearson $r$ ($+0.26$ and $+0.22$, neither reaching
significance), indicating that the linear fit is carried in part by magnitude differences
that the rank correlation does not capture.

\paragraph{Leverage point in the AUT-score correlations.}
The LOO analysis identifies \textsc{Gemma-3-270M-it} as the influential observation for both AUT-score correlations. It is the smallest model in our set, so it anchors the low end of both the size and AUT-score ranges. Removing it shifts the DMN correlation from
$r=+0.52$ to $+0.40$ and the FPN correlation from $+0.46$ to $+0.29$, preserving the positive direction while reducing the magnitude. A denser sampling of small models would likely stabilise these estimates.

\paragraph{Multiple comparisons.}
Our full grid spans model $\times$ network $\times$ brain task $\times$ model task
$\times$ stage $\times$ token pooling, and reporting uncorrected $p$-values across all
cells would inflate the false-positive rate. We pre-specify a primary family of three
correlations corresponding to the positive effects claimed in the main text: alignment
against model size and AUT score at the prompt stage in the matched AUT condition, in the
two creativity-relevant networks. After Bonferroni correction over this family, the
DMN--size ($p_{\mathrm{bonf}}=0.016$) and DMN--AUT-score ($p_{\mathrm{bonf}}=0.036$)
correlations remain significant, while the FPN--AUT-score correlation does not ($p_{\mathrm{bonf}}=0.079$).

\paragraph{Control conditions.}
The control cells behave as predicted. In the matched non-creative OCT condition, the
DMN--size correlation at the prompt stage is $r=+0.09$ ($p=0.66$) with last-token pooling
and $r=+0.22$ ($p=0.30$) with mean-token pooling, compared with $r=+0.55$ ($p=0.005$) in
the matched AUT condition; no FPN--OCT cell reaches significance
($-0.34 \le r \le +0.23$, all $p>0.10$). The FPN--size correlations in the AUT condition,
which we do not claim as a positive effect, are likewise null across all four stage and
pooling combinations ($-0.27 \le r \le +0.05$, all $p>0.20$). Of the eight Somatomotor
control cells, seven are non-significant; the exception is a negative correlation with
model size at the response stage under mean-token pooling ($r=-0.42$, $p=0.046$), which
is opposite in direction to the effects we report.

\section{Within-Family Scaling}
\label{sec:within-family}
The correlations reported in the main text pool models across families, which confounds parameter count with differences in pretraining data, tokenizer, and architecture. As a complementary analysis that holds these factors approximately fixed, we report
within-family scaling scans for the two families in our set with sufficient size coverage: Gemma-3 ($270$M--$27$B, $n=5$) and Qwen2.5 ($0.5$B--$72$B, $n=6$). All
correlations are computed in the DMN on the AUT task, using last-token pooling for model size and mean-token pooling for AUT score, matching the main analyses.

\paragraph{Gemma-3.}
Alignment increases monotonically with model size across the Gemma-3 family at both stages (Figures \ref{fig:alignment-aut-yeo-dmn-prompt-gemma} and \ref{fig:alignment-aut-yeo-dmn-resp-gemma}). At the prompt stage, $r=0.90$ ($p=0.04$) for model size and $r=0.85$ ($p=0.07$) for AUT score;
at the prompt+response stage, $r=0.95$ ($p=0.01$) and $r=0.97$ ($p=0.01$) respectively.

\paragraph{Qwen2.5.}
The Qwen2.5 family shows the same positive scaling relationship at the prompt stage (Figure \ref{fig:alignment-aut-yeo-dmn-resp-qwen}): $r=0.89$ ($p=0.02$) for model size and $r=0.74$
($p=0.09$) for AUT score. We note that parameter counts in this family are unevenly distributed, with four of six models below $10$B and the remaining two at $32$B and
$72$B, so the size correlation is sensitive to the two largest models.
}

\begin{figure}[h]
\includegraphics[width=\linewidth]{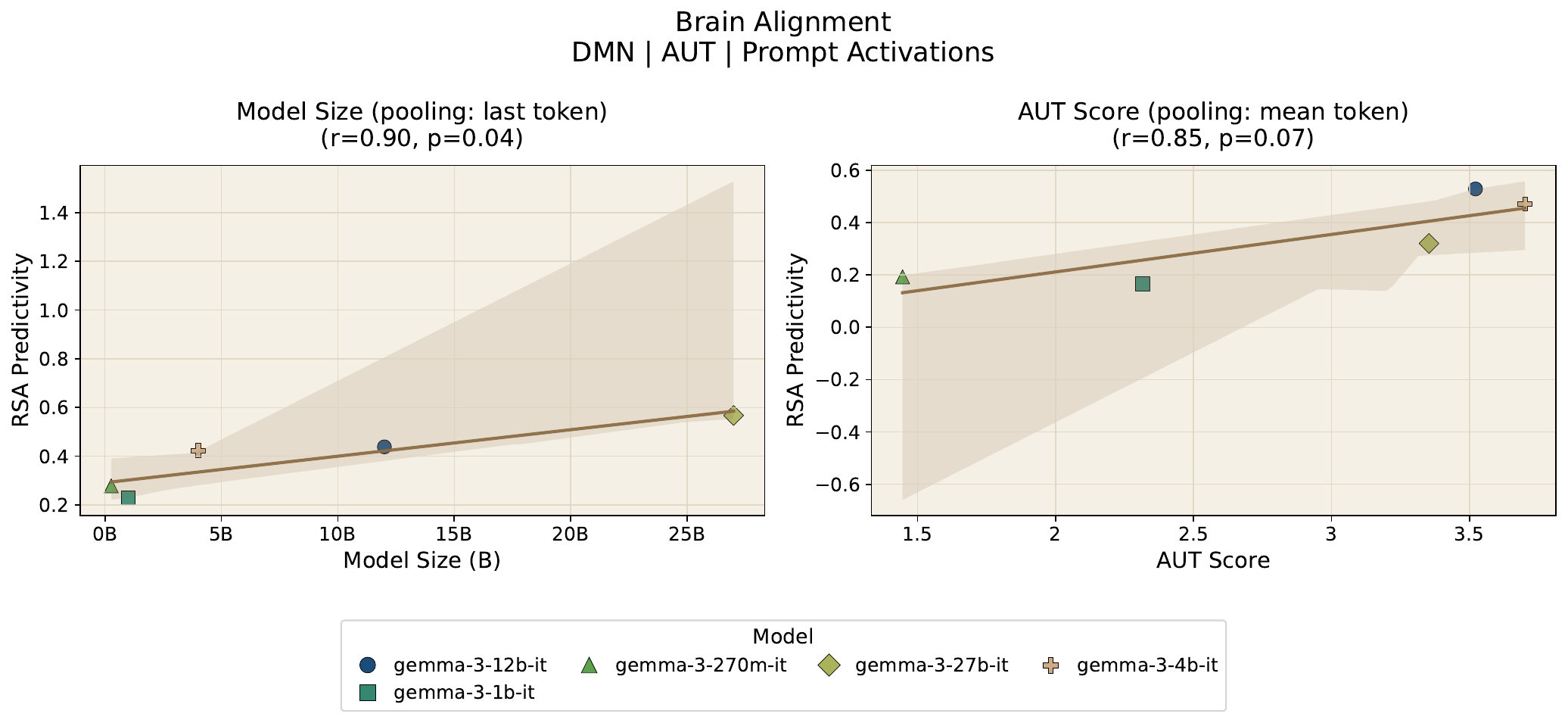}
\centering
\caption{\textbf{Default Mode Network (DMN) AUT Gemma family brain alignment results by model size and task performance using prompt-only model activations.} $r$ and $p$ correspond to the Pearson correlation coefficient and p-value.}
\label{fig:alignment-aut-yeo-dmn-prompt-gemma}
\end{figure}

\begin{figure}[h]
\includegraphics[width=\linewidth]{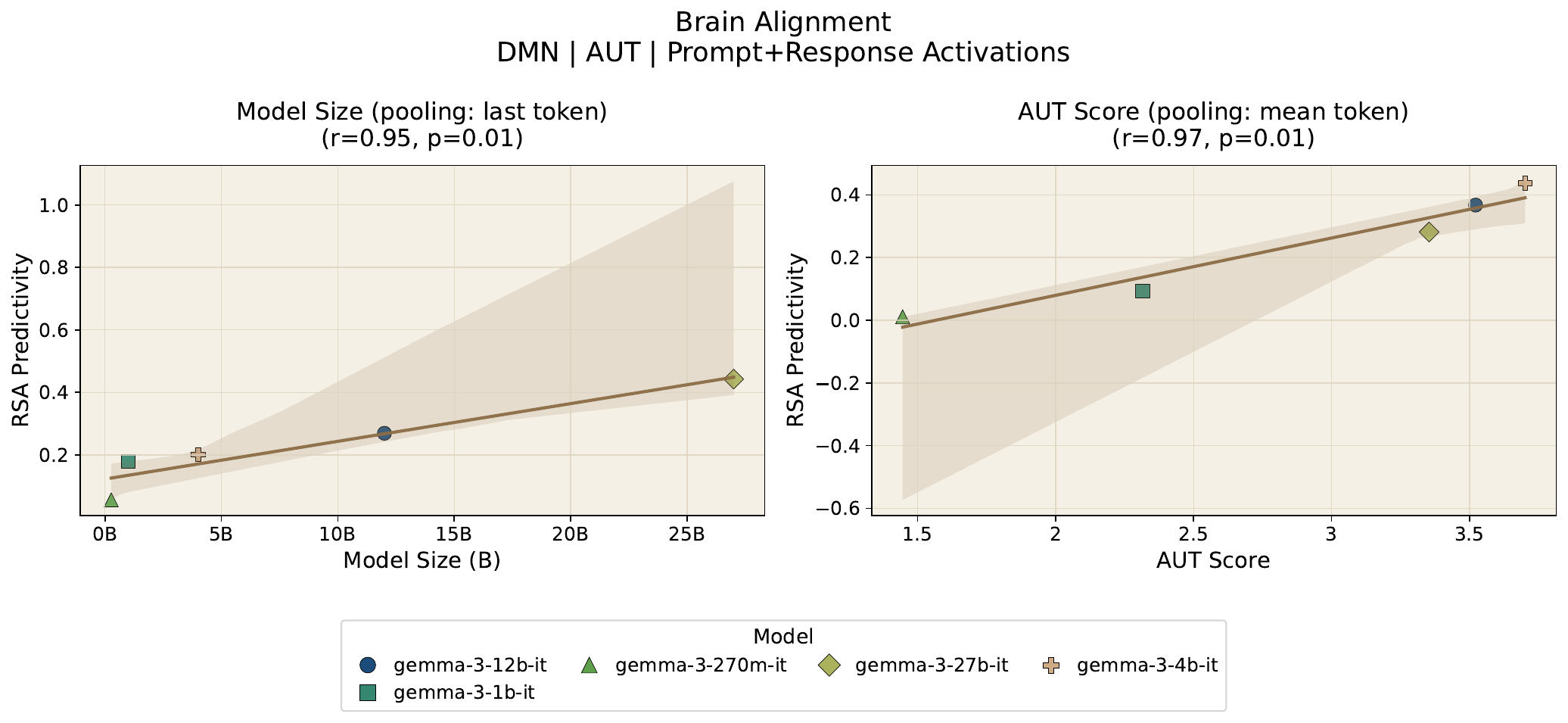}
\centering
\caption{\textbf{Default Mode Network (DMN) AUT Gemma family brain alignment results by model size and task performance using prompt+response model activations.} $r$ and $p$ correspond to the Pearson correlation coefficient and p-value.}
\label{fig:alignment-aut-yeo-dmn-resp-gemma}
\end{figure}

\begin{figure}[h]
\includegraphics[width=\linewidth]{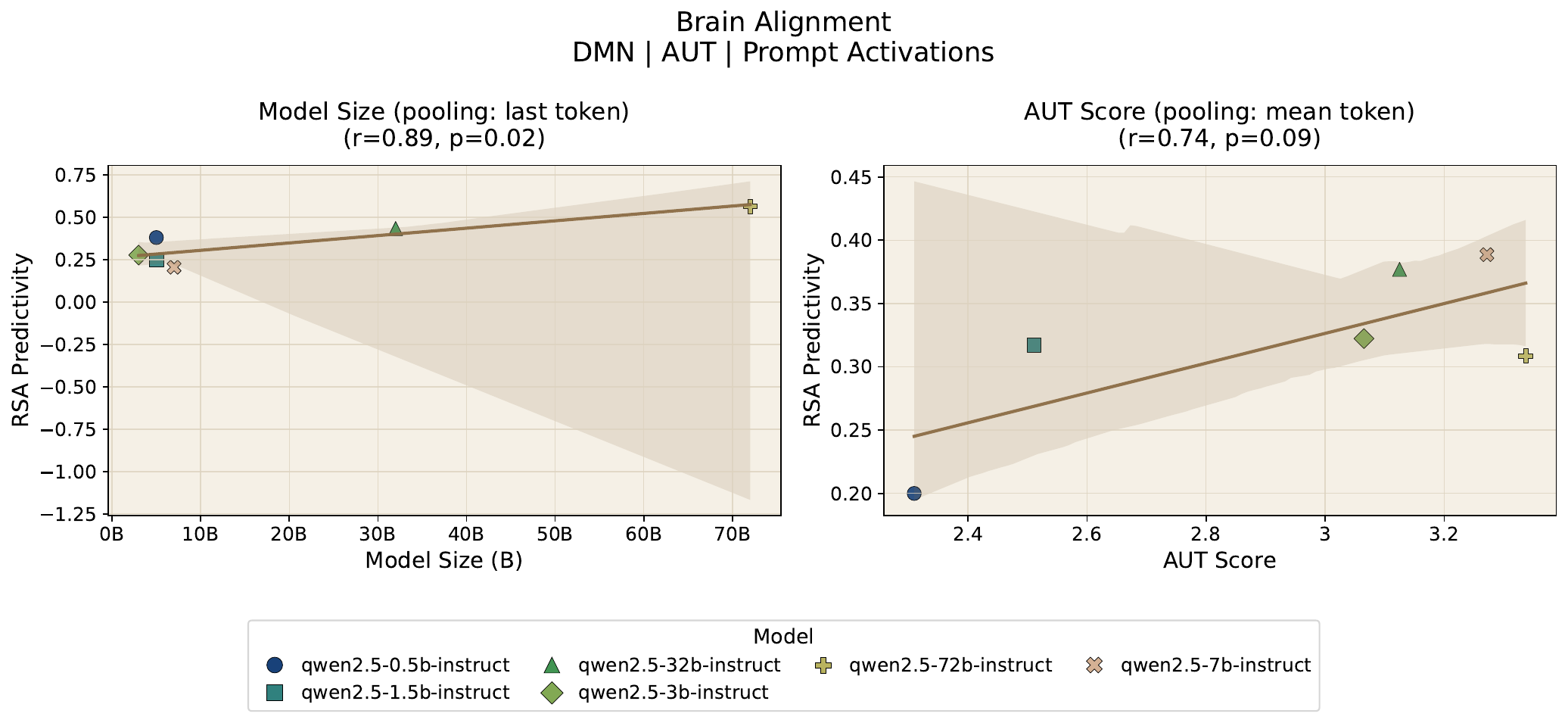}
\centering
\caption{\textbf{Default Mode Network (DMN) AUT Qwen family brain alignment results by model size and task performance using prompt-only model activations.} $r$ and $p$ correspond to the Pearson correlation coefficient and p-value.}
\label{fig:alignment-aut-yeo-dmn-resp-qwen}
\end{figure}

\input{113_per_stim_subj_analysis}
\input{116_absolute_alignment}
\input{111_post_training_results}
\input{114_full_main_results}
\input{115_cv_layer_analysis}

\begin{figure}[h]
\includegraphics[width=\linewidth]{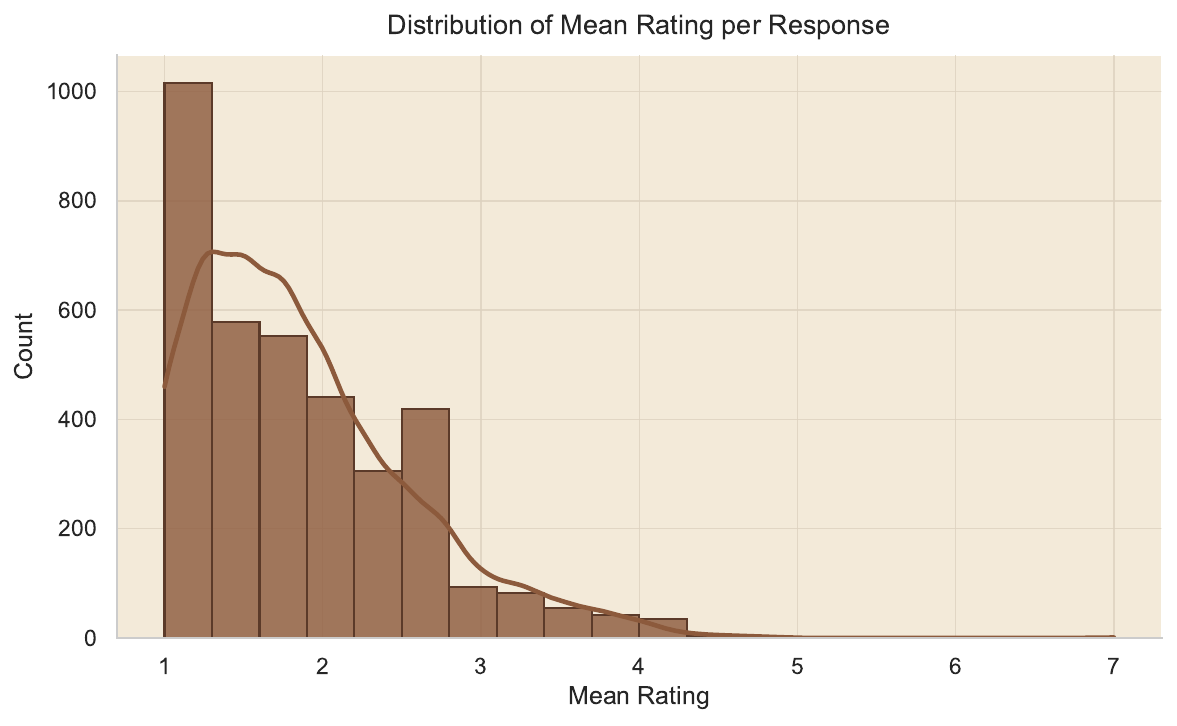}
\centering
\caption{Distribution of mean AUT ratings per human response.}
\label{fig:dist-mean-rating}
\end{figure}

\begin{prompt}[title={Modified AUT Instructions for scoring}, label=prompt:aut_instructions_scoring]
\begin{Verbatim}[breaklines, breaksymbol={}]
Think of creative uses for objects. Report only the most original idea in a few words.
\end{Verbatim}
\end{prompt}

\begin{prompt}[title={Modified OCT Instructions for scoring}, label=prompt:oct_instructions_scoring]
\begin{Verbatim}[breaklines, breaksymbol={}]
Think of the physical properties of objects. Report only the most prominent characteristic in a few words.
\end{Verbatim}
\end{prompt}

\input{112_gen_lengths}

%% file: 08_limitations.tex
\section{Limitations}

\paragraph{Correlational rather than causal evidence.}
We emphasize that our findings are correlational rather than causal. The model variants we compare differ along multiple dimensions beyond their post-training objective --- including pretraining data, optimizer, fine-tuning corpus, and generation length --- and a causal attribution would require training a single fixed base model under each objective with matched data and generation budgets, which is beyond the scope of the
present work. Several of the patterns we report also hold only partially across model families: the negative high-creativity alignment observed for the Llama reasoning distill is not reproduced in its Qwen counterpart, where only the positive low-creativity alignment replicates. Our claims should therefore be read as identifying systematic associations between post-training objectives and brain alignment, not as establishing that these objectives cause the observed representational shifts.

\paragraph{No mechanistic account of the observed effects.}
Our analyses establish that different post-training objectives are associated with different alignment patterns, but they do not explain why. One behavioral explanation could be that creativity-optimized models produce responses rated more creative than those of
reasoning-distilled models
However, a full mechanistic account would require identifying where and how representations diverge
from their shared base model: layer-wise comparisons localizing the divergence within the network, probing analyses testing whether creativity-relevant features are linearly decodable from model activations, or component-level analyses attributing the shift to specific attention heads or MLP blocks. We regard these as the natural next steps.

\paragraph{Instruction tuning and reasoning budget are not isolated.}
Two design factors remain uncontrolled in our setup. First, we restrict our analyses to instruction-tuned models, as the AUT is an active task requiring instruction following, and rely on prior work \citep{aw2023instruction} for the effect of instruction tuning itself; we do not report a within-model comparison of the same base model before and after instruction tuning, nor a layer-wise account of how tuning transforms generic prompt representations into task-specific ones. Second, our generation setup caps output
at 1024 tokens but does not control the reasoning budget: the models in our set are non-thinking models by default, with the exception of the reasoning-distilled variants,
and we do not include models with an explicit thinking mode. A systematic comparison of thinking versus non-thinking models under matched reasoning budgets would be required to disentangle these factors.

Finally, our post-training comparison does not include a model fine-tuned for domain-general cognitive modelling, which would provide a useful reference point against the creativity- and reasoning-optimised objectives we do examine.

%% file: 113_per_stim_subj_analysis.tex
\begin{figure}[h]
\centering
\includegraphics[width=\textwidth]{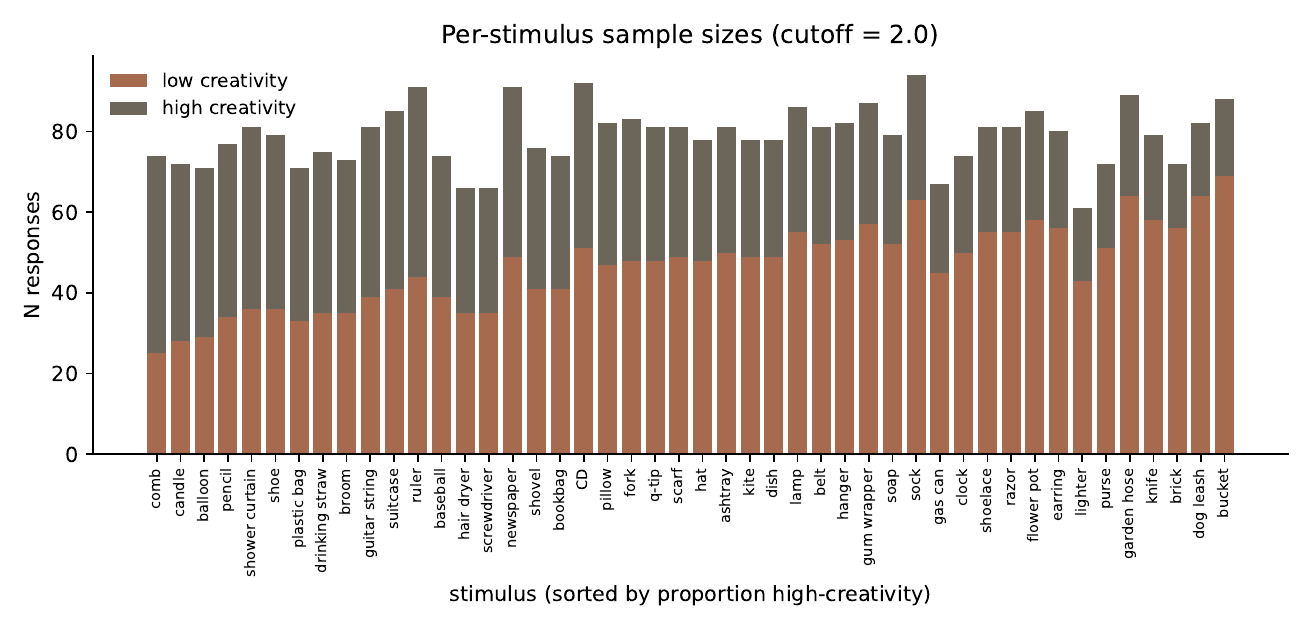}
\caption{Number of low- and high-creativity responses per AUT stimulus at the $2.0$
cutoff, sorted by the proportion of high-creativity responses. All 46 stimuli contribute
to both buckets.}
\label{fig:split-per-stimulus}
\end{figure}
\begin{figure}[h]
\centering
\includegraphics[width=0.55\textwidth]{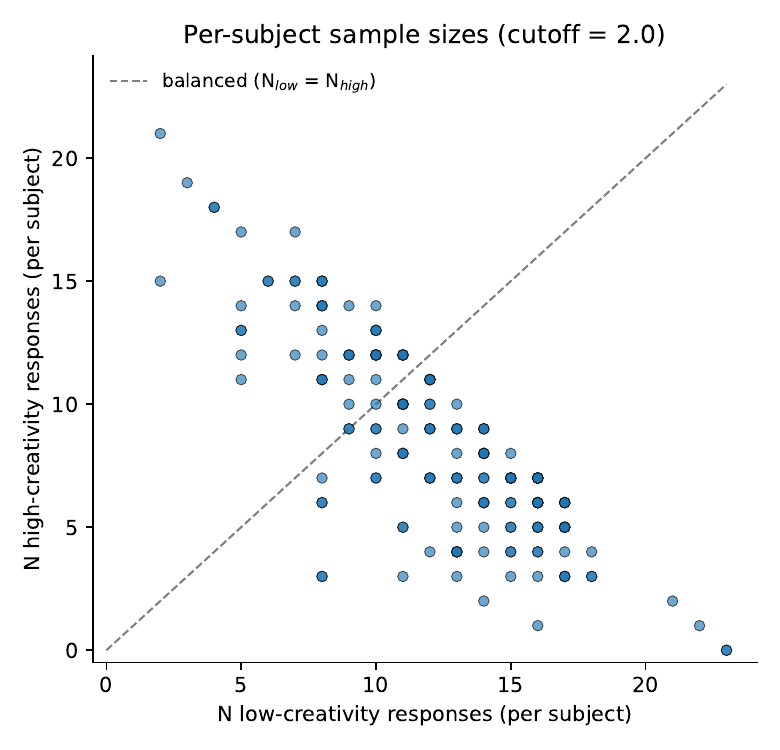}
\caption{Per-subject sample sizes in the low- and high-creativity buckets at the $2.0$
cutoff. Each point is one subject; the dashed line marks equal counts. The anti-diagonal
band reflects the fixed number of AUT trials per subject.}
\label{fig:split-per-subject}
\end{figure}

\camready{
\section{Per-Stimulus and Per-Subject Sample Sizes for the Response Creativity Partitioning}
\label{sec:split-sizes}
Because alignment is computed per subject and aggregated by taking the median across subjects (Section \ref{sec:brain-alignment}), the high/low creativity partition can in principle produce subjects or stimuli that are represented in only one of the two
populations. We therefore report the distribution of responses across the two buckets at both levels. Figures \ref{fig:split-per-stimulus} and \ref{fig:split-per-subject} show these distributions at the $2.0$ cutoff used in the main text.

\paragraph{Per-stimulus distribution.}
Figure \ref{fig:split-per-stimulus} shows the number of low- and high-creativity responses for each of the 46 AUT stimuli, sorted by the proportion of high-creativity responses. Total responses per stimulus range from approximately 61 to 94, and every stimulus contributes responses to both buckets. The proportion of high-creativity responses varies from roughly $0.66$ (\textit{comb}, \textit{candle}, \textit{balloon}) to roughly $0.22$ (\textit{bucket}, \textit{dog leash}, \textit{brick}), indicating that some objects more readily elicit original uses than others, but that no stimulus is confined to a single bucket. The partition is therefore not driven by a subset of stimuli falling entirely on one side of the cutoff.

\paragraph{Per-subject distribution.}
Figure \ref{fig:split-per-subject} plots, for each subject, the number of low-creativity responses against the number of high-creativity responses. Because each subject completed a fixed number of AUT trials, the points lie along an anti-diagonal
band rather than scattering freely: a subject with more low creativity responses necessarily has fewer high-creativity ones. Most subjects fall below the identity line, consistent with the overall imbalance at this cutoff ($N_{low}=2{,}147$ vs. $N_{high}=1{,}481$).

The relevant caveat is visible at the extremes of this distribution. While the majority of subjects contribute between 5 and 18 responses to each bucket, a small number contribute very few to one of them: several subjects have fewer than three high-creativity responses, and at least one has none. For these subjects, the RDM in the sparser bucket is estimated from very few stimuli and the corresponding alignment value is correspondingly noisy. Because we aggregate using the median rather than the mean,
these subjects have limited influence on the reported values, but we note their presence explicitly.
}

%% file: 116_absolute_alignment.tex
\camready{
\section{Absolute Alignment Levels Across Conditions}
\label{sec:condition-comparisons}

Our main analyses concern how alignment relates to model properties. As a complementary
descriptive analysis, we also compare absolute alignment levels across conditions. Since
the same models are evaluated in every condition, all comparisons are paired over models:
we compute one difference per model and test it against zero with a Wilcoxon signed-rank
test, with a sign test as a distribution-free cross-check and a bootstrap CI on the median
difference. Twelve comparisons are reported; we note where results survive Bonferroni
correction across them.

\paragraph{Creativity-relevant versus control network.}
Comparing the DMN with the Somatomotor control network on the AUT task, alignment is
reliably higher in the DMN at the prompt stage under both pooling strategies: median
difference $+0.15$ (CI $[+0.12, +0.21]$; 18 of 23 models; $p=0.002$) with last-token
pooling, and $+0.16$ (CI $[+0.07, +0.27]$; 22 of 23 models; $p<10^{-6}$) with mean-token
pooling. Both survive correction. Neither response-stage cell shows a reliable difference
($p=0.94$ and $p=0.34$).

\paragraph{Creative versus non-creative task.}
The corresponding comparison between the matched AUT and OCT conditions is less
consistent. Of eight cells, one shows a clear AUT advantage and survives correction (DMN,
prompt stage, mean-token pooling: median difference $+0.14$, CI $[+0.09, +0.18]$; 23 of 24
models; $p<10^{-6}$). Two are nominally significant in opposite directions and neither
survives correction (FPN, response stage, last-token pooling: $+0.12$, $p=0.034$; DMN,
response stage, mean-token pooling: $-0.11$, $p=0.039$). The remaining five are null,
including the DMN at the prompt stage under last-token pooling, where median alignment is
$0.44$ for AUT and $0.34$ for OCT but models split evenly (12 of 24 in each direction,
$p=0.46$).

\paragraph{Summary.}
Absolute alignment in the DMN is reliably higher than in the Somatomotor control network
at the prompt stage, while absolute alignment during the non-creative OCT task is
comparable in magnitude to the creative AUT task in most configurations. These
comparisons concern alignment magnitudes and are complementary to the correlational
analyses reported in the main text, which concern the relationship between alignment and
model properties.
}

%% file: 111_post_training_results.tex
\begin{figure}[h]
\includegraphics[width=\linewidth]{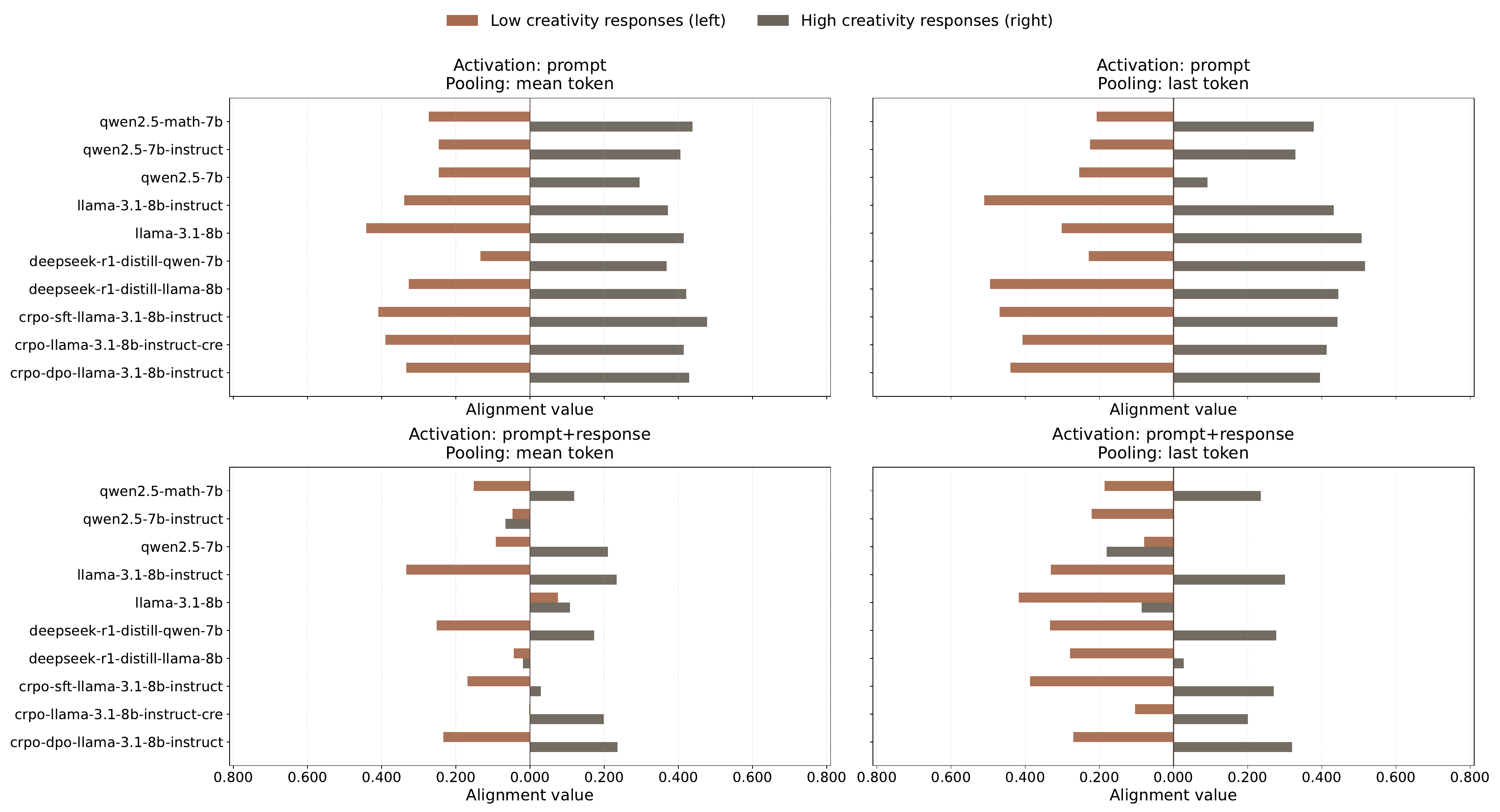}
\centering
\caption{\textbf{Default Mode Network (DMN) AUT brain alignment results for high and low creativity response populations (rating cutoff = $1.75$).} The left red and right gray bars indicate positive alignment with high and low creativity responses, respectively. Flipped bars indicate negative alignment.}
\label{fig:alignment-high-low-175}
\end{figure}

\begin{figure}[h]
\includegraphics[width=\linewidth]{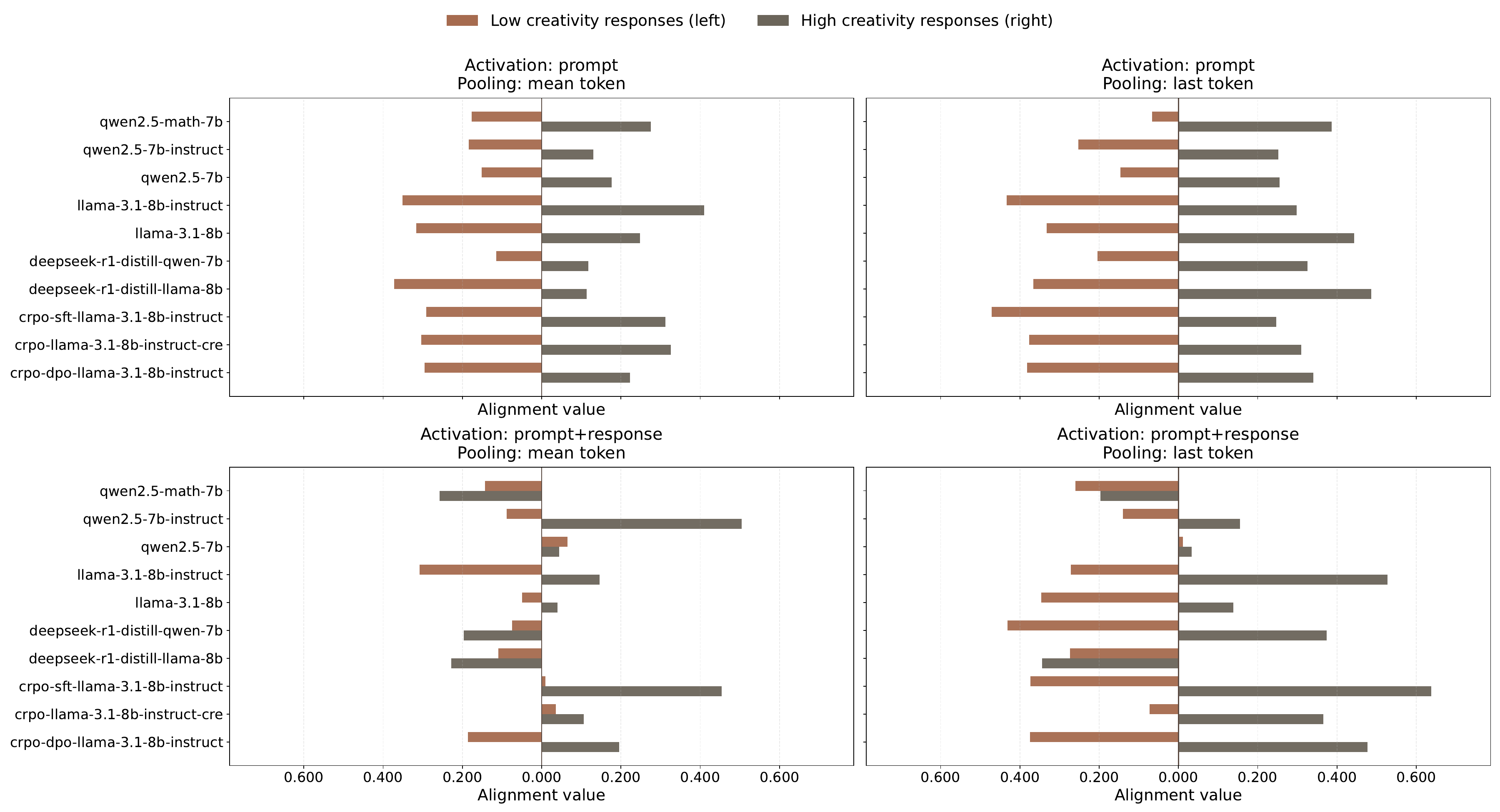}
\centering
\caption{\textbf{Default Mode Network (DMN) AUT brain alignment results for high and low creativity response populations (rating cutoff = $2.25$).} The left red and right gray bars indicate positive alignment with high and low creativity responses, respectively. Flipped bars indicate negative alignment.}
\label{fig:alignment-high-low-225}
\end{figure}

\begin{table}[h]
\centering
\small
\scalebox{0.8}{
\begin{tabular}{llcccccc}
\toprule
 &  & \multicolumn{3}{c}{last-token} & \multicolumn{3}{c}{mean-token} \\
\cmidrule(lr){3-5} \cmidrule(lr){6-8}
\textbf{model} & \textbf{pop} & \textbf{median} & \textbf{95\% CI} & \textbf{alignment} & \textbf{median} & \textbf{95\% CI} & \textbf{alignment} \\
\midrule
crpo-dpo-llama-3.1-8b-instruct & high & +0.320 & [-0.119,+0.507] & $\pm$ & +0.104 & [-0.208,+0.384] & $\pm$ \\
crpo-dpo-llama-3.1-8b-instruct & low & +0.281 & [+0.040,+0.530] & $+$ & +0.159 & [-0.127,+0.453] & $\pm$ \\
\midrule
crpo-llama-3.1-8b-instruct-cre & high & +0.158 & [-0.200,+0.460] & $\pm$ & +0.132 & [-0.189,+0.415] & $\pm$ \\
crpo-llama-3.1-8b-instruct-cre & low & +0.104 & [-0.162,+0.336] & $\pm$ & +0.007 & [-0.206,+0.097] & $-$ \\
\midrule
crpo-sft-llama-3.1-8b-instruct & high & +0.270 & [+0.010,+0.542] & $+$ & +0.018 & [-0.273,+0.200] & $\pm$ \\
crpo-sft-llama-3.1-8b-instruct & low & +0.300 & [+0.045,+0.623] & $+$ & +0.124 & [-0.294,+0.617] & $\pm$ \\
\midrule
deepseek-r1-distill-llama-8b & high & +0.027 & [-0.173,+0.170] & $\pm$ & -0.019 & [-0.358,+0.117] & $\pm$ \\
deepseek-r1-distill-llama-8b & low & +0.175 & [-0.248,+0.404] & $\pm$ & +0.027 & [-0.328,+0.269] & $\pm$ \\
\midrule
deepseek-r1-distill-qwen-7b & high & +0.217 & [+0.047,+0.556] & $+$ & +0.173 & [-0.109,+0.386] & $\pm$ \\
deepseek-r1-distill-qwen-7b & low & +0.272 & [+0.023,+0.555] & $+$ & +0.259 & [+0.075,+0.439] & $+$ \\
\midrule
llama-3.1-8b & high & -0.104 & [-0.409,+0.276] & $\pm$ & +0.073 & [-0.297,+0.280] & $\pm$ \\
llama-3.1-8b & low & +0.322 & [-0.113,+0.596] & $\pm$ & -0.153 & [-0.394,+0.141] & $\pm$ \\
\midrule
llama-3.1-8b-instruct & high & +0.176 & [-0.225,+0.404] & $\pm$ & +0.233 & [+0.003,+0.422] & $+$ \\
llama-3.1-8b-instruct & low & +0.326 & [+0.151,+0.461] & $+$ & +0.291 & [+0.054,+0.546] & $+$ \\
\midrule
qwen2.5-7b & high & -0.181 & [-0.549,+0.193] & $\pm$ & +0.210 & [-0.128,+0.434] & $\pm$ \\
qwen2.5-7b & low & +0.079 & [-0.150,+0.240] & $\pm$ & +0.081 & [-0.190,+0.376] & $\pm$ \\
\midrule
qwen2.5-7b-instruct & high & -0.002 & [-0.183,+0.307] & $\pm$ & -0.066 & [-0.298,+0.427] & $\pm$ \\
qwen2.5-7b-instruct & low & +0.182 & [-0.055,+0.343] & $+$ & +0.050 & [-0.213,+0.237] & $\pm$ \\
\midrule
qwen2.5-math-7b & high & +0.235 & [-0.020,+0.444] & $+$ & +0.119 & [-0.160,+0.476] & $\pm$ \\
qwen2.5-math-7b & low & +0.192 & [+0.001,+0.523] & $+$ & +0.114 & [-0.082,+0.306] & $+$ \\
\bottomrule
\end{tabular}
}
\caption{Default Mode Network (DMN) AUT alignment results (prompt+response activations) per model and creativity population with rating cutoff=$1.75$.}
\label{tab:hl_alignment_model_resp_nc0_r1.75}
\end{table}

\begin{table}[h]
\centering
\small
\scalebox{0.8}{
\begin{tabular}{llcccccc}
\toprule
 &  & \multicolumn{3}{c}{last-token} & \multicolumn{3}{c}{mean-token} \\
\cmidrule(lr){3-5} \cmidrule(lr){6-8}
\textbf{model} & \textbf{pop} & \textbf{median} & \textbf{95\% CI} & \textbf{alignment} & \textbf{median} & \textbf{95\% CI} & \textbf{alignment} \\
\midrule
crpo-dpo-llama-3.1-8b-instruct & high & +0.307 & [+0.023,+0.673] & $+$ & +0.312 & [+0.114,+0.620] & $+$ \\
crpo-dpo-llama-3.1-8b-instruct & low & +0.348 & [+0.040,+0.726] & $+$ & +0.159 & [-0.167,+0.453] & $\pm$ \\
\midrule
crpo-llama-3.1-8b-instruct-cre & high & +0.284 & [-0.050,+0.495] & $+$ & +0.080 & [-0.125,+0.244] & $\pm$ \\
crpo-llama-3.1-8b-instruct-cre & low & +0.077 & [-0.256,+0.298] & $\pm$ & +0.007 & [-0.271,+0.096] & $-$ \\
\midrule
crpo-sft-llama-3.1-8b-instruct & high & +0.581 & [+0.303,+0.789] & $+$ & +0.375 & [-0.009,+0.770] & $+$ \\
crpo-sft-llama-3.1-8b-instruct & low & +0.403 & [+0.071,+0.710] & $+$ & +0.234 & [-0.237,+0.678] & $\pm$ \\
\midrule
deepseek-r1-distill-llama-8b & high & -0.092 & [-0.364,+0.172] & $\pm$ & -0.172 & [-0.388,-0.019] & $-$ \\
deepseek-r1-distill-llama-8b & low & +0.166 & [-0.248,+0.426] & $\pm$ & -0.015 & [-0.336,+0.290] & $\pm$ \\
\midrule
deepseek-r1-distill-qwen-7b & high & +0.338 & [-0.014,+0.645] & $+$ & +0.132 & [-0.075,+0.460] & $+$ \\
deepseek-r1-distill-qwen-7b & low & +0.244 & [-0.026,+0.612] & $+$ & +0.267 & [+0.129,+0.442] & $+$ \\
\midrule
llama-3.1-8b & high & +0.003 & [-0.332,+0.207] & $\pm$ & +0.039 & [-0.245,+0.310] & $\pm$ \\
llama-3.1-8b & low & +0.424 & [-0.089,+0.605] & $+$ & -0.114 & [-0.315,+0.258] & $\pm$ \\
\midrule
llama-3.1-8b-instruct & high & +0.113 & [-0.099,+0.421] & $+$ & +0.221 & [+0.025,+0.549] & $+$ \\
llama-3.1-8b-instruct & low & +0.326 & [+0.157,+0.494] & $+$ & +0.314 & [+0.012,+0.547] & $+$ \\
\midrule
qwen2.5-7b & high & -0.042 & [-0.316,+0.216] & $\pm$ & -0.326 & [-0.564,-0.079] & $-$ \\
qwen2.5-7b & low & +0.079 & [-0.178,+0.256] & $\pm$ & +0.109 & [-0.183,+0.406] & $\pm$ \\
\midrule
qwen2.5-7b-instruct & high & +0.086 & [-0.215,+0.361] & $\pm$ & +0.072 & [-0.334,+0.359] & $\pm$ \\
qwen2.5-7b-instruct & low & +0.145 & [-0.121,+0.285] & $\pm$ & +0.035 & [-0.216,+0.268] & $\pm$ \\
\midrule
qwen2.5-math-7b & high & +0.051 & [-0.225,+0.432] & $\pm$ & +0.175 & [-0.098,+0.444] & $+$ \\
qwen2.5-math-7b & low & +0.182 & [-0.072,+0.477] & $+$ & +0.076 & [-0.083,+0.331] & $+$ \\
\bottomrule
\end{tabular}
}
\caption{Default Mode Network (DMN) AUT alignment results (prompt+response activations) per model and creativity population with rating cutoff=$2.0$.}
\label{tab:hl_alignment_model_resp_nc0_r2}
\end{table}

\begin{table}[h]
\centering
\small
\scalebox{0.8}{
\begin{tabular}{llcccccc}
\toprule
 &  & \multicolumn{3}{c}{last-token} & \multicolumn{3}{c}{mean-token} \\
\cmidrule(lr){3-5} \cmidrule(lr){6-8}
\textbf{model} & \textbf{pop} & \textbf{median} & \textbf{95\% CI} & \textbf{alignment} & \textbf{median} & \textbf{95\% CI} & \textbf{alignment} \\
\midrule
crpo-dpo-llama-3.1-8b-instruct & high & +0.467 & [-0.049,+1.000] & $+$ & +0.091 & [-0.282,+0.479] & $\pm$ \\
crpo-dpo-llama-3.1-8b-instruct & low & +0.454 & [+0.115,+0.862] & $+$ & +0.253 & [-0.184,+0.670] & $\pm$ \\
\midrule
crpo-llama-3.1-8b-instruct-cre & high & +0.366 & [-0.228,+0.666] & $\pm$ & +0.088 & [-0.135,+0.629] & $\pm$ \\
crpo-llama-3.1-8b-instruct-cre & low & +0.112 & [-0.183,+0.360] & $\pm$ & -0.040 & [-0.252,+0.077] & $-$ \\
\midrule
crpo-sft-llama-3.1-8b-instruct & high & +0.633 & [+0.324,+0.848] & $+$ & +0.454 & [-0.173,+0.769] & $\pm$ \\
crpo-sft-llama-3.1-8b-instruct & low & +0.331 & [+0.170,+0.485] & $+$ & -0.126 & [-0.506,+0.279] & $\pm$ \\
\midrule
deepseek-r1-distill-llama-8b & high & -0.486 & [-0.833,-0.320] & $-$ & -0.248 & [-0.524,+0.024] & $-$ \\
deepseek-r1-distill-llama-8b & low & +0.297 & [+0.124,+0.550] & $+$ & +0.100 & [-0.246,+0.301] & $\pm$ \\
\midrule
deepseek-r1-distill-qwen-7b & high & +0.263 & [-0.206,+0.835] & $\pm$ & -0.197 & [-0.520,+0.157] & $\pm$ \\
deepseek-r1-distill-qwen-7b & low & +0.161 & [-0.082,+0.527] & $+$ & -0.020 & [-0.281,+0.428] & $\pm$ \\
\midrule
llama-3.1-8b & high & +0.088 & [-0.238,+0.470] & $\pm$ & +0.040 & [-0.500,+0.318] & $\pm$ \\
llama-3.1-8b & low & +0.452 & [+0.146,+1.000] & $+$ & +0.019 & [-0.428,+0.268] & $\pm$ \\
\midrule
llama-3.1-8b-instruct & high & +0.495 & [-0.041,+0.862] & $+$ & +0.094 & [-0.324,+0.503] & $\pm$ \\
llama-3.1-8b-instruct & low & +0.289 & [+0.155,+0.595] & $+$ & +0.197 & [-0.020,+0.472] & $+$ \\
\midrule
qwen2.5-7b & high & +0.033 & [-0.394,+0.234] & $\pm$ & +0.044 & [-0.338,+0.558] & $\pm$ \\
qwen2.5-7b & low & +0.034 & [-0.206,+0.309] & $\pm$ & -0.016 & [-0.189,+0.163] & $\pm$ \\
\midrule
qwen2.5-7b-instruct & high & +0.068 & [-0.440,+0.369] & $\pm$ & +0.351 & [-0.112,+0.748] & $\pm$ \\
qwen2.5-7b-instruct & low & -0.034 & [-0.308,+0.254] & $\pm$ & +0.061 & [-0.186,+0.266] & $\pm$ \\
\midrule
qwen2.5-math-7b & high & -0.197 & [-0.572,+0.179] & $\pm$ & -0.258 & [-0.657,+0.033] & $-$ \\
qwen2.5-math-7b & low & +0.404 & [+0.086,+0.755] & $+$ & +0.268 & [-0.049,+0.432] & $+$ \\
\bottomrule
\end{tabular}
}
\caption{Default Mode Network (DMN) AUT alignment results (prompt+response activations) per model and creativity population with rating cutoff=$2.25$.}
\label{tab:hl_alignment_model_resp_nc0_r2.25}
\end{table}

%% file: 114_full_main_results.tex
\begin{figure}[h]
\includegraphics[width=\linewidth]{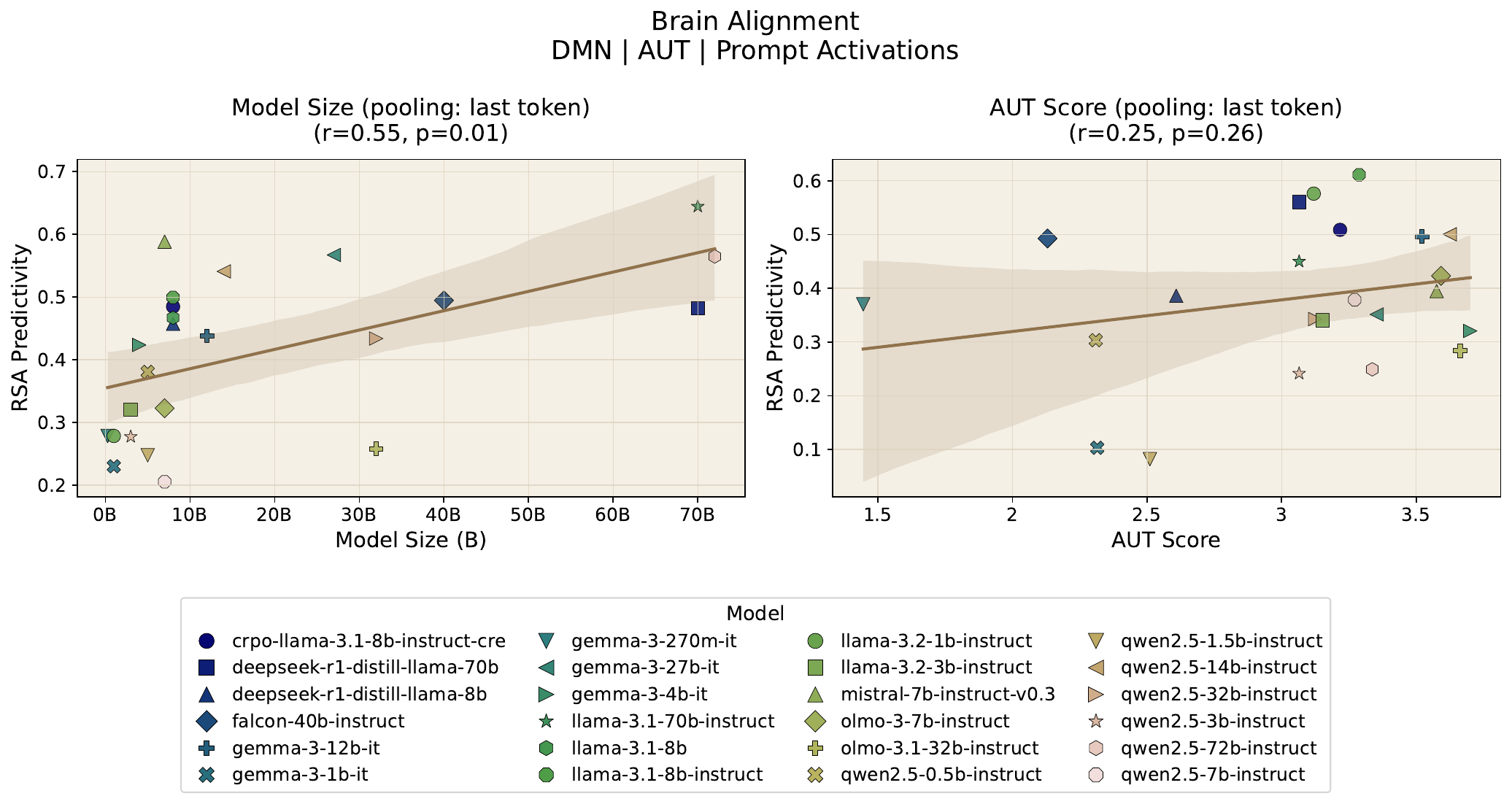}
\centering
\caption{\textbf{Default Mode Network (DMN) AUT brain alignment results by model size and task performance using prompt-only model activations with last-token pooling.} $r$ and $p$ correspond to the Pearson correlation coefficient and p-value.}
\label{fig:alignment-aut-yeo-dmn-prompt-lt}
\end{figure}

\begin{figure}[h]
\includegraphics[width=\linewidth]{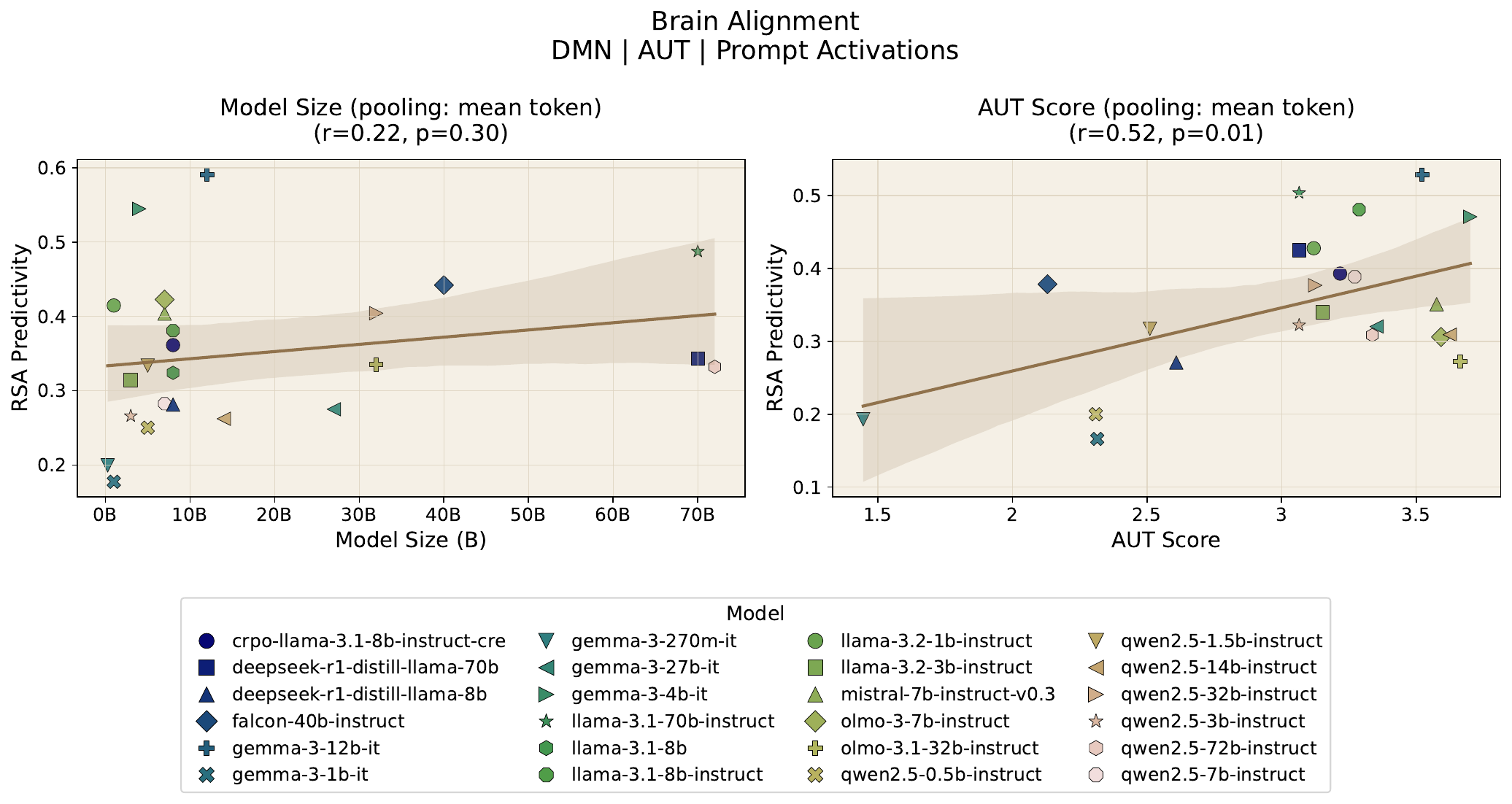}
\centering
\caption{\textbf{Default Mode Network (DMN) AUT brain alignment results by model size and task performance using prompt-only model activations with mean-token pooling.} $r$ and $p$ correspond to the Pearson correlation coefficient and p-value.}
\label{fig:alignment-aut-yeo-dmn-prompt-mt}
\end{figure}

\begin{figure}[h]
\includegraphics[width=\linewidth]{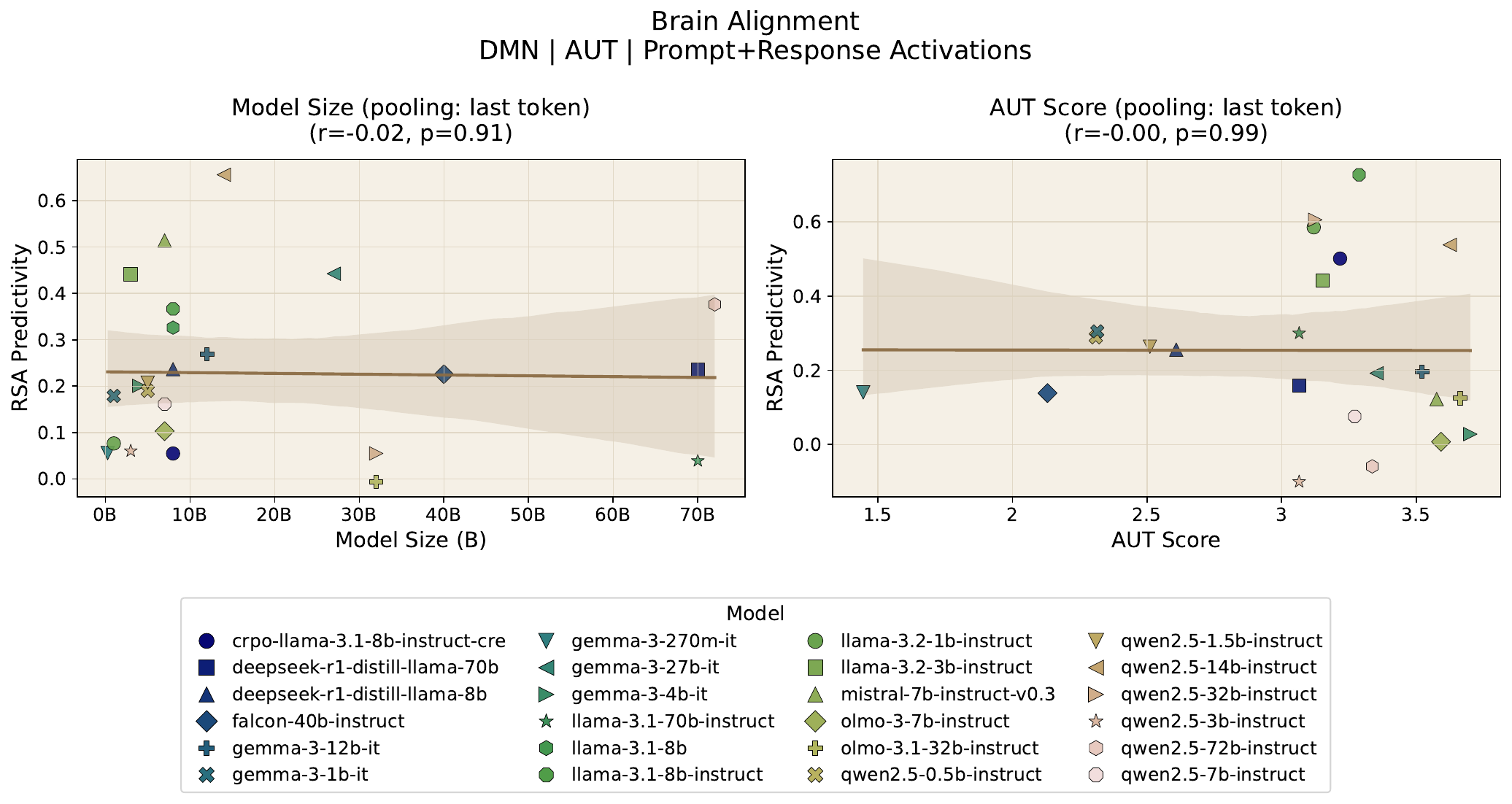}
\centering
\caption{\textbf{Default Mode Network (DMN) AUT brain alignment results by model size and task performance using prompt+response model activations with last-token pooling.} $r$ and $p$ correspond to the Pearson correlation coefficient and p-value.}
\label{fig:alignment-aut-yeo-dmn-resp-lt}
\end{figure}

\begin{figure}[h]
\includegraphics[width=\linewidth]{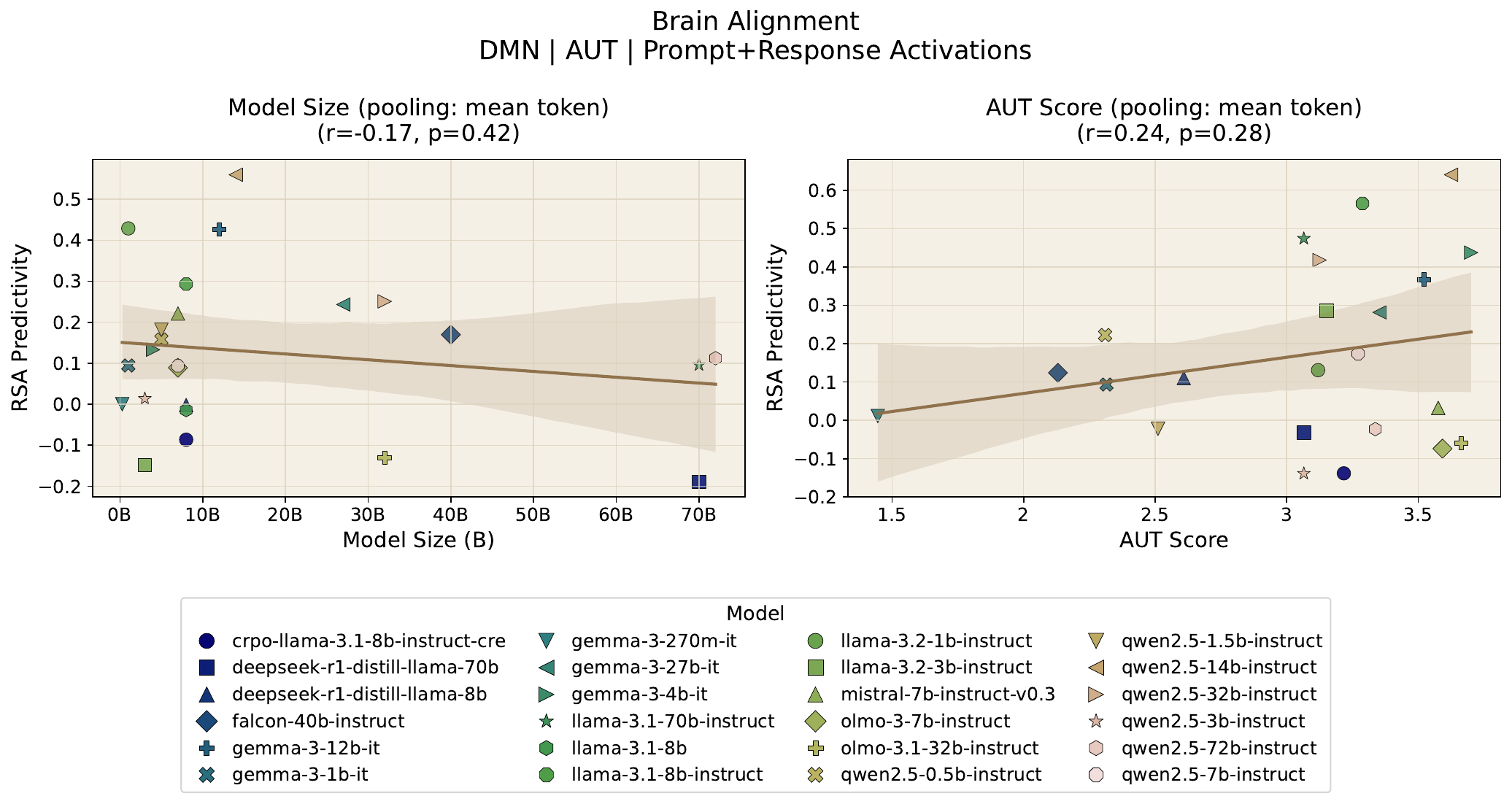}
\centering
\caption{\textbf{Default Mode Network (DMN) AUT brain alignment results by model size and task performance using prompt+response model activations with mean-token pooling.} $r$ and $p$ correspond to the Pearson correlation coefficient and p-value.}
\label{fig:alignment-aut-yeo-dmn-resp-mt}
\end{figure}

\begin{figure}[h]
\includegraphics[width=\linewidth]{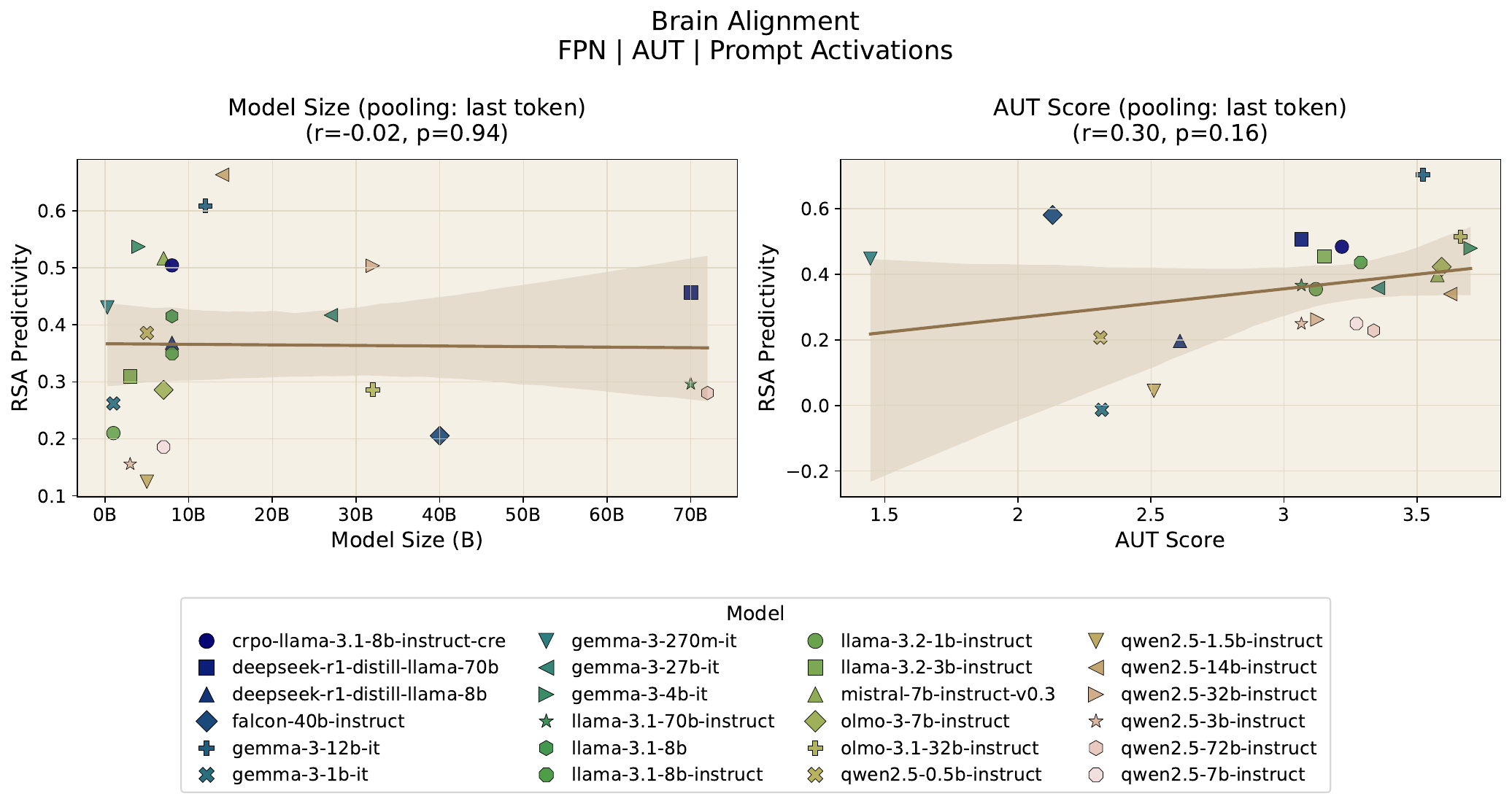}
\centering
\caption{\textbf{Frontoparietal Network (FPN) AUT brain alignment results by model size and task performance using prompt-only model activations with last-token pooling.} $r$ and $p$ correspond to the Pearson correlation coefficient and p-value.}
\label{fig:alignment-aut-yeo-fpn-prompt-lt}
\end{figure}

\begin{figure}[h]
\includegraphics[width=\linewidth]{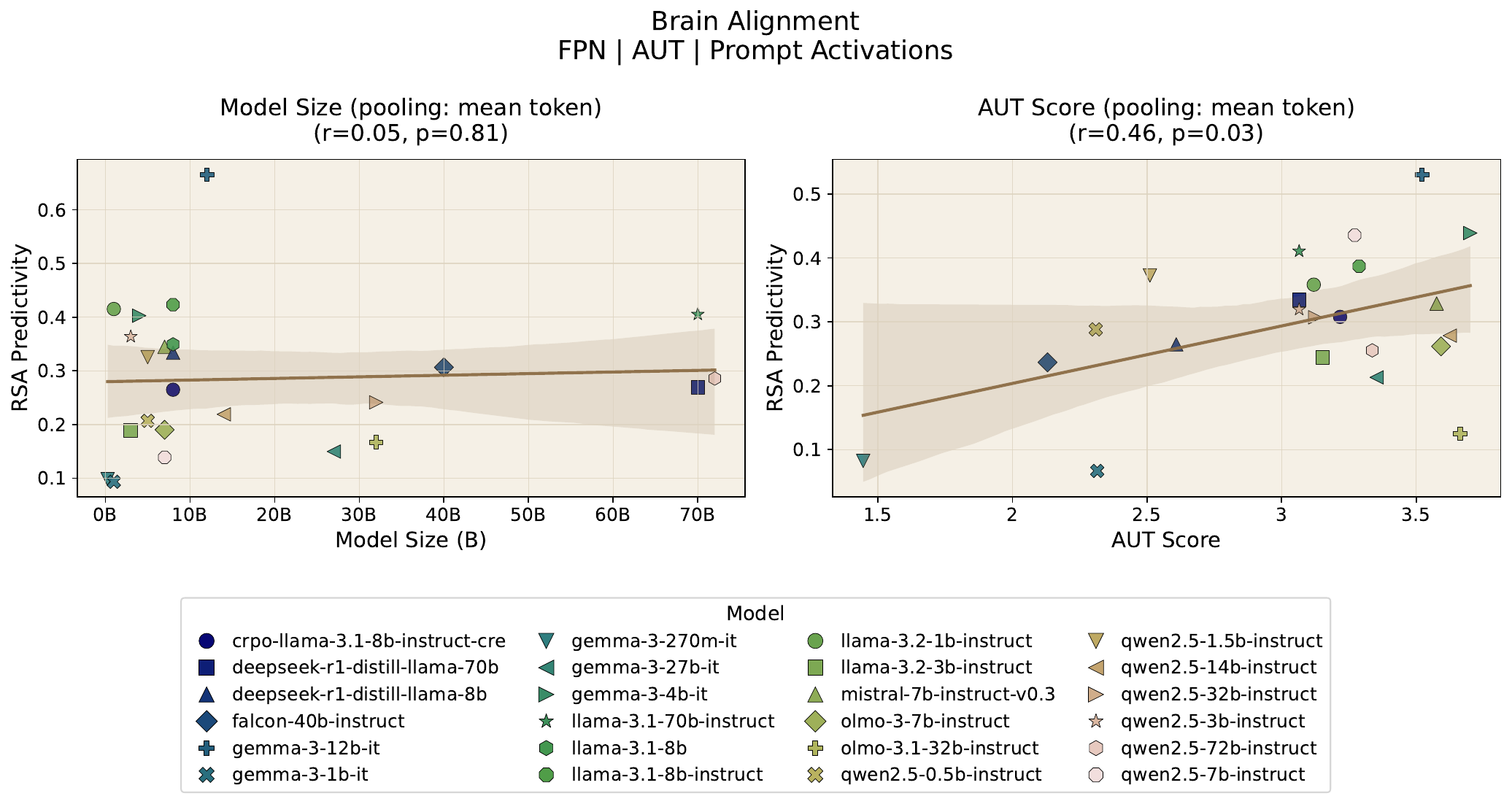}
\centering
\caption{\textbf{Frontoparietal Network (FPN) AUT brain alignment results by model size and task performance using prompt-only model activations with mean-token pooling.} $r$ and $p$ correspond to the Pearson correlation coefficient and p-value.}
\label{fig:alignment-aut-yeo-fpn-prompt-mt}
\end{figure}

\begin{figure}[h]
\includegraphics[width=\linewidth]{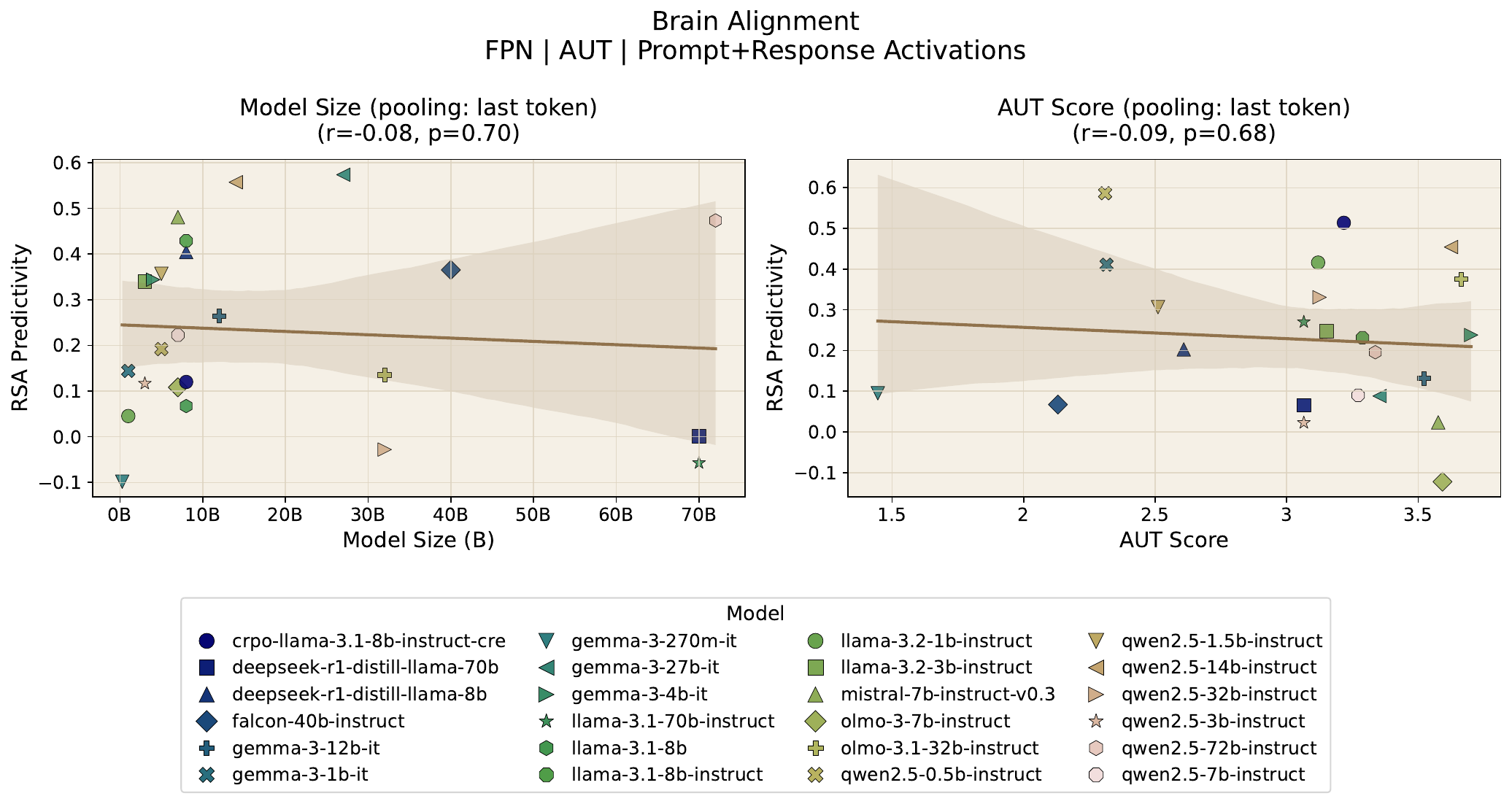}
\centering
\caption{\textbf{Frontoparietal Network (FPN) AUT brain alignment results by model size and task performance using prompt+response model activations with last-token pooling.} $r$ and $p$ correspond to the Pearson correlation coefficient and p-value.}
\label{fig:alignment-aut-yeo-fpn-resp-lt}
\end{figure}

\begin{figure}[h]
\includegraphics[width=\linewidth]{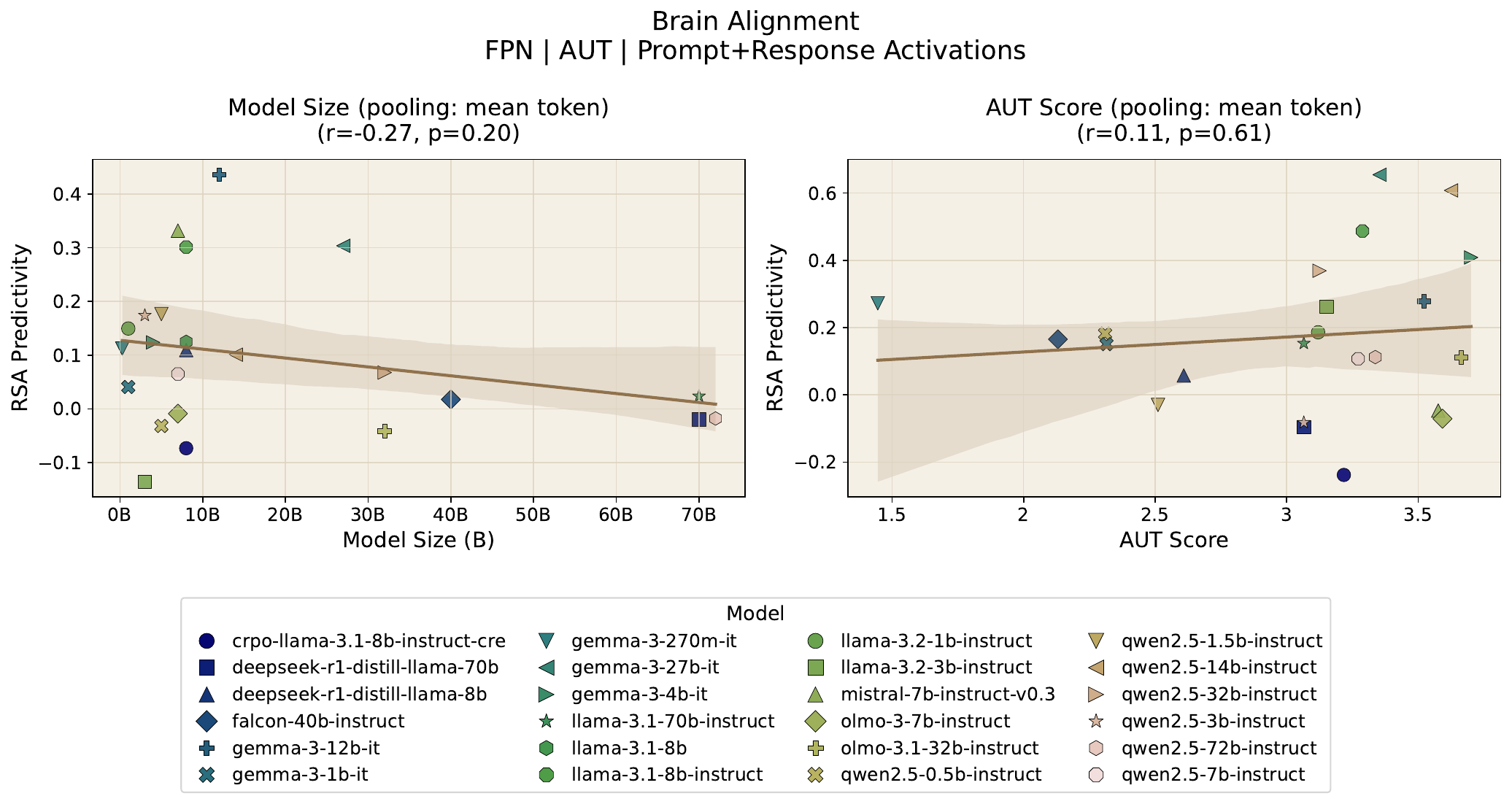}
\centering
\caption{\textbf{Frontoparietal Network (FPN) AUT brain alignment results by model size and task performance using prompt+response model activations with mean-token pooling.} $r$ and $p$ correspond to the Pearson correlation coefficient and p-value.}
\label{fig:alignment-aut-yeo-fpn-resp-mt}
\end{figure}

\begin{figure}[h]
\includegraphics[width=\linewidth]{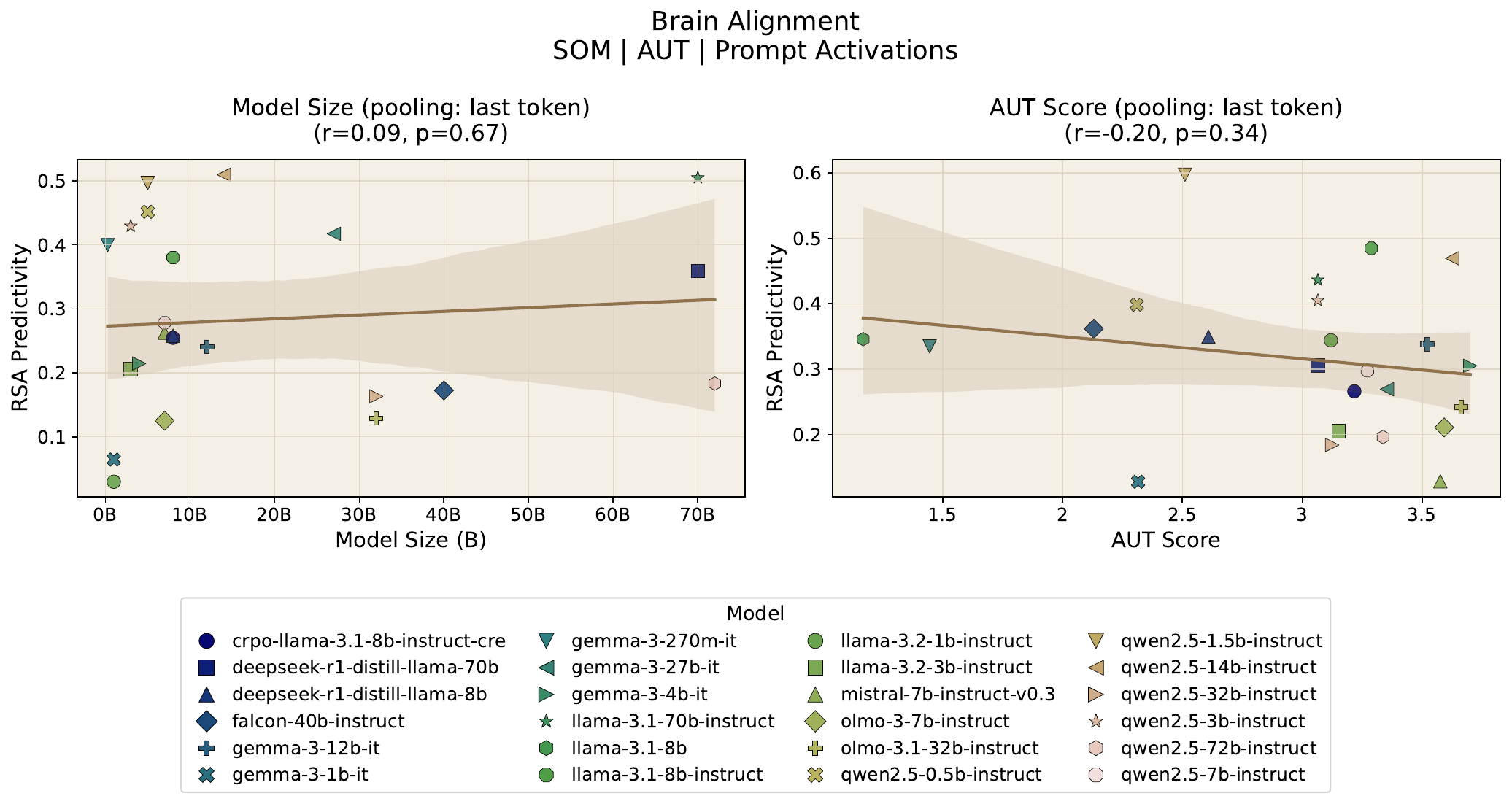}
\centering
\caption{\textbf{Somatomotor Network (SOM) AUT brain alignment results by model size and task performance using prompt-only model activations with last-token pooling.} $r$ and $p$ correspond to the Pearson correlation coefficient and p-value.}
\label{fig:alignment-aut-yeo-som-prompt-lt}
\end{figure}

\begin{figure}[h]
\includegraphics[width=\linewidth]{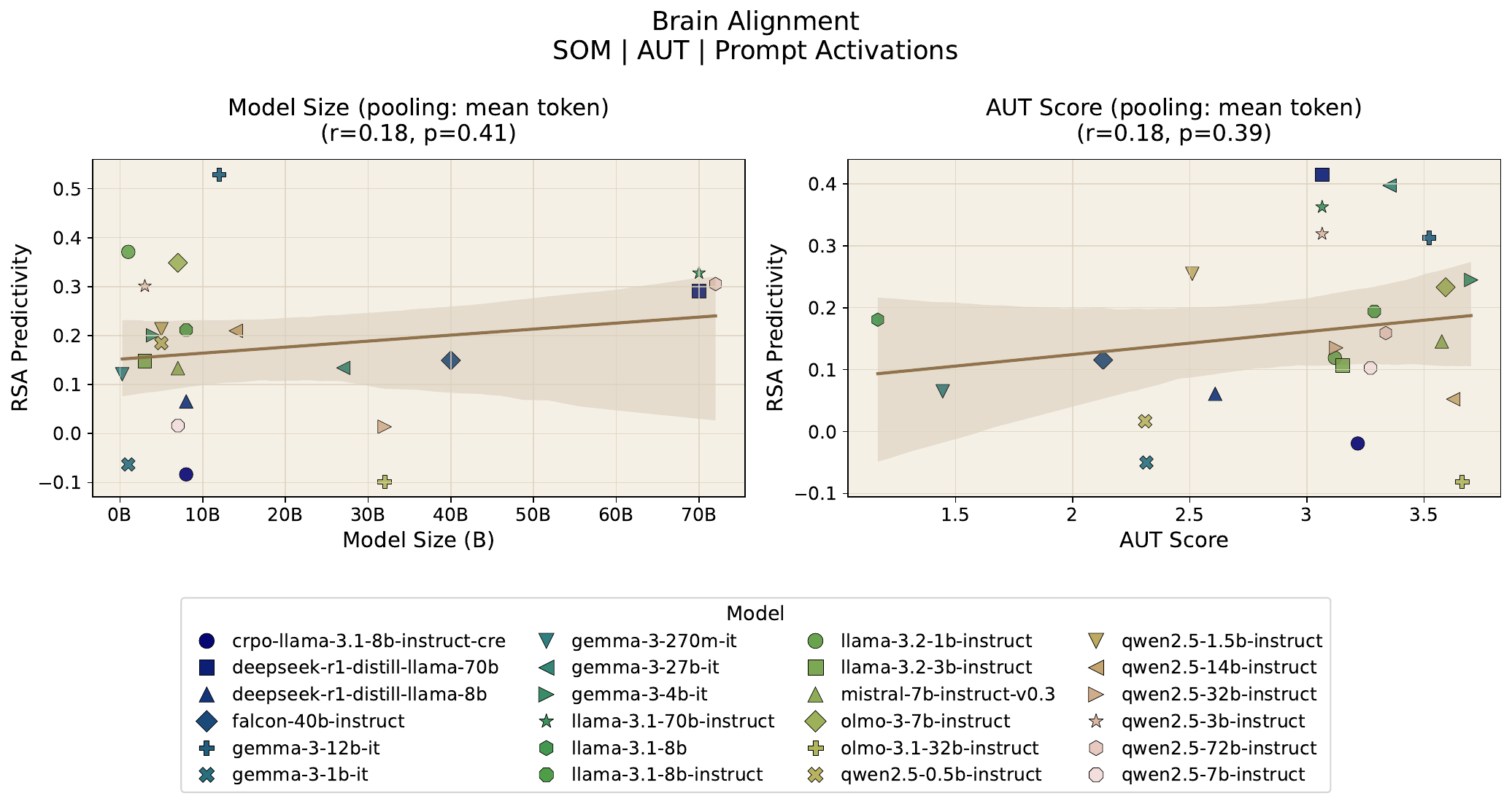}
\centering
\caption{\textbf{Somatomotor Network (SOM) AUT brain alignment results by model size and task performance using prompt-only model activations with mean-token pooling.} $r$ and $p$ correspond to the Pearson correlation coefficient and p-value.}
\label{fig:alignment-aut-yeo-som-prompt-mt}
\end{figure}

\begin{figure}[h]
\includegraphics[width=\linewidth]{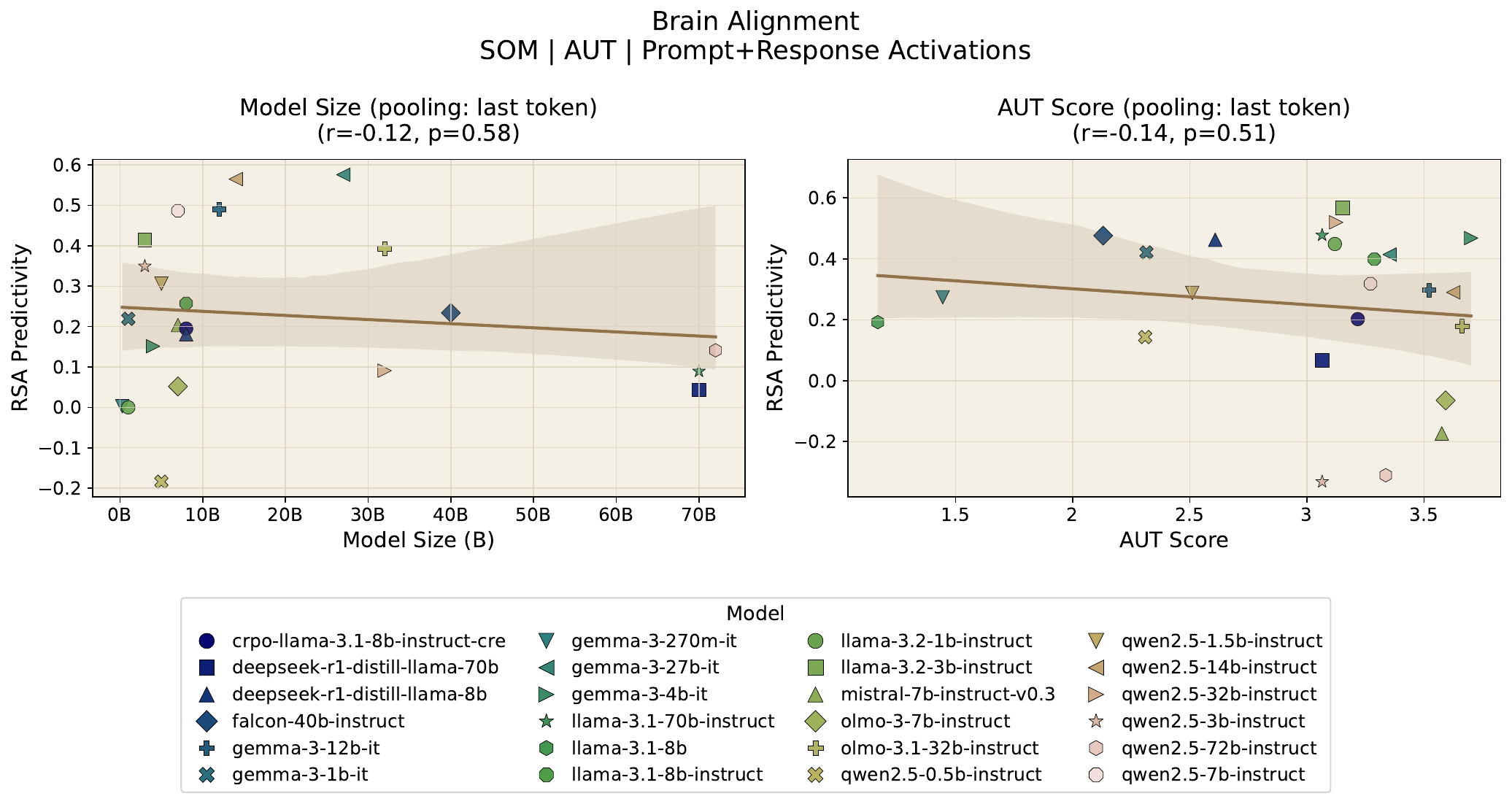}
\centering
\caption{\textbf{Somatomotor Network (SOM) AUT brain alignment results by model size and task performance using prompt+response model activations with last-token pooling.} $r$ and $p$ correspond to the Pearson correlation coefficient and p-value.}
\label{fig:alignment-aut-yeo-som-resp-lt}
\end{figure}

\begin{figure}[h]
\includegraphics[width=\linewidth]{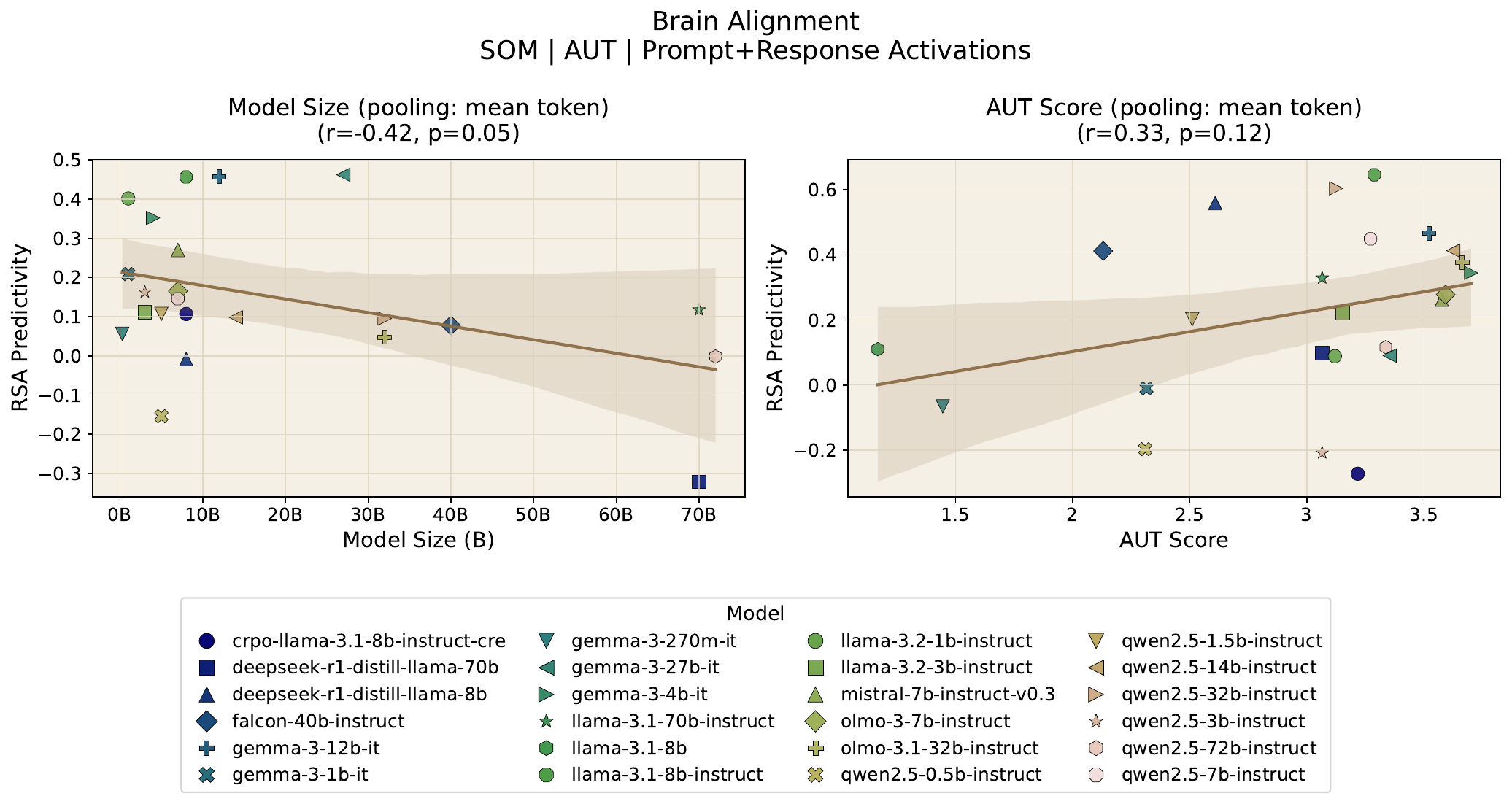}
\centering
\caption{\textbf{Somatomotor Network (SOM) AUT brain alignment results by model size and task performance using prompt+response model activations with mean-token pooling.} $r$ and $p$ correspond to the Pearson correlation coefficient and p-value.}
\label{fig:alignment-aut-yeo-som-resp-mt}
\end{figure}

\begin{figure}[h]
\includegraphics[width=\linewidth]{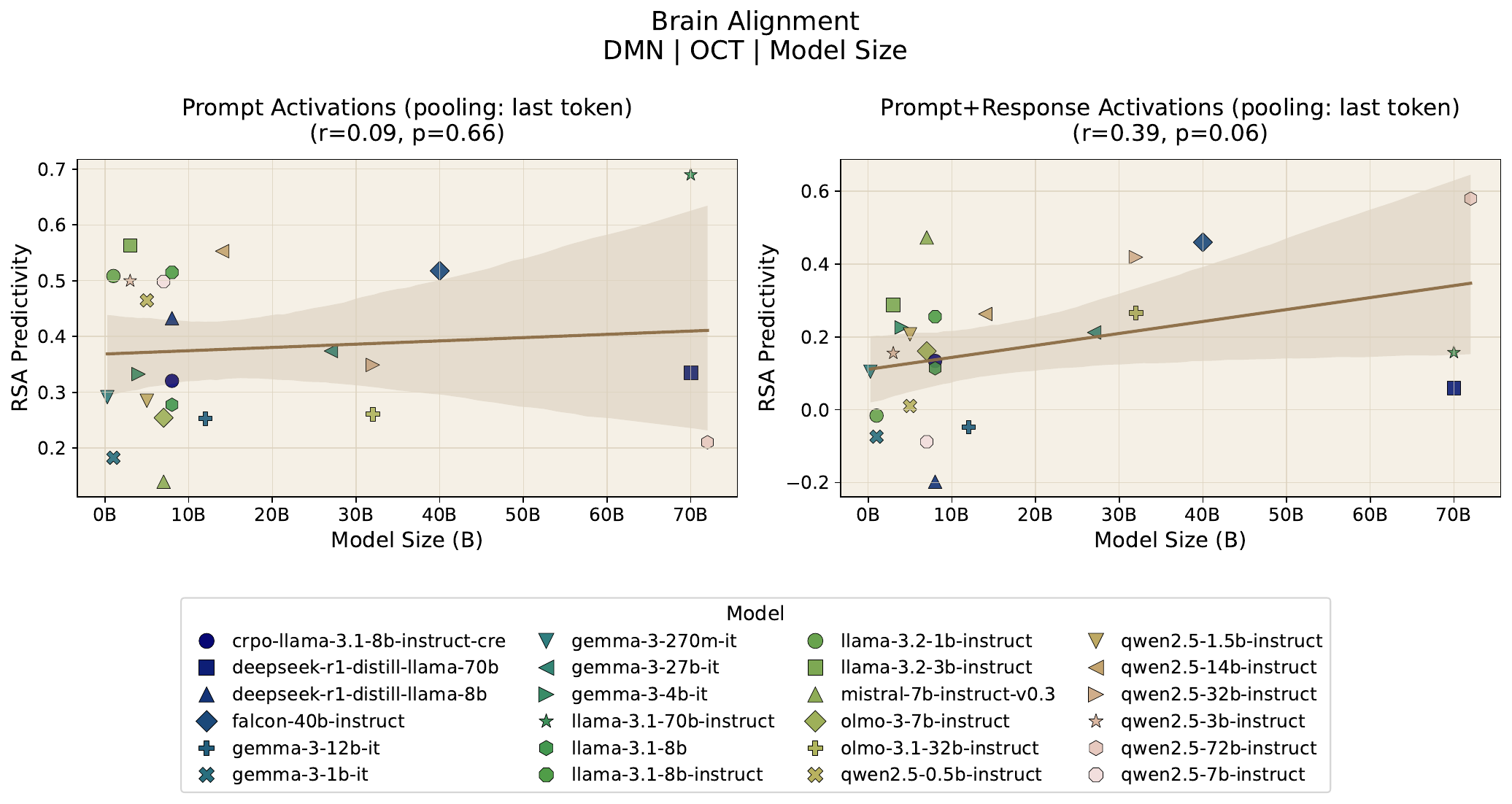}
\centering
\caption{\textbf{Default Mode Network (DMN) OCT brain alignment results by model size with last-token pooling.} $r$ and $p$ correspond to the Pearson correlation coefficient and p-value.}
\label{fig:alignment-oct-yeo-dmn-lt}
\end{figure}

\begin{figure}[h]
\includegraphics[width=\linewidth]{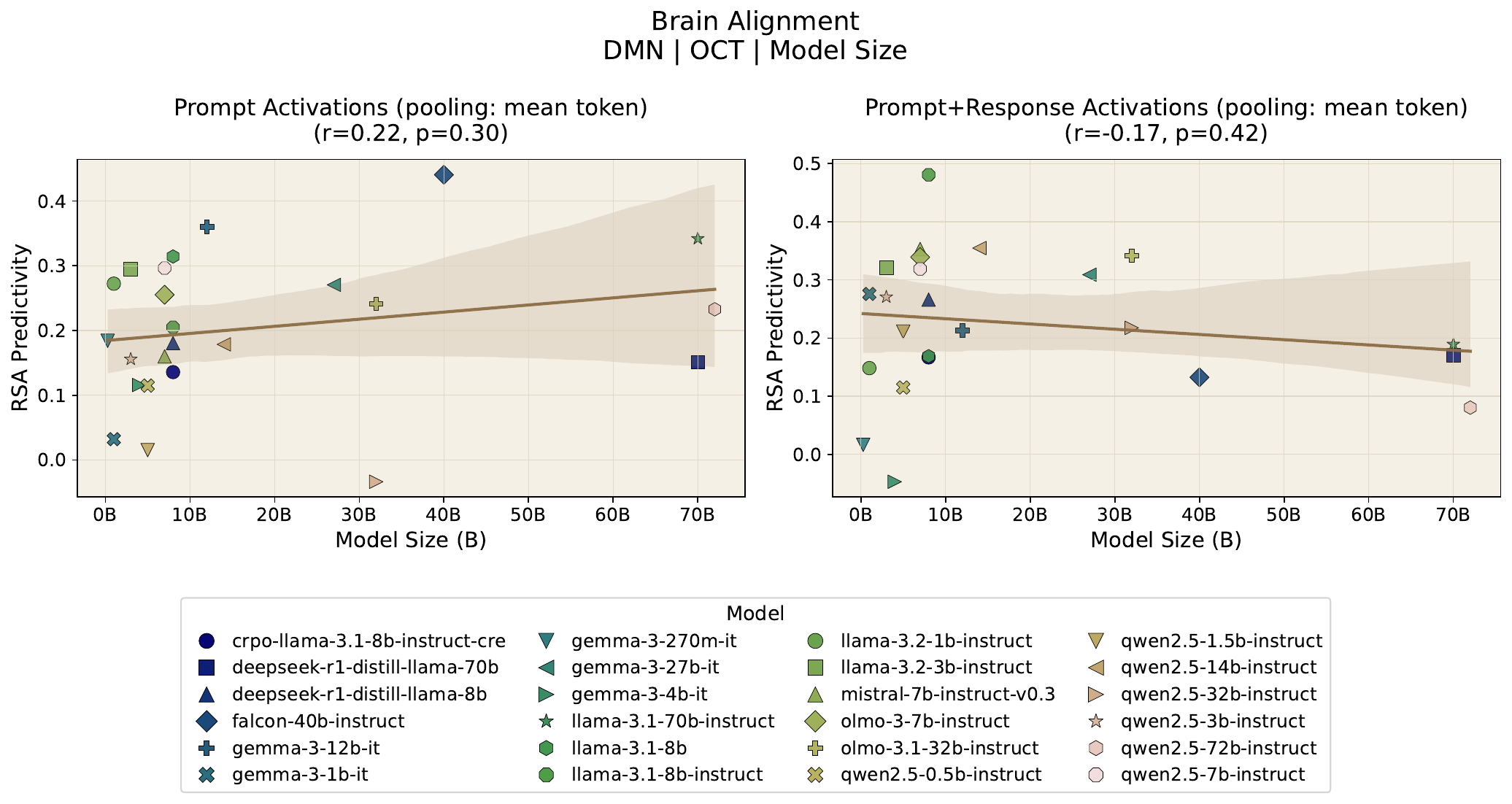}
\centering
\caption{\textbf{Default Mode Network (DMN) OCT brain alignment results by model size with mean-token pooling.} $r$ and $p$ correspond to the Pearson correlation coefficient and p-value.}
\label{fig:alignment-oct-yeo-dmn-mt}
\end{figure}

\begin{figure}[h]
\includegraphics[width=\linewidth]{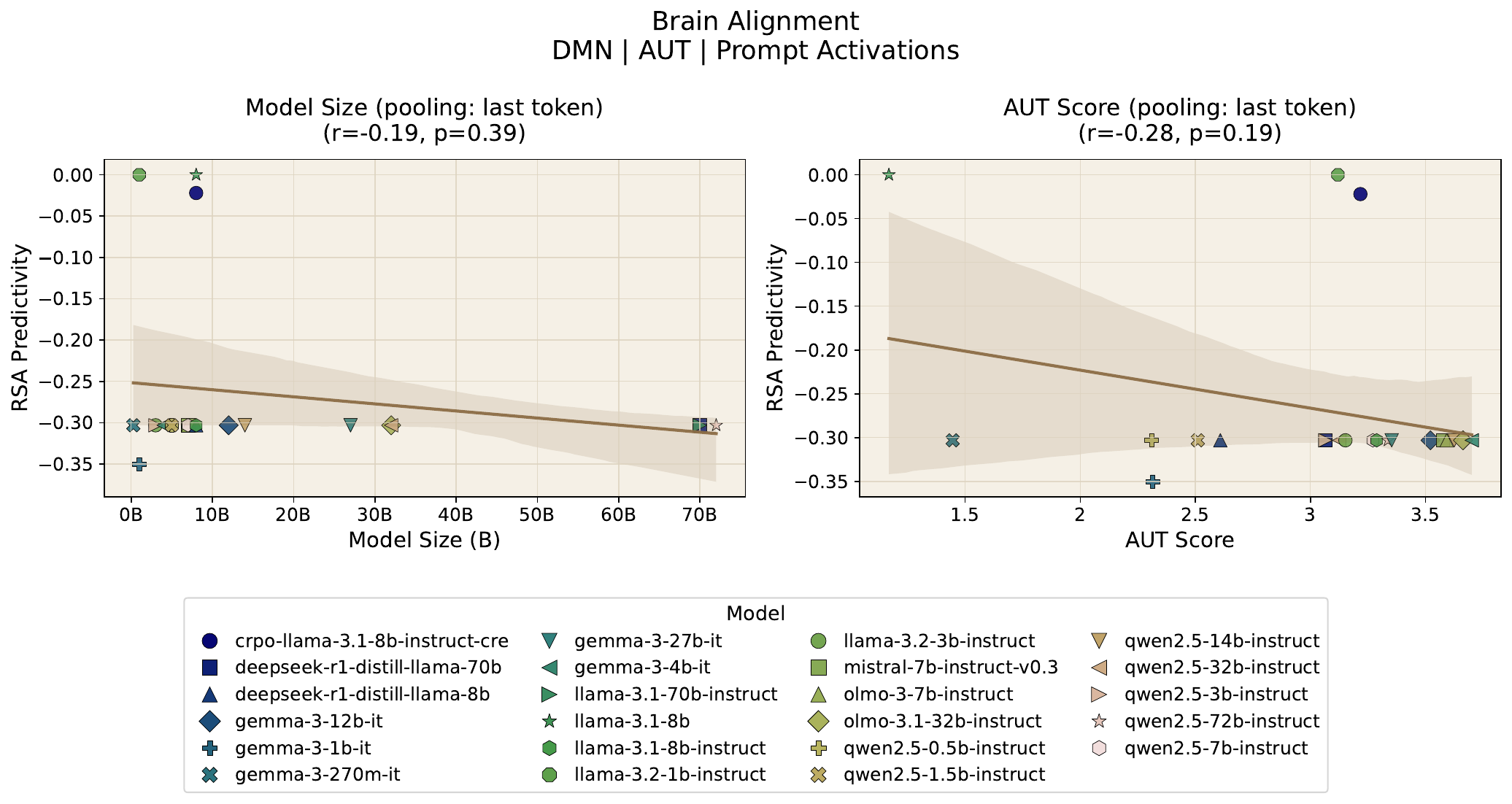}
\centering
\caption{\textbf{Default Mode Network (DMN) AUT brain alignment results by model size and task performance using empty prompt-only model activations with last-token pooling.} $r$ and $p$ correspond to the Pearson correlation coefficient and p-value. Note that \textsc{Falcon-40B-Instruct} result is missing due to the failure to extract representations from an empty prompt.}
\label{fig:alignment-aut-yeo-dmn-empty-prompt-lt}
\end{figure}

\begin{figure}[h]
\includegraphics[width=\linewidth]{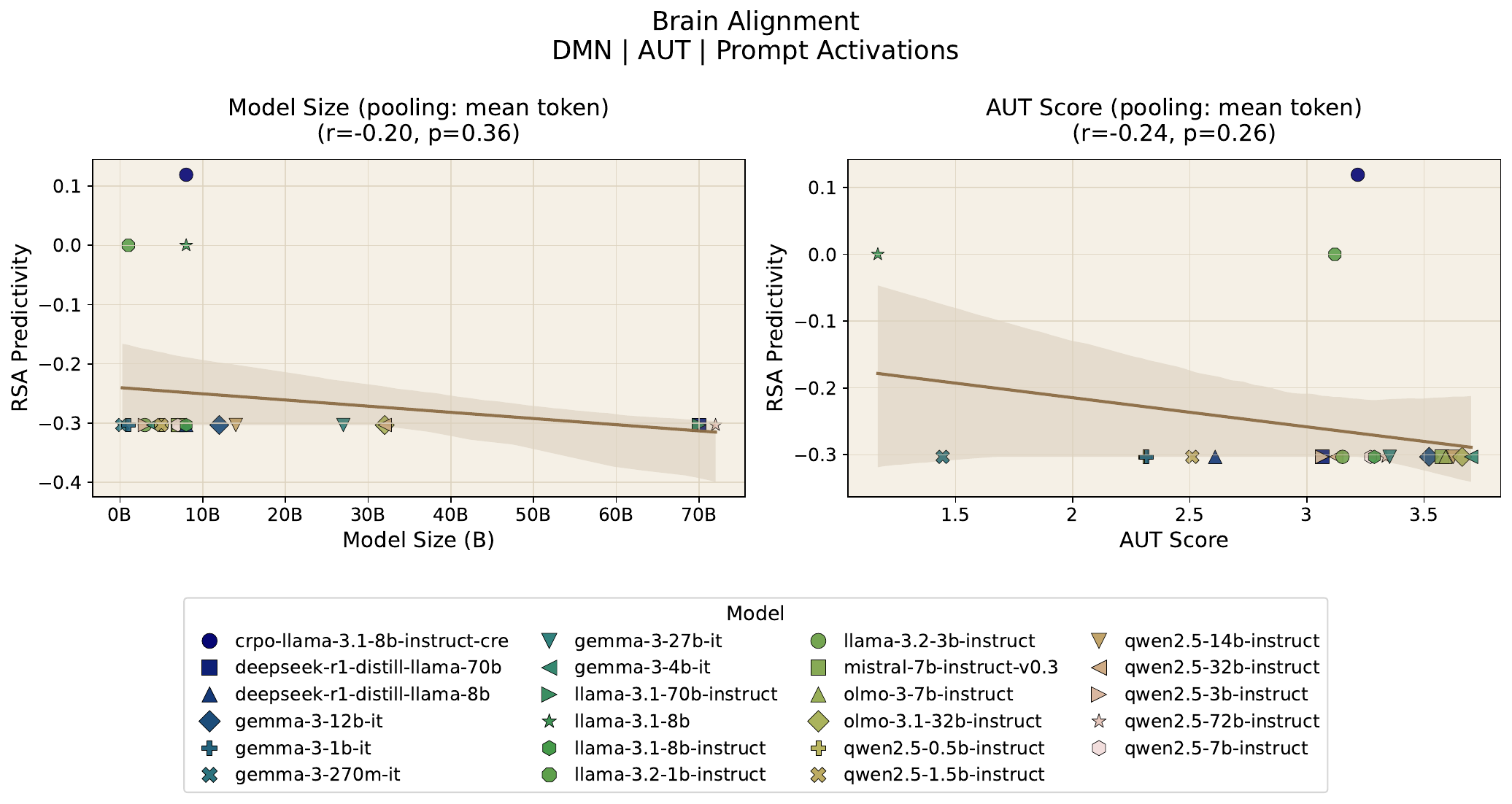}
\centering
\caption{\textbf{Default Mode Network (DMN) AUT brain alignment results by model size and task performance using empty prompt-only model activations with mean-token pooling.} $r$ and $p$ correspond to the Pearson correlation coefficient and p-value. Note that \textsc{Falcon-40B-Instruct} result is missing due to the failure to extract representations from an empty prompt.}
\label{fig:alignment-aut-yeo-dmn-empty-prompt-mt}
\end{figure}

\begin{figure}[h]
\includegraphics[width=\linewidth]{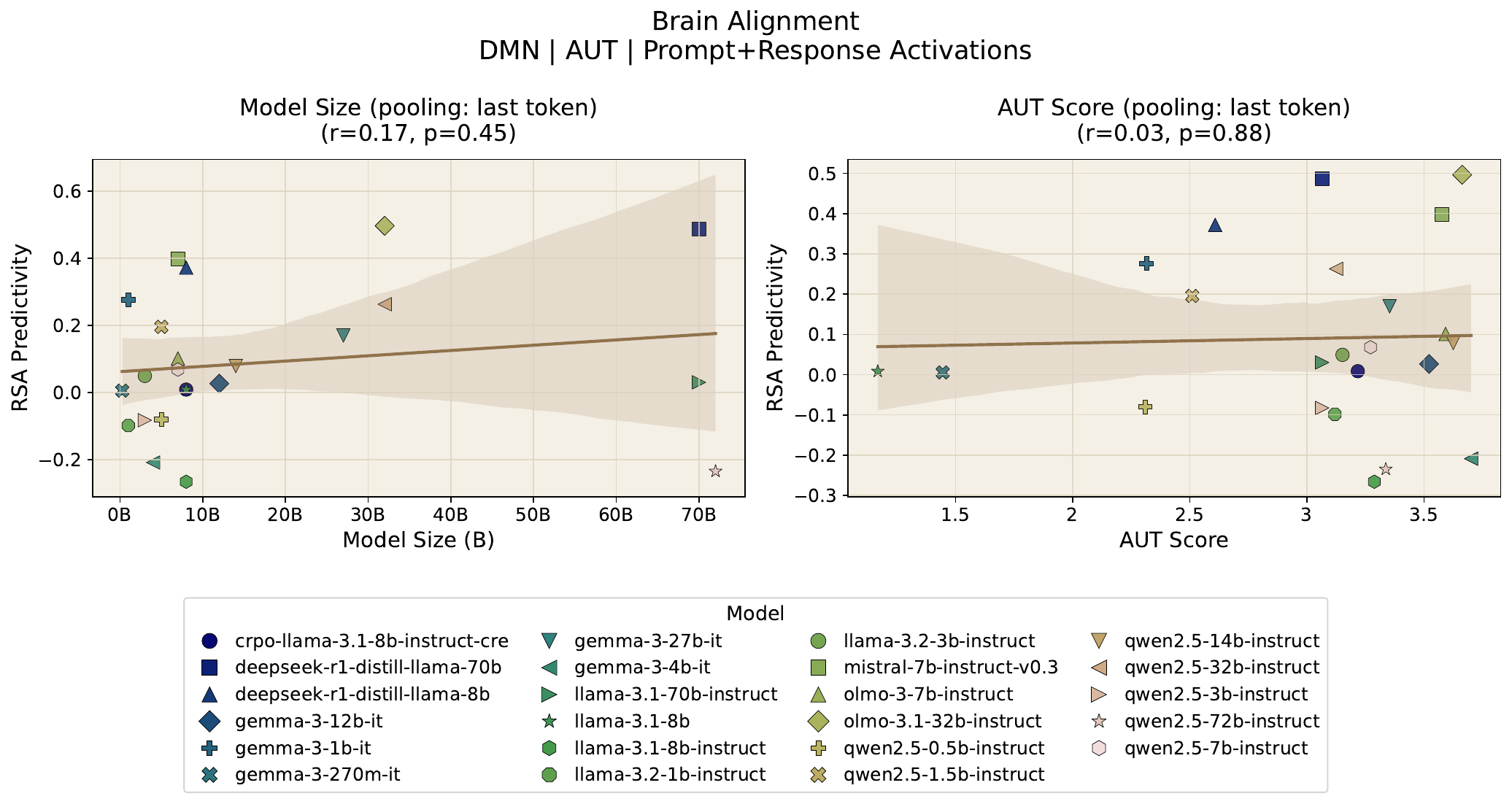}
\centering
\caption{\textbf{Default Mode Network (DMN) AUT brain alignment results by model size and task performance using empty prompt+response model activations with last-token pooling.} $r$ and $p$ correspond to the Pearson correlation coefficient and p-value. Note that \textsc{Falcon-40B-Instruct} result is missing due to the failure to extract representations from an empty prompt.}
\label{fig:alignment-aut-yeo-dmn-empty-prompt-resp-lt}
\end{figure}

\begin{figure}[h]
\includegraphics[width=\linewidth]{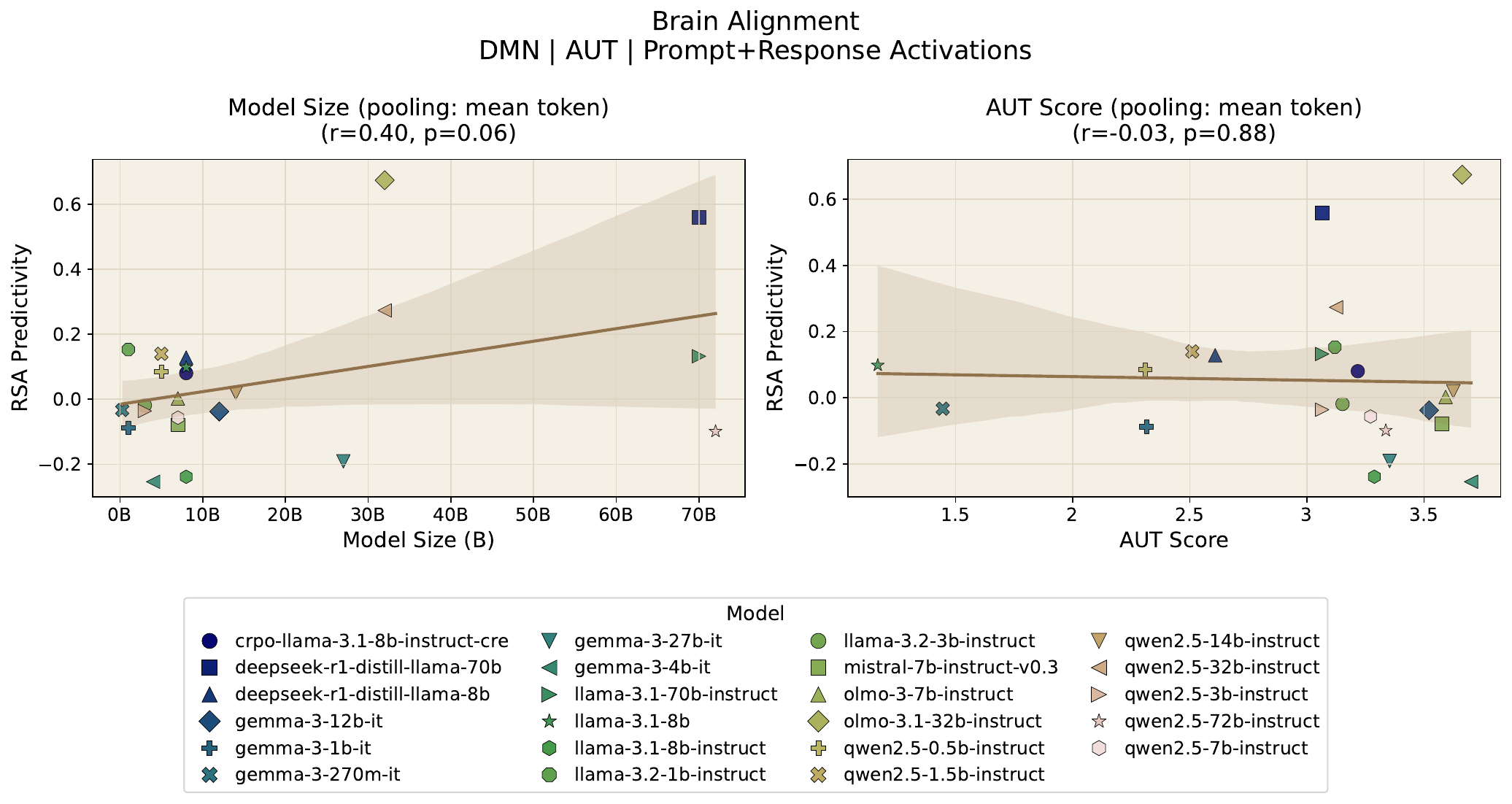}
\centering
\caption{\textbf{Default Mode Network (DMN) AUT brain alignment results by model size and task performance using empty prompt+response model activations with mean-token pooling.} $r$ and $p$ correspond to the Pearson correlation coefficient and p-value. Note that \textsc{Falcon-40B-Instruct} result is missing due to the failure to extract representations from an empty prompt.}
\label{fig:alignment-aut-yeo-dmn-empty-prompt-resp-mt}
\end{figure}

\begin{figure}[h]
\includegraphics[width=\linewidth]{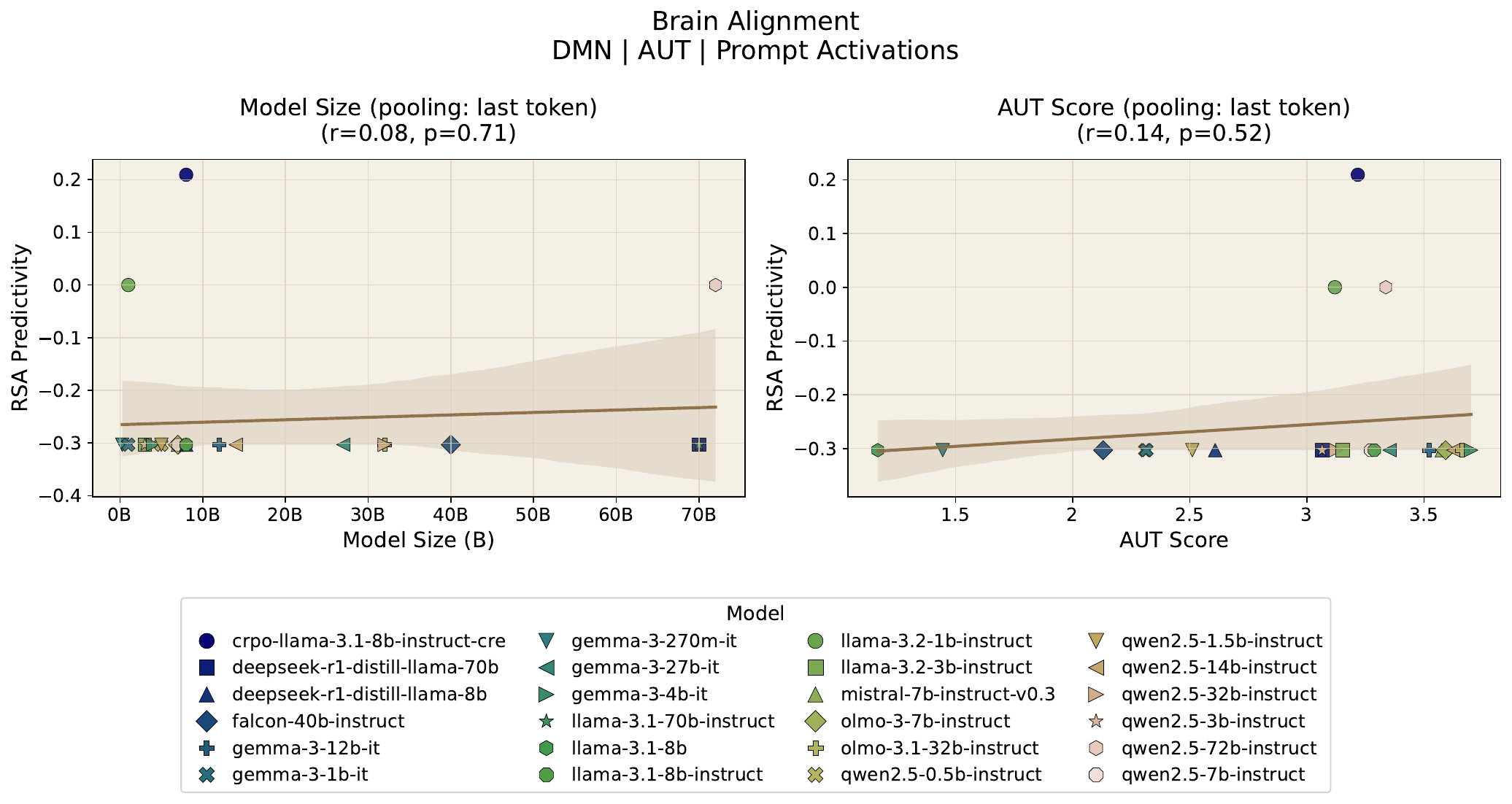}
\centering
\caption{\textbf{Default Mode Network (DMN) AUT brain alignment results by model size and task performance using placeholder prompt-only model activations with last-token pooling.} $r$ and $p$ correspond to the Pearson correlation coefficient and p-value.}
\label{fig:alignment-aut-yeo-dmn-plch-prompt-lt}
\end{figure}

\begin{figure}[h]
\includegraphics[width=\linewidth]{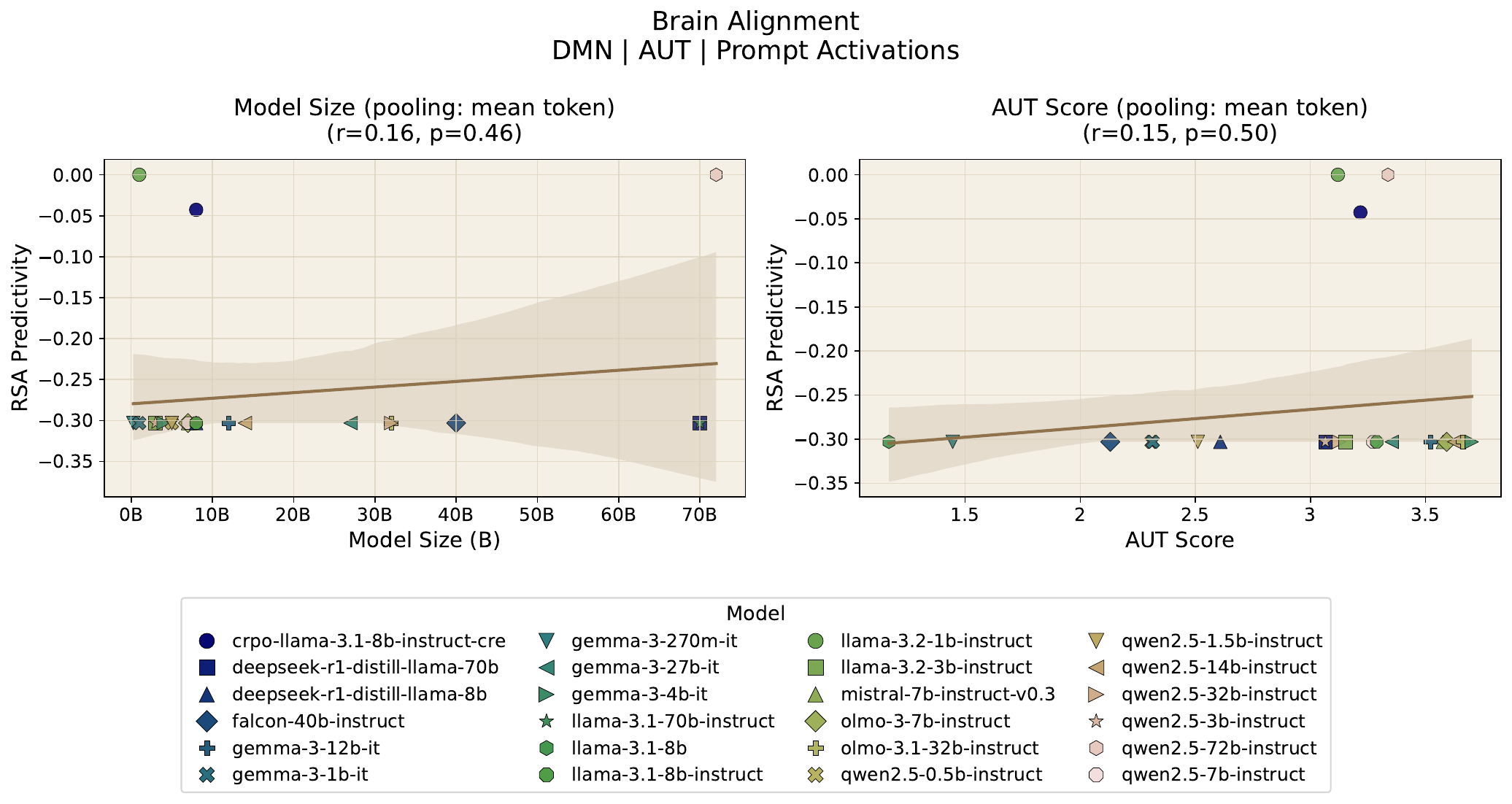}
\centering
\caption{\textbf{Default Mode Network (DMN) AUT brain alignment results by model size and task performance using placeholder prompt-only model activations with mean-token pooling.} $r$ and $p$ correspond to the Pearson correlation coefficient and p-value.}
\label{fig:alignment-aut-yeo-dmn-plch-prompt-mt}
\end{figure}

\begin{figure}[h]
\includegraphics[width=\linewidth]{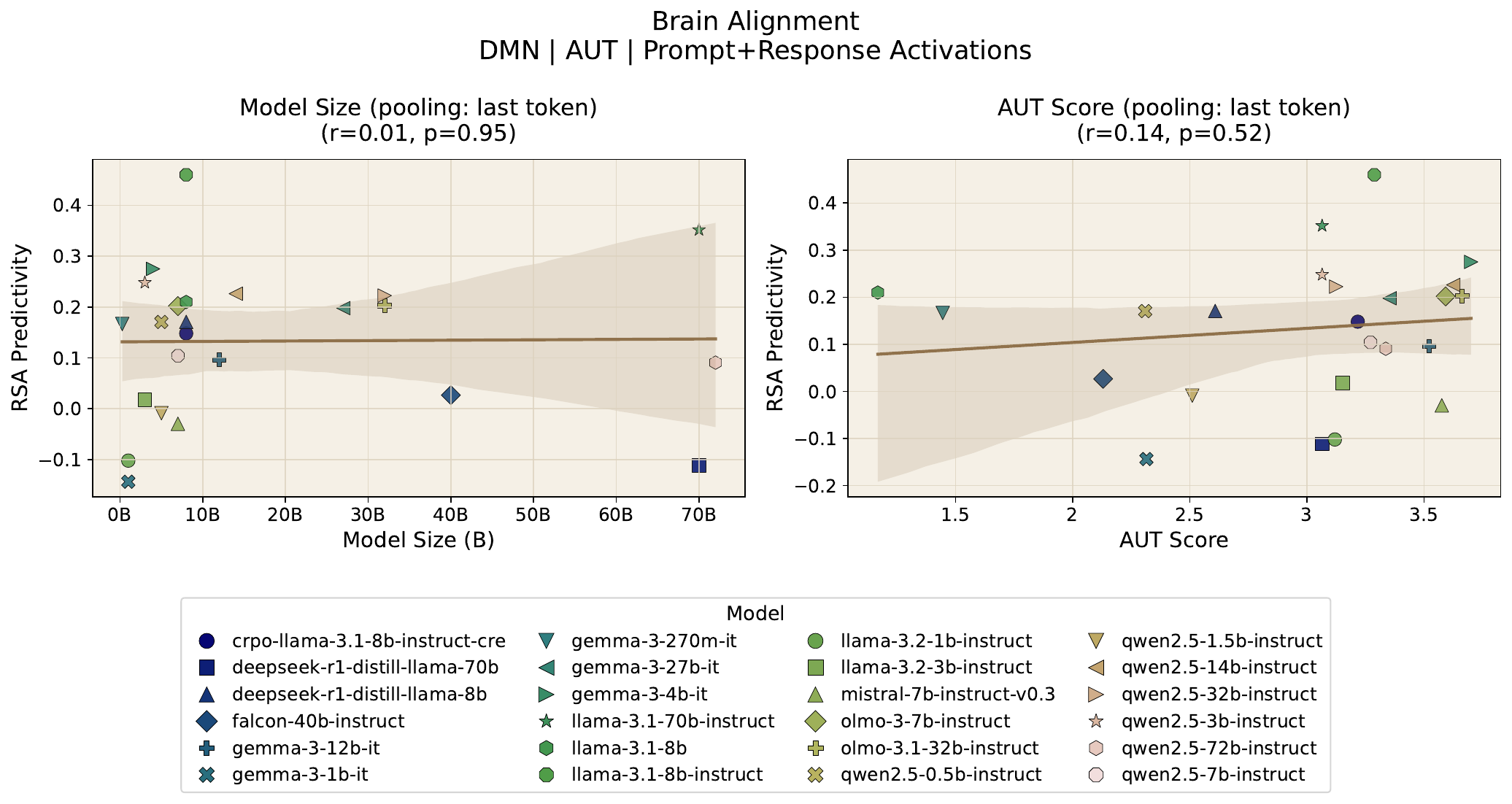}
\centering
\caption{\textbf{Default Mode Network (DMN) AUT brain alignment results by model size and task performance using placeholder prompt+response model activations with last-token pooling.} $r$ and $p$ correspond to the Pearson correlation coefficient and p-value.}
\label{fig:alignment-aut-yeo-dmn-plch-prompt-resp-lt}
\end{figure}

\begin{figure}[h]
\includegraphics[width=\linewidth]{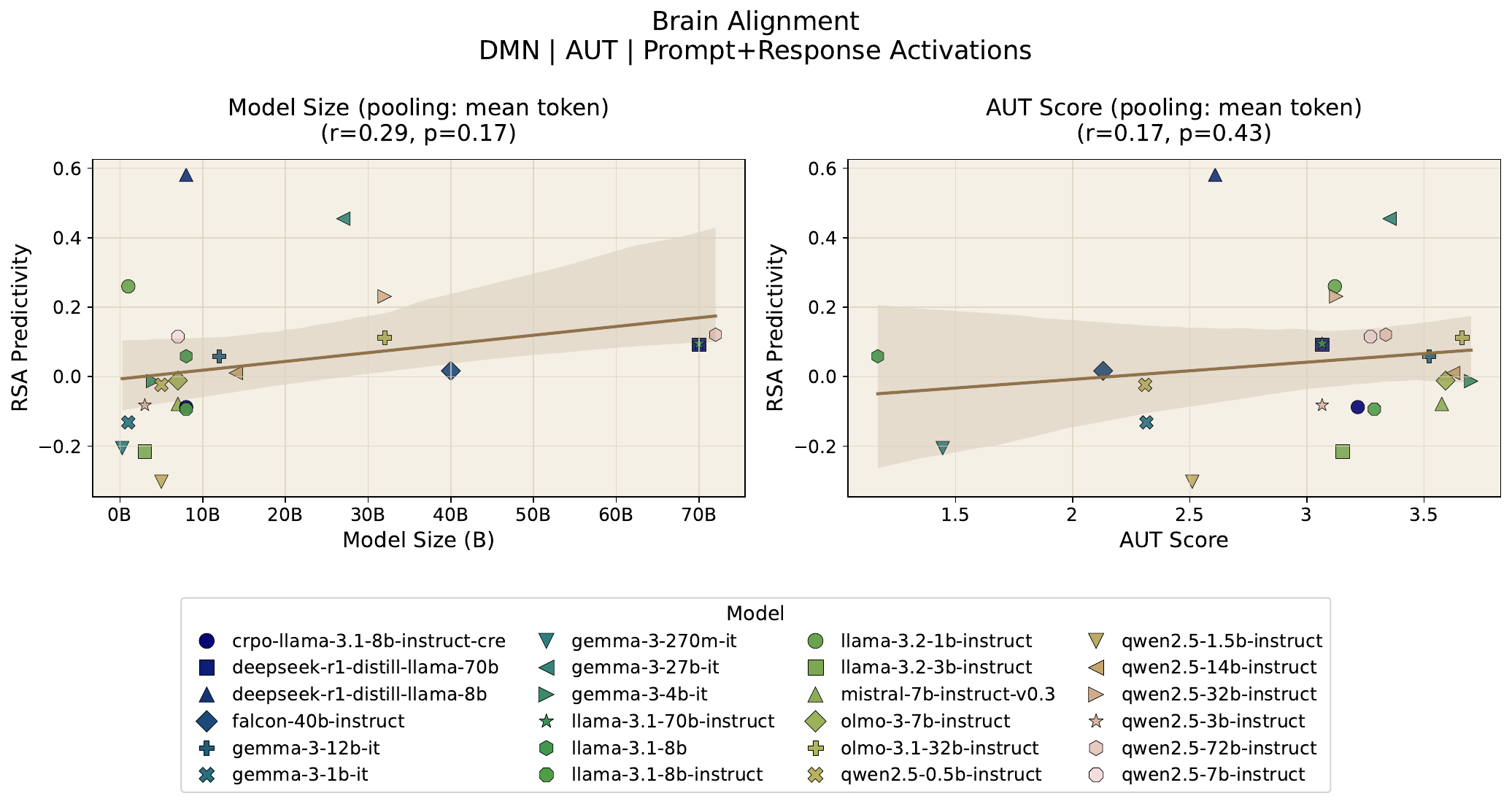}
\centering
\caption{\textbf{Default Mode Network (DMN) AUT brain alignment results by model size and task performance using placeholder prompt+response model activations with mean-token pooling.} $r$ and $p$ correspond to the Pearson correlation coefficient and p-value.}
\label{fig:alignment-aut-yeo-dmn-plch-prompt-resp-mt}
\end{figure}

\begin{figure}[t]
\includegraphics[width=\linewidth]{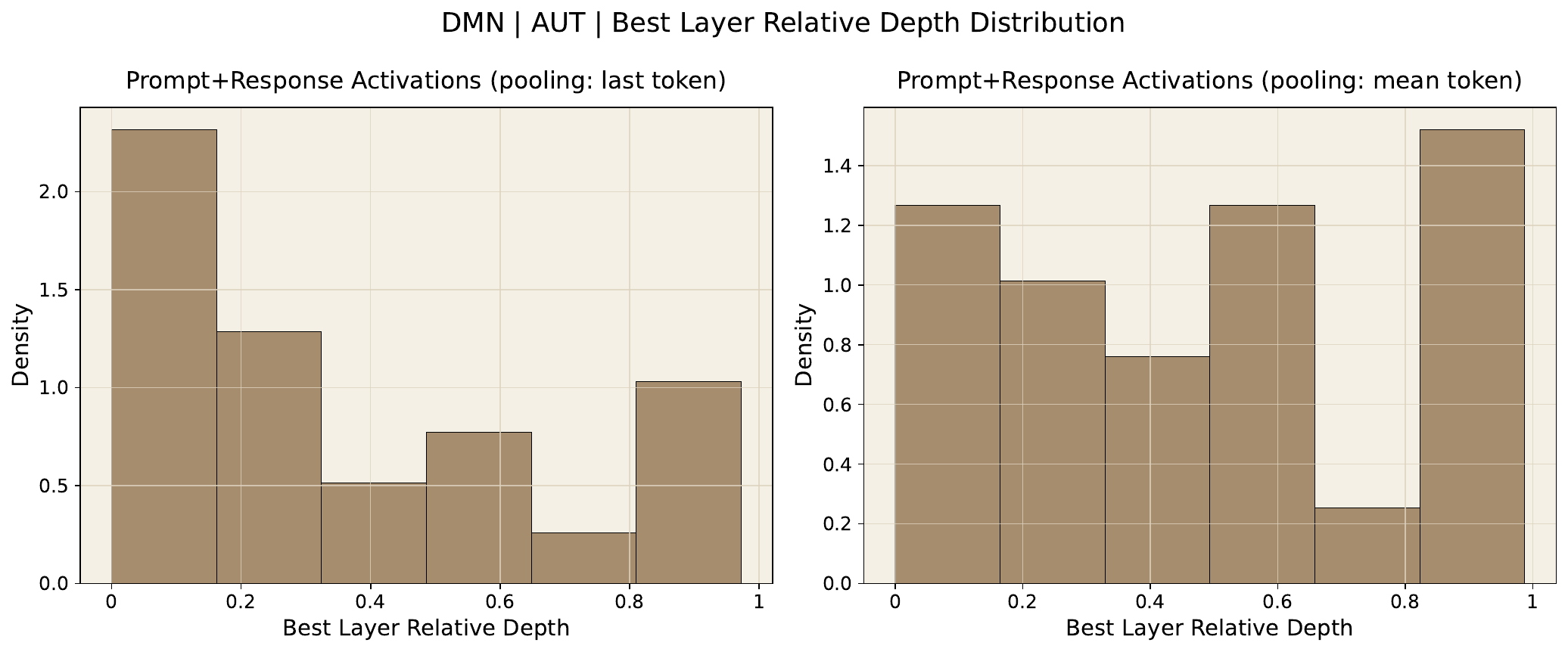}
\centering
\caption{The distribution of best model layer (measured by relative depth) for DMN AUT brain alignment on prompt+response model activations. }
\label{fig:alignment-best-layer-dist-aut-yeo-dmn-resp}
\end{figure}

\begin{figure}[t]
\includegraphics[width=\linewidth]{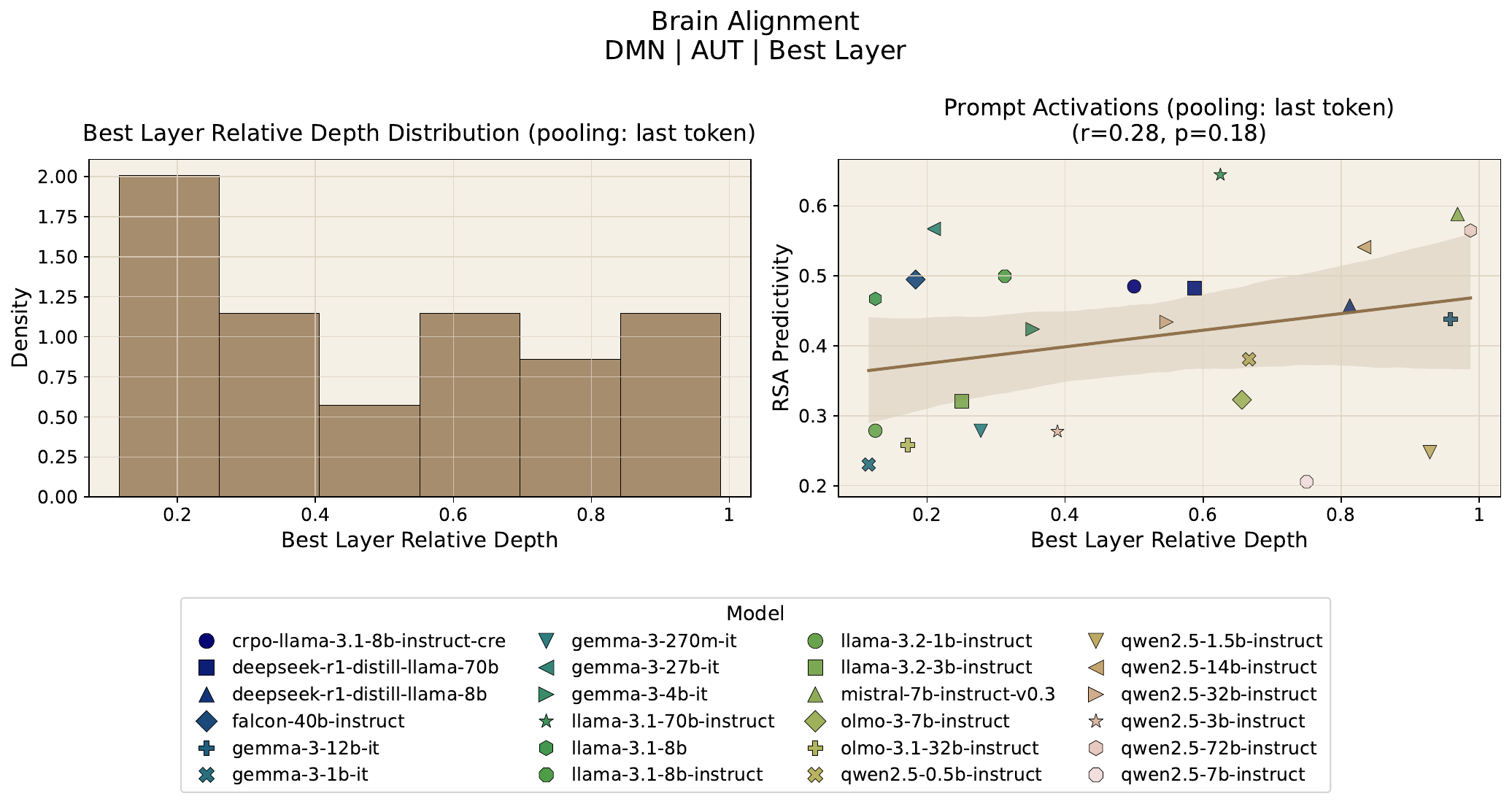}
\centering
\caption{The distribution of best model layer (measured by relative depth) and its correlation with DMN AUT brain alignment on prompt-only model activations with last-token pooling.}
\label{fig:alignment-best-layer-aut-yeo-dmn-lt}
\end{figure}

\begin{figure}[t]
\includegraphics[width=\linewidth]{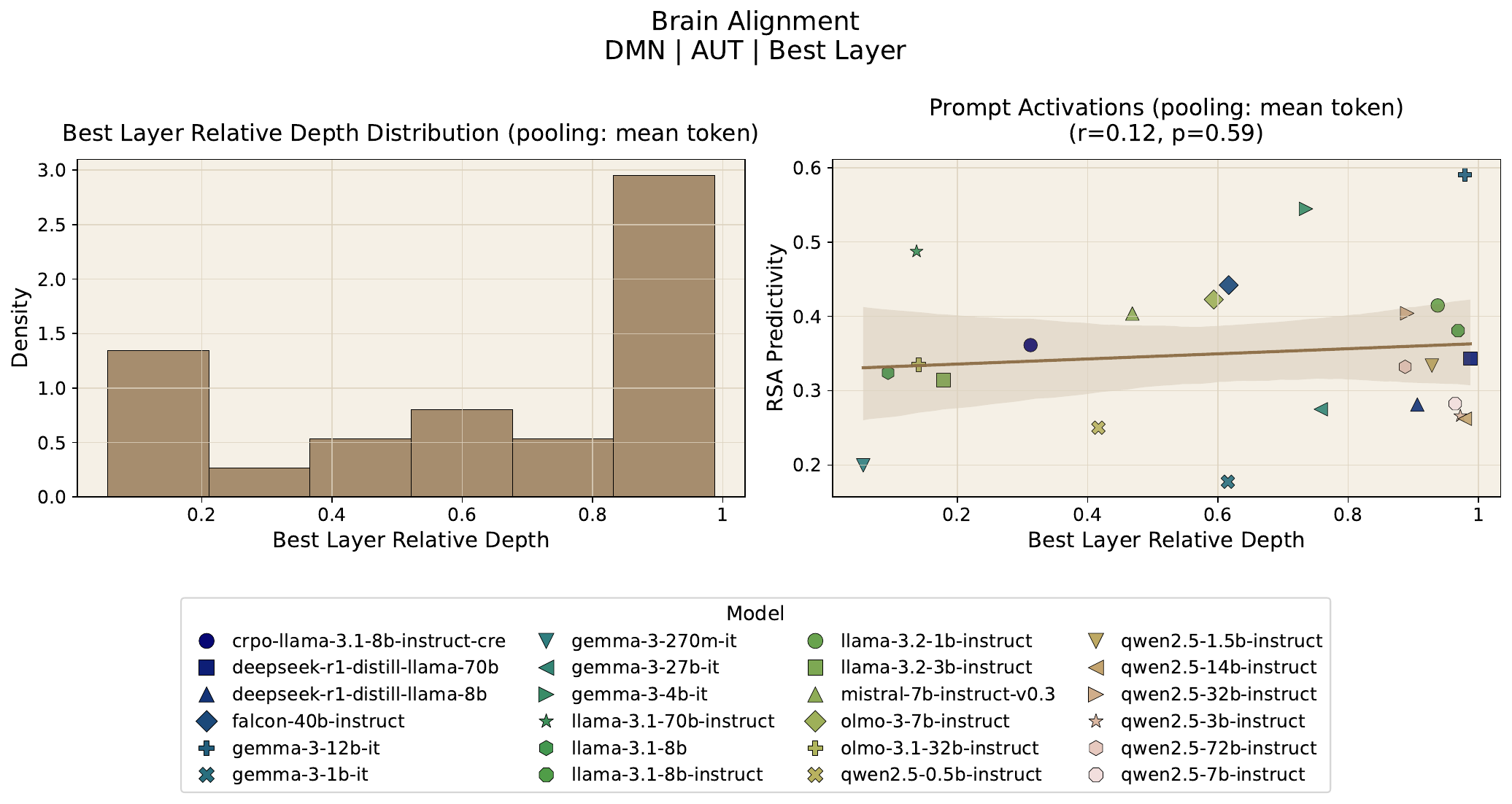}
\centering
\caption{The distribution of best model layer (measured by relative depth) and its correlation with DMN AUT brain alignment on prompt-only model activations with mean-token pooling.}
\label{fig:alignment-best-layer-aut-yeo-dmn-mt}
\end{figure}

\begin{figure}[t]
\includegraphics[width=\linewidth]{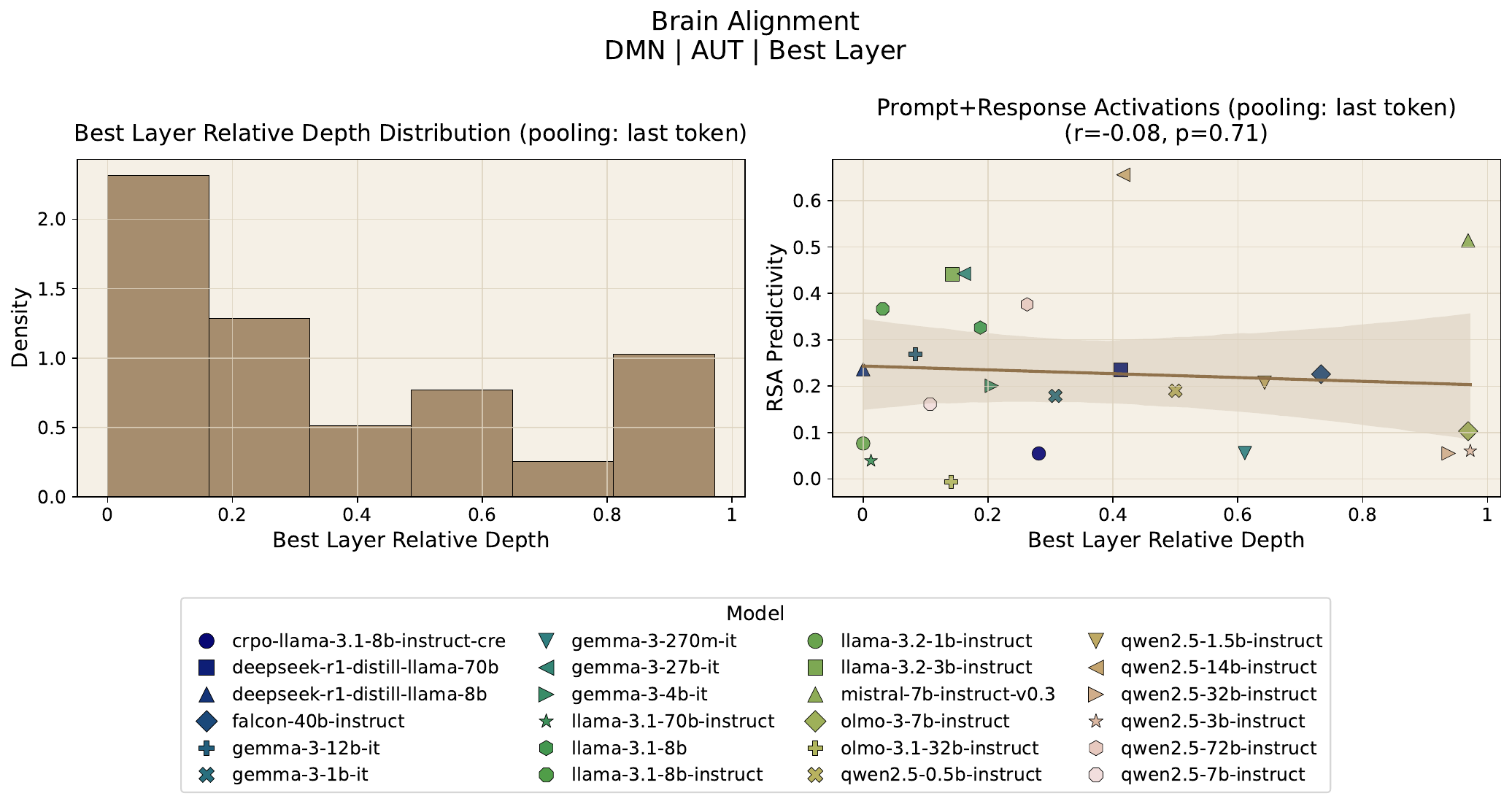}
\centering
\caption{The distribution of best model layer (measured by relative depth) and its correlation with DMN AUT brain alignment on prompt+response model activations with last-token pooling.}
\label{fig:alignment-best-layer-aut-yeo-dmn-resp-lt}
\end{figure}

\begin{figure}[t]
\includegraphics[width=\linewidth]{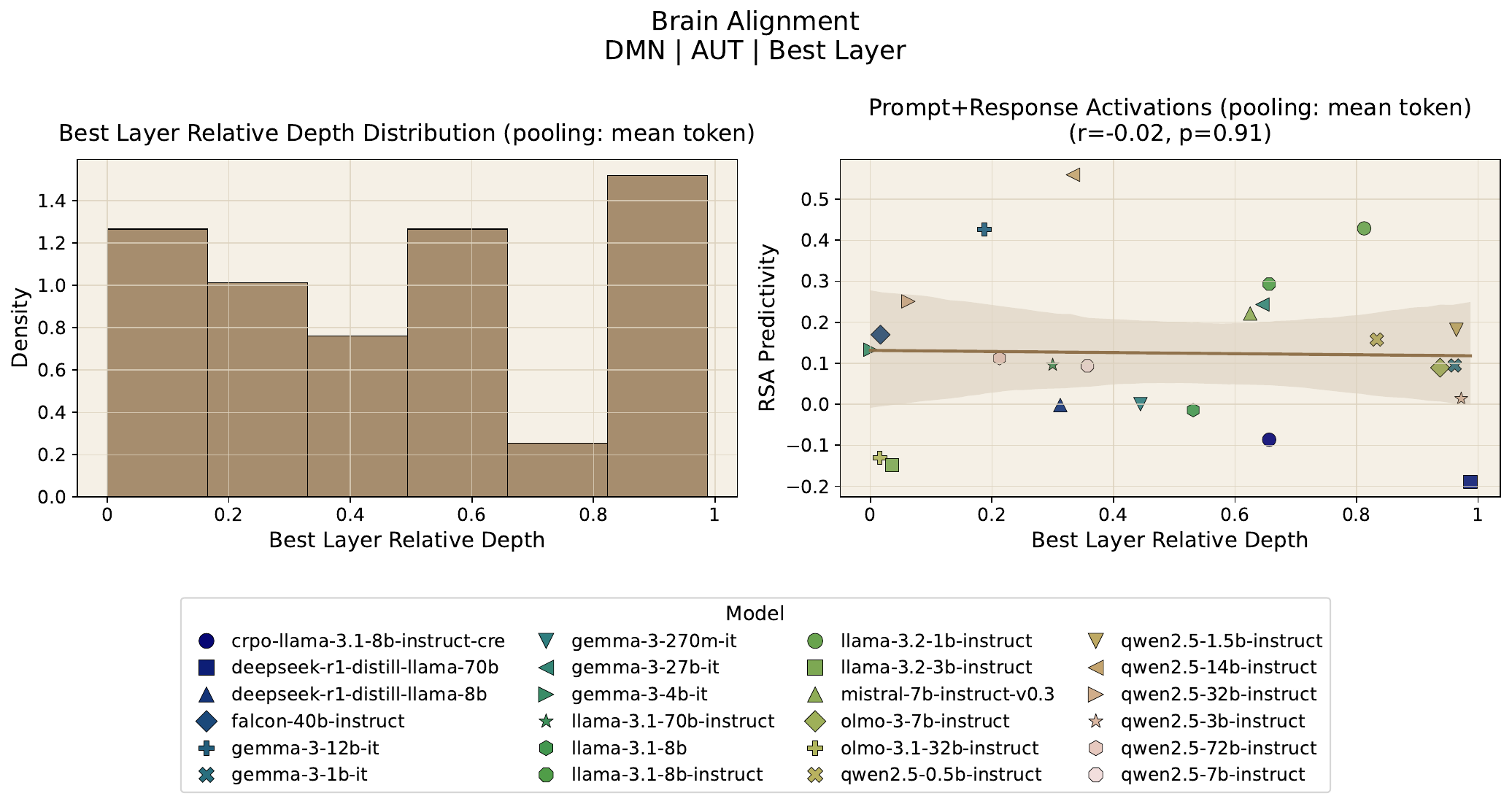}
\centering
\caption{The distribution of best model layer (measured by relative depth) and its correlation with DMN AUT brain alignment on prompt+response model activations with mean-token pooling.}
\label{fig:alignment-best-layer-aut-yeo-dmn-resp-mt}
\end{figure}

\begin{figure}[h]
\includegraphics[width=\linewidth]{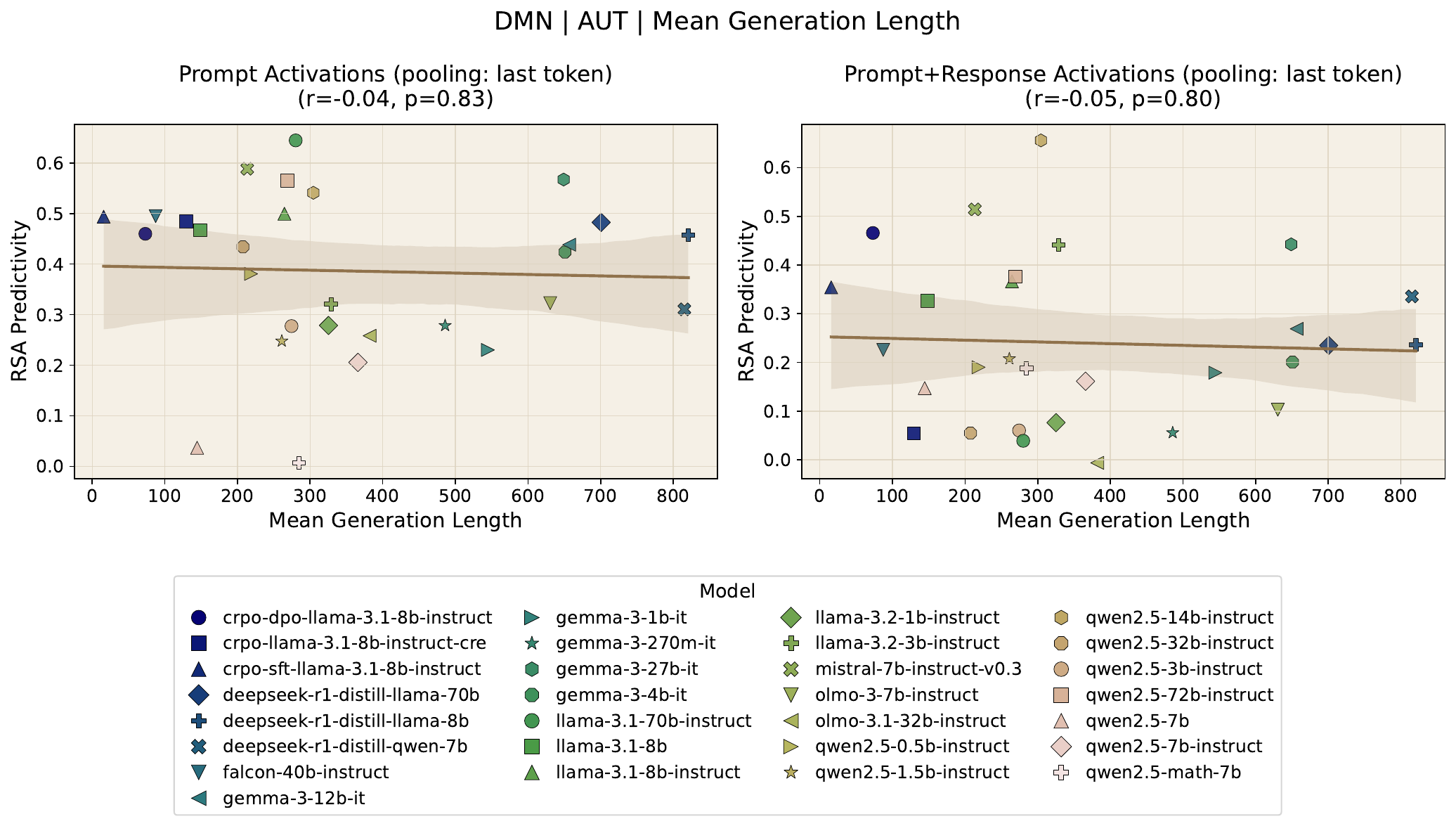}
\centering
\caption{\textbf{Default Mode Network (DMN) AUT brain alignment results by mean response generation length using prompt+response model activations with last-token pooling.} $r$ and $p$ correspond to the Pearson correlation coefficient and p-value.}
\label{fig:alignment-aut-yeo-dmn-mean-genlen-lt}
\end{figure}

\begin{figure}[h]
\includegraphics[width=\linewidth]{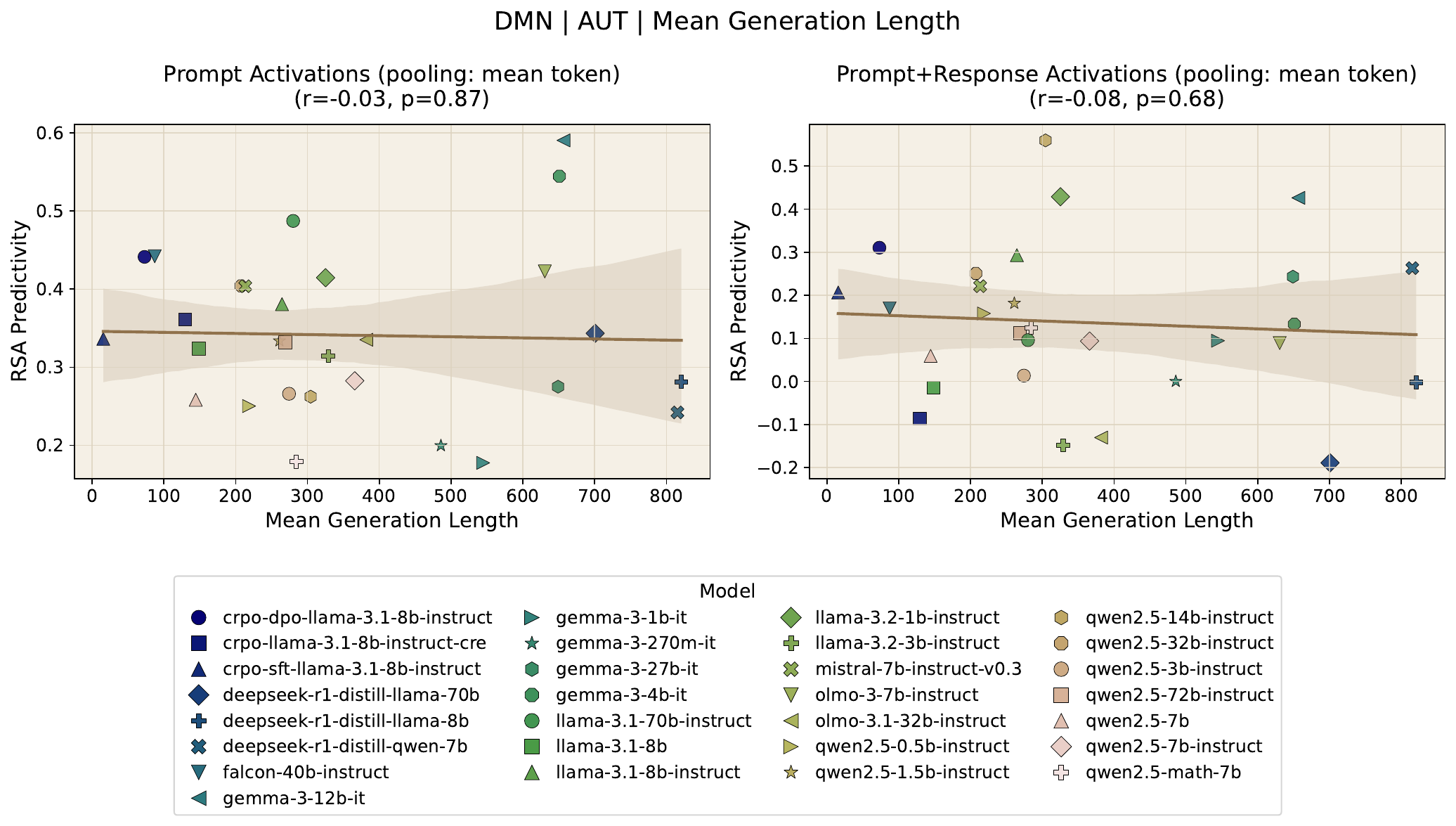}
\centering
\caption{\textbf{Default Mode Network (DMN) AUT brain alignment results by mean response generation length using prompt+response model activations with mean-token pooling.} $r$ and $p$ correspond to the Pearson correlation coefficient and p-value.}
\label{fig:alignment-aut-yeo-dmn-mean-genlen-mt}
\end{figure}

\begin{figure}[h]
\includegraphics[width=\linewidth]{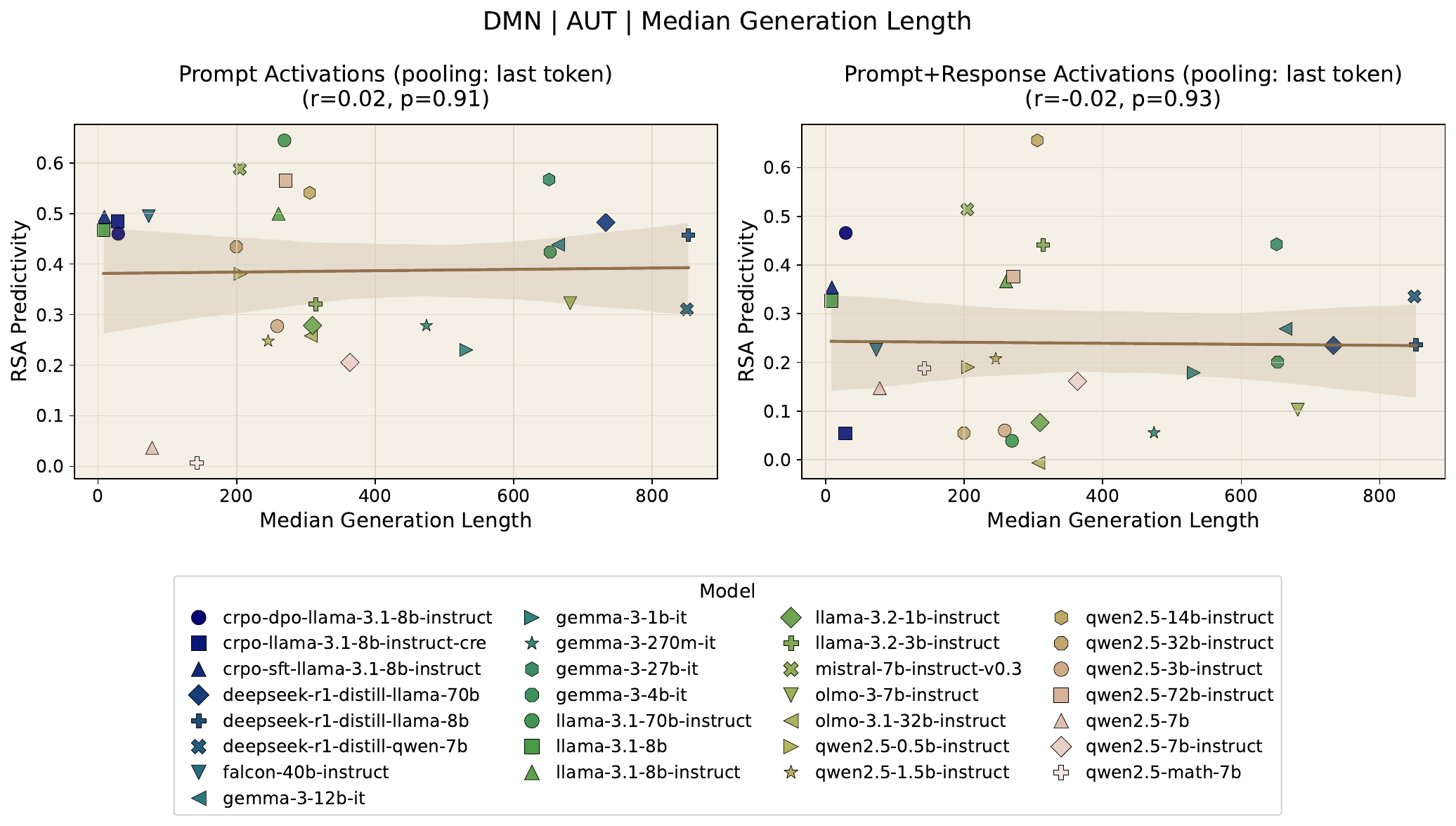}
\centering
\caption{\textbf{Default Mode Network (DMN) AUT brain alignment results by median response generation length using prompt+response model activations with last-token pooling.} $r$ and $p$ correspond to the Pearson correlation coefficient and p-value.}
\label{fig:alignment-aut-yeo-dmn-median-genlen-lt}
\end{figure}

\begin{figure}[h]
\includegraphics[width=\linewidth]{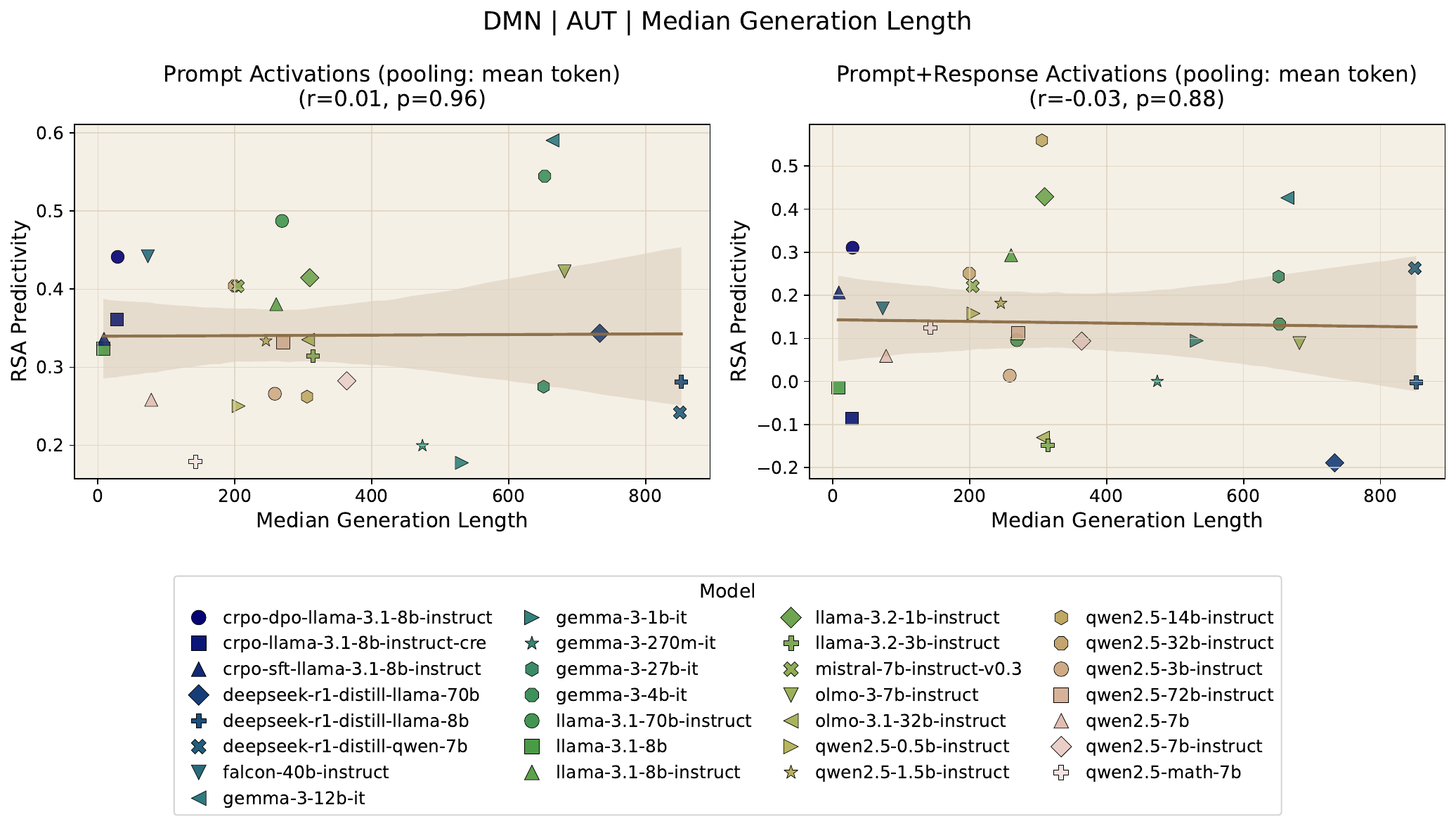}
\centering
\caption{\textbf{Default Mode Network (DMN) AUT brain alignment results by median response generation length using prompt+response model activations with mean-token pooling.} $r$ and $p$ correspond to the Pearson correlation coefficient and p-value.}
\label{fig:alignment-aut-yeo-dmn-median-genlen-mt}
\end{figure}

\begin{figure}[h]
\includegraphics[width=\linewidth]{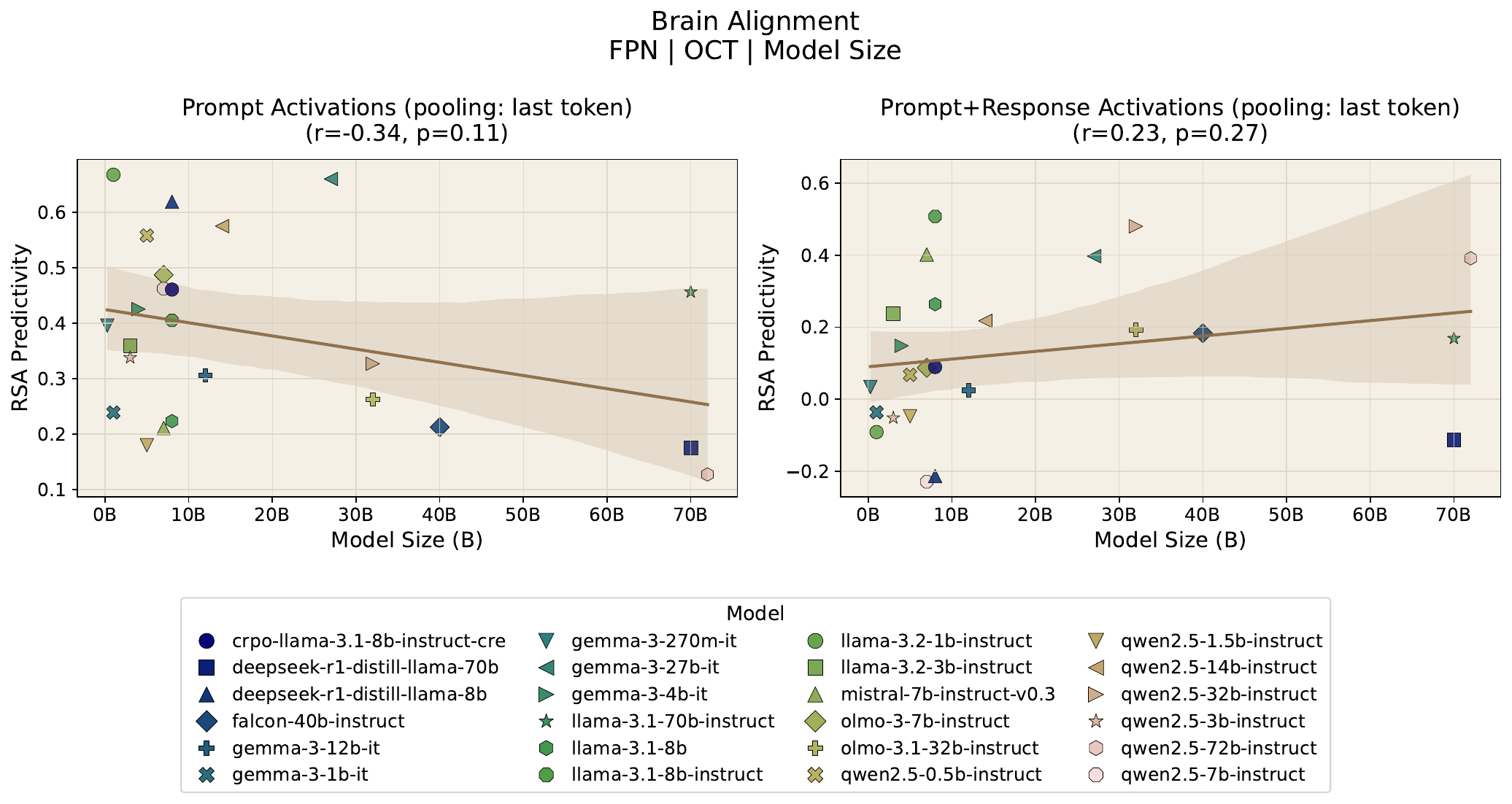}
\centering
\caption{\textbf{Frontoparietal network (FPN) OCT brain alignment results by model size with last-token pooling.} $r$ and $p$ correspond to the Pearson correlation coefficient and p-value.}
\label{fig:alignment-oct-yeo-fpn-lt}
\end{figure}

\begin{figure}[h]
\includegraphics[width=\linewidth]{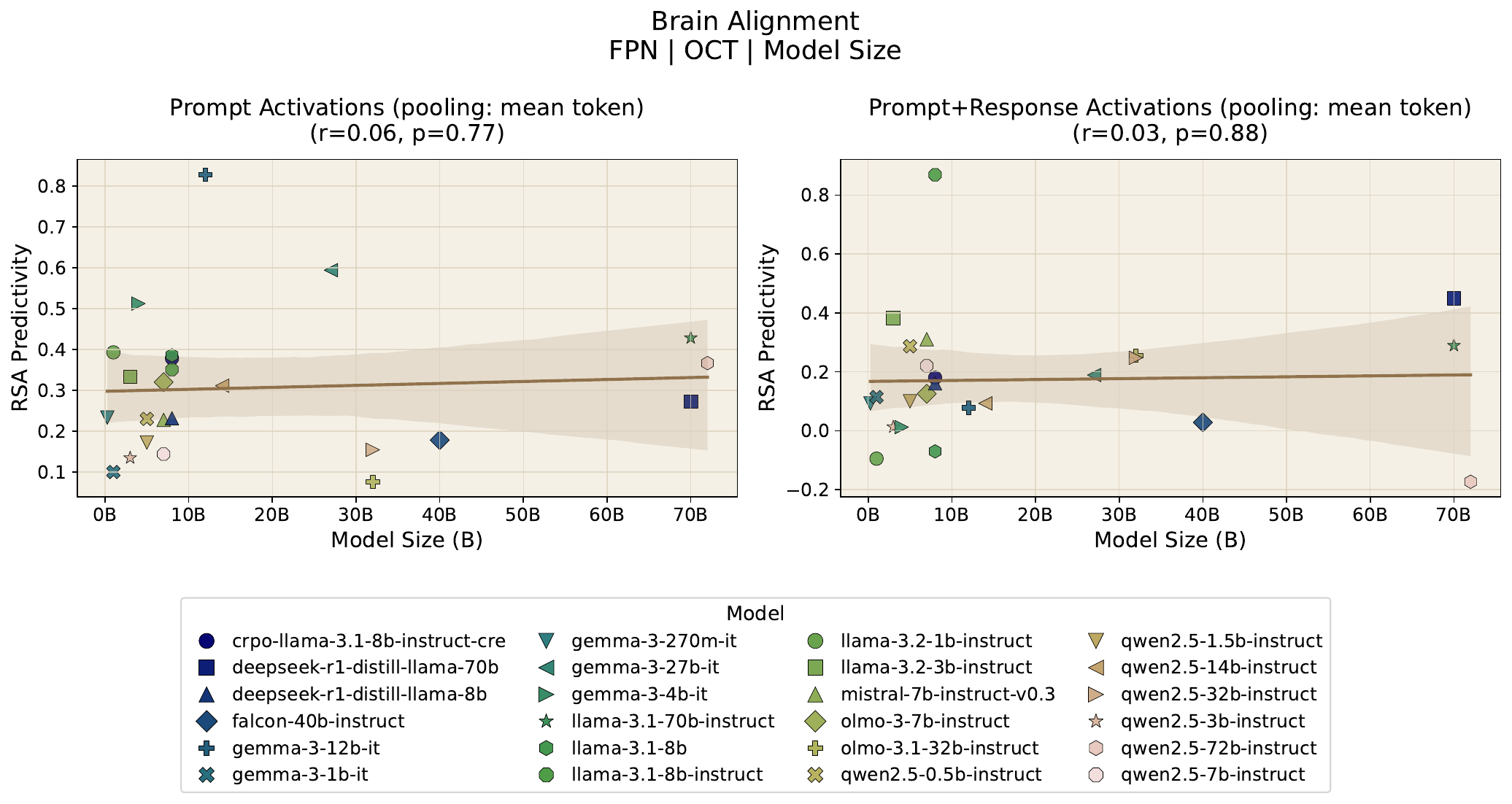}
\centering
\caption{\textbf{Frontparietal network (FPN) OCT brain alignment results by model size with mean-token pooling.} $r$ and $p$ correspond to the Pearson correlation coefficient and p-value.}
\label{fig:alignment-oct-yeo-fpn-mt}
\end{figure}

%% file: 115_cv_layer_analysis.tex
\camready{
\section{Cross-Validated Layer Selection}
\label{sec:layer-selection}

Our reported alignment scores take the maximum across layers of each model \citep{schrimpf2018brain}. Because the expected maximum over more candidates is
higher, this estimator is biased upward, and in principle the bias could scale with
layer count. Since layer count is strongly collinear with parameter count in our set
($r=0.94$), such a bias could in principle produce a size--alignment correlation with no representational basis. We therefore quantify it directly.

\paragraph{Procedure.}
For each model we hold out subjects when selecting the layer. Subjects are partitioned
into five folds; for each fold, the best layer is chosen as the argmax of the median
alignment across the remaining four folds, and alignment is then evaluated on the
held-out fold at that layer. Every subject thus receives a score from a layer selected
without them. We repeat the partition 20 times and average. Selection and evaluation
never share subjects, so the maximum operation cannot inflate the evaluated value.

\paragraph{The bias is substantial but does not scale with layer count.}
Cross-validation reduces alignment scores by a median of $0.13$ (median $33\%$ of the
naive value; range $0.007$--$0.267$), confirming that best-layer scores are upper bounds
rather than unbiased estimates. However, the reduction is \emph{negatively} related to
layer count ($r=-0.48$, $p=0.018$), the opposite of what the confound requires. What it
tracks instead is how much genuine layer structure a model has: the gap is strongly
negatively related to cross-validated alignment ($r=-0.81$, $p<0.001$). Models with a
reliable alignment peak lose almost nothing under cross-validation
(\textsc{Qwen2.5-72B-Instruct}: $0.565 \rightarrow 0.557$), whereas models without one
lose most of their apparent alignment (\textsc{Gemma-3-270M-it}:
$0.278 \rightarrow 0.011$), because in the latter case the maximum reflects noise rather
than signal.

\paragraph{The size correlation survives.}
Because the bias is larger for weakly-aligned models, which are predominantly the
smaller ones, it attenuates rather than manufactures a positive size relationship.
Recomputing the DMN--size correlation on cross-validated alignment values leaves it
essentially unchanged, and marginally stronger: $r=+0.56$ ($p=0.004$) versus $r=+0.55$
($p=0.006$) under the naive estimator. We conclude that this correlation is not an
artifact of layer selection.
}

%% file: 112_gen_lengths.tex
\begin{table}[h]
\centering
\small
\begin{tabular}{lccc}
\toprule
\textbf{Model} & \textbf{Mean} & \textbf{Median} & \textbf{Std} \\
\midrule
llama-3.2-1b-instruct & 325.27 & 309.50 & 87.26 \\
llama-3.2-3b-instruct & 328.85 & 314.00 & 91.82 \\
llama-3.1-8b & 148.34 & 8.00 & 292.45 \\
llama-3.1-8b-instruct & 264.61 & 260.50 & 61.98 \\
crpo-llama-3.1-8b-instruct-cre & 129.39 & 28.00 & 273.30 \\
crpo-sft-llama-3.1-8b-instruct & 15.71 & 9.00 & 65.67 \\
crpo-dpo-llama-3.1-8b-instruct & 73.09 & 29.00 & 178.99 \\
deepseek-r1-distill-llama-8b & 820.72 & 852.00 & 85.06 \\
deepseek-r1-distill-llama-70b & 700.77 & 733.00 & 149.62 \\
llama-3.1-70b-instruct & 280.10 & 269.00 & 81.04 \\
gemma-3-270m-it & 485.91 & 474.00 & 127.26 \\
gemma-3-1b-it & 544.76 & 531.50 & 78.84 \\
gemma-3-4b-it & 650.92 & 652.50 & 52.31 \\
gemma-3-12b-it & 656.64 & 664.00 & 40.42 \\
gemma-3-27b-it & 649.09 & 651.00 & 27.51 \\
olmo-3-7b-instruct & 630.65 & 681.50 & 153.70 \\
olmo-3.1-32b-instruct & 381.91 & 307.00 & 203.53 \\
falcon-40b-instruct & 87.33 & 73.00 & 75.00 \\
qwen2.5-0.5b-instruct & 218.70 & 205.50 & 85.65 \\
qwen2.5-1.5b-instruct & 261.00 & 245.50 & 120.88 \\
qwen2.5-3b-instruct & 274.38 & 258.50 & 98.06 \\
deepseek-r1-distill-qwen-7b & 815.35 & 850.00 & 100.00 \\
qwen2.5-math-7b & 284.51 & 142.50 & 281.23 \\
qwen2.5-7b & 144.43 & 78.00 & 161.71 \\
qwen2.5-7b-instruct & 365.88 & 363.50 & 85.31 \\
qwen2.5-14b-instruct & 304.46 & 305.50 & 56.72 \\
qwen2.5-32b-instruct & 207.46 & 199.50 & 58.42 \\
qwen2.5-72b-instruct & 268.90 & 270.50 & 100.12 \\
mistral-7b-instruct-v0.3 & 213.41 & 204.50 & 54.92 \\
\bottomrule
\end{tabular}
\caption{Generation length (in words) statistics per model over the AUT responses.}
\label{tab:generation_lengths}
\end{table}